\documentclass[11pt,letterpaper]{article}
\usepackage{graphicx}
\usepackage[utf8]{inputenc}
\usepackage[T1]{fontenc}
\input{epsf}
\usepackage{epsfig}
\usepackage{epstopdf}
\usepackage{float}
\usepackage{arcs}
\usepackage{subfigure}
\usepackage{cite}
\usepackage{mathtools}
\usepackage[dvipsnames]{xcolor}
\usepackage{tensor}
\usepackage{multirow}
\usepackage{braket}
\usepackage{amssymb,amsmath,mathrsfs,bm,feynmf,setspace} 

 \usepackage[
 colorlinks=true,
 linkcolor=darkblue,  
 urlcolor=blue,    
 filecolor=blue,     
 citecolor=pink,
 linktocpage=true,
 pdfstartview=FitV,
 bookmarksopen=true  
 ]{hyperref}
 \definecolor{darkblue}{rgb}{0.2, 0, 0.8} 
 \definecolor{pink}{rgb}{1, 0.15, 0.15}  

\usepackage{amssymb}
\usepackage{pifont}
\usepackage{booktabs, makecell, tabularx}
\setcellgapes{4pt}
\usepackage{siunitx}

\usepackage{amsmath}
\newcommand{\be}{\begin{equation}}
\newcommand{\ee}{\end{equation}}

\usepackage{tikz}
\usetikzlibrary{calc}   
\usetikzlibrary{topaths}
\usepackage{pgfplots}
\newcommand{\labell}[1]{\label{#1}}

\newcommand{\bea}{\begin{eqnarray}}
\newcommand{\eea}{\end{eqnarray}}
\newcommand{\ba}{\begin{eqnarray}}
\newcommand{\ea}{\end{eqnarray}}

\newcommand{\beq}{\begin{equation}}
\newcommand{\eeq}{\end{equation}}
\newcommand{\beqa}{\begin{eqnarray}}
\newcommand{\eeqa}{\end{eqnarray}}
\newcommand{\beqar}{\begin{eqnarray*}}
\newcommand{\eeqar}{\end{eqnarray*}}

\newcommand{\ssc}{\scriptscriptstyle}
\newcommand{\eg}{{\it e.g.,}\ }
\newcommand{\ie}{{\it i.e.,}\ }

\newcommand{\fin}{f_\infty}

\newcommand{\req}[1]{(\ref{#1})} 

\newcommand{\rh}{{r_{b}}}
\begin{document}  
\begin{titlepage}
\vspace*{2.3cm}

\begin{center}
{\LARGE \bf NUTs and bolts beyond Lovelock} \\

\vspace*{1.2cm}
{\bf Pablo Bueno$^{1}$, Pablo A. Cano$^{2}$, Robie A. Hennigar$^{3}$ and Robert B. Mann$^{3}$}

\bigskip

$^{1}$Instituut voor Theoretische Fysica, KU Leuven,\\
Celestijnenlaan 200D, B-3001 Leuven, Belgium
\bigskip

$^{2}$Instituto de F\'isica Te\'orica UAM/CSIC,\\
C/ Nicol\'as Cabrera, 13-15, C.U. Cantoblanco, 28049 Madrid, Spain
\bigskip

$^{3}$Department of Physics and Astronomy, University of Waterloo,\\
Waterloo, Ontario, Canada, N2L 3G1
\bigskip
\vspace*{0.2cm}

pablo@itf.fys.kuleuven.be, pablo.cano@uam.es, rhenniga@uwaterloo.ca, rbmann@uwaterloo.ca  \\
\end{center}
\vspace*{0.1cm}

\begin{abstract}  
We construct a plethora of new Euclidean AdS-Taub-NUT and bolt solutions of several four- and six-dimensional  higher-curvature theories of gravity  with various base spaces $\mathcal{B}$. In $D=4$, we consider Einsteinian cubic gravity, for which we construct solutions with $\mathcal{B}=\mathbb{S}^2,\mathbb{T}^2$. These represent the first generalizations of the Einstein gravity Taub-NUT/bolt solutions for any higher-curvature theory  in four dimensions.
 In $D=6$, we show that no new solutions are allowed for any Generalized quasi-topological gravity at cubic order. They exist however when we consider quartic Quasi-topological and Generalized quasi-topological terms, for which we construct new solutions with $\mathcal{B}=\mathbb{CP}^2,\mathbb{S}^2\times\mathbb{S}^2,\mathbb{S}^2\times\mathbb{T}^2,\mathbb{T}^2\times\mathbb{T}^2$.
 In all cases, the solutions are characterized by a single metric function, and they reduce to the corresponding ones
  in Einstein gravity   when the  higher-curvature couplings are set to zero.  While the explicit profiles must be constructed numerically (except for a few cases), we  obtain fully analytic expressions for the thermodynamic properties of all solutions. The new solutions present important differences with respect to Einstein gravity, including regular bolts for arbitrary values of the NUT charge, critical points, and re-entrant phase transitions. 
\end{abstract}

\end{titlepage}

\setcounter{tocdepth}{2}
{\small
\setlength\parskip{-0.5mm} 
\tableofcontents
}
%
%
%
%
%

\section{Introduction}\labell{Introduction}

 Higher curvature theories of gravity are proving to be increasingly useful in providing us with knowledge connected with fundamental questions in gravitational physics. Quadratic curvature theories are renormalizable
\cite{Stelle:1976gc}, and it has long been realized that corrections
to the Einstein-Hilbert action that go like powers in the curvature generically arise as low energy corrections from a UV complete theory of gravity, e.g. string theory~\cite{Zwiebach:1985uq}.  Furthermore, in the context of the AdS/CFT correspondence conjecture,   higher curvature corrections correspond to $1/N$ corrections in the large $N$ limit of the dual CFT, allowing investigation of a much broader class of CFTs~\cite{Buchel:2008vz, Hofman:2009ug,Buchel:2009sk, Camanho:2009hu,deBoer:2009gx,Myers:2010jv}. 

 Imposing the condition that  higher-curvature gravity be ghost-free (even if only on a constant-curvature background) severely limits the number of sensible theories available, since they generically contain such excitations. The most well-known class of higher curvature theories that are ghost free on maximally symmetric backgrounds is \textit{Lovelock gravity}~\cite{Lovelock:1971yv}, in which the dimensionally extended Euler densities are included in the gravitational action. In $D \ge 2k+1$, the $k^{th}$ order Euler density is non-trivial;
in this sense Lovelock gravity is a natural extension of Einstein gravity, and is the unique higher curvature theory maintaining second order field equations for the metric.

However  other ghost-free higher  curvature theories exist,  and one class of particular interest has recently  been identified \cite{PabloPablo,Hennigar:2016gkm,PabloPablo2,Hennigar:2017ego,PabloPablo3,Ahmed:2017jod,PabloPablo4,Feng:2017tev}.
The Lagrangian of this new family can be written schematically as
\begin{equation}
\mathcal{L}(g^{ab},R_{cdef})=\frac{1}{16\pi G}\left[\frac{(D-1)(D-2)}{L^2}+R+\sum_{n=2} \mu_n L^{2(n-1)}\mathfrak{R}_{(n)}\right]\, ,
\end{equation}
where $L$ is some length scale, $\mu_n$ are independent dimensionless couplings, and $\mathfrak{R}_{(n)}$ are certain  linear combinations of order-$n$ densities constructed from contractions of the metric and the Riemann tensor. The theories are characterized by the following properties:  i) they have second-order equations of motion when linearized around any maximally symmetric spacetime, \ie just like Einstein gravity, they only propagate a massless and traceless graviton on such backgrounds;
 ii) they possess a continuous and well-defined Einstein gravity limit corresponding to  $\mu_n\rightarrow 0$ for all $\mu_n$; iii) they admit generalizations of the Schwarzschild-(A)dS black hole --- so they reduce to it in the Einstein gravity limit --- characterized by a single function, 
\begin{equation}\label{FFbh}
ds^2=-f(r) dt^2+\frac{dr^2}{f(r)}+r^2 d\Sigma_{(D-2)}^2\, ,
\end{equation}
where $d\Sigma_{(D-2)}^2$ is the metric of the horizon cross sections (not necessarily spherical); iv) the function $f(r)$ is determined from (at most) a second-order differential equation which, for a fixed set of $\mu_{n}$, admits a unique black-hole solution completely characterized by its ADM energy and which, at least in the spherically symmetric case, describes the exterior field of matter distributions with that symmetry \cite{PabloPablo3}; v) the thermodynamic properties of such black holes can be obtained from a system of algebraic equations with no free parameters.

The above class of theories can be subdivided if one considers more restrictive criteria. In particular, replacing i) by the requirement that the full non-linear equations of the theory are second order selects the Lovelock family \cite{Lovelock1,Lovelock:1971yv,Concha:2017nca}. Keeping i) as it is, but replacing instead iv) by the requirement that $f(r)$ is determined by an algebraic equation, selects a more general class of theories, known as Quasi-topological gravities \cite{Quasi,Quasi2,Dehghani:2011vu,Cisterna:2017umf} (which of course include Lovelock as particular cases). More generally, the family of higher-curvature gravities satisfying i)-v) is larger than the Quasi-topological one. The missing theories posses black holes whose  metric function is determined by second-order differential equations, and the full set has been coined ``Generalized Quasi-topological Gravity'' (GQTG) in \cite{Hennigar:2017ego}. 

One intriguing feature of these new theories, in contradistinction to those belonging to the Quasi-topological subset (except for Einstein gravity itself) is that some of them are nontrivial in $D=4$. The simplest possible case of that kind, and the first to be identified, corresponds to a single additional cubic term and goes by the name of four-dimensional Einsteinian cubic gravity\footnote{The $D$-dimensional version of ECG was originally obtained as the most general cubic theory defined in a dimension-independent way --- \ie so that the relative coefficients of the cubic invariants involved do not depend on $D$ --- possessing second-order linearized equations on general maximally symmetric backgrounds \cite{PabloPablo}. However, it is only for $D=4$ that ECG additionally satisfies properties ii)-v) \cite{Hennigar:2016gkm,PabloPablo2}.} (ECG) \cite{PabloPablo}, whose action is given in \req{ECG} below. Many examples of GQTG theories in general dimensions have now been constructed, and their respective black hole solutions studied and characterized \cite{Hennigar:2016gkm,PabloPablo2,Hennigar:2017ego,PabloPablo3,Ahmed:2017jod,PabloPablo4,Hennigar:2017umz,Feng:2017tev,Hennigar:2018hza,Colleaux:2017ibe,HoloECG,Peng:2018vbe,Dey:2016pei}.
 
A different class of exact static solutions of Einstein gravity is given by the Taub-NUT family. The Euclidean section of the corresponding metrics can be written as
\begin{equation}\label{FFnut}
ds^2=V_{\mathcal{B}}(r) (d\tau+ n A_{\mathcal{B}})^2+\frac{dr^2}{V_{\mathcal{B}}(r)}+(r^2-n^2)d\sigma_{\mathcal{B}}^2\, ,
\end{equation}
which, in even dimensions, can be understood as $U(1)$ fibrations over $(D-2)$-dimensional K\"ahler-Einstein base spaces $\mathcal{B}$ with metric $g_{\mathcal{B}}$. In \req{FFnut}, $\tau$ is a periodic coordinate parametrizing the $\mathbb{S}^1$, and $J=dA_{\mathcal{B}}$ is the K\"ahler form on ${\mathcal{B}}$. The non-triviality of the fibration is controlled by the presence of a non-zero parameter $n$, customarily called ``NUT charge''. Depending on the dimension of the set of fixed points of the $U(1)$ isometry --- namely those for which $V_{\mathcal{B}}(r)=0$ --- the solution is said to be a ``NUT'' or a ``bolt''. Taub-bolt solutions are characterized by $(D-2)$-dimensional fixed-point sets, whereas smaller dimensionalities give rise to Taub-NUT solutions.  

 It has been known for some time that NUT-charged solutions exist
in Lovelock gravity \cite{Dehghani:2005zm,Dehghani:2006aa,Dotti:2007az,Hendi:2008wq}. A  broad understanding of their thermodynamics remains an ongoing subject of investigation \cite{Clarkson:2002uj,Mann:2004mi,Dehghani:2006dk,Ghezelbash:2007kw,Ghezelbash:2008zz,KhodamMohammadi:2008fh,Nashed:2012ud,Johnson:2014xza,Johnson:2014pwa,Lee:2014tma,Nashed:2015pga,Pradhan:2015jia,Lee:2015wua}
since it was realized that their contribution to the entropy does not obey the area law, even in Einstein gravity
\cite{Hawking:1998ct,Garfinkle:2000ms}.  A recent review of Taub-NUT spacetimes and their symmetries has appeared \cite{Frolov:2017kze}.

On general grounds, one expects two independent functions to be required to describe Taub-NUT solutions in general higher-curvature gravities.
The relevant observation for us is that both for Einstein gravity and Gauss-Bonnet,  all Taub-NUT solutions are characterized by a single function for each choice of base space.
 This is analogous to the situation encountered for static black-hole solutions.  One is then naturally led to wonder whether the rest of GQTG theories also admit generalizations of the Einstein gravity Taub-NUT solutions characterized by a single function, $V_{\mathcal{B}}(r)$, just like they admit generalizations of the Schwarzschild black hole with that property. The answer turns out to be yes and, as we show here, a plethora of new Taub-NUT and Taub-bolt solutions of the form \req{FFnut} can be constructed in various dimensions and for different base spaces.

Each choice of $D$, base space and Taub-NUT class has its peculiarities, but some aspects of our construction can be explained in general. First of all, and in a similar fashion to what occurs with black holes, inserting ansatz \req{FFnut} in the equations of motion of the corresponding GQTG theory we will observe that, whenever the corresponding theory admits  Taub-NUT solutions of that form\footnote{Not all GQTG theories will admit all possible Taub-NUT solutions of the form \req{FFnut} for all possible base spaces.}, they reduce to a single  third-order equation for $V_{\mathcal{B}}(r)$. Interestingly, this equation always admits a simple integrable factor which allows us to integrate it once. Hence, in each case we are left with a single equation of the form
\begin{equation}\label{mastere}
\mathcal{E}_{\mathcal{B}}[V_{\mathcal{B}},V'_{\mathcal{B}},V''_{\mathcal{B}},r]=C\, ,
\end{equation}
where $C$ is an integration constant related to the ADM energy of the solution. 
Also, in analogy with the black-hole case, one of the integration constants of this second-order differential equation will always be fixed by imposing the solutions to be locally asymptotically AdS. With regards to the second, recall that the defining properties characterizing $V_{\mathcal{B}}(r)$ are, respectively,
\begin{align}\label{smooth}
&V_{\mathcal{B}}(r)|_{r=n}=0\, , \quad V_{\mathcal{B}}'(r)|_{r=n}=4\pi/ \beta_{\tau}\,, \quad \text{for NUT}\, ,\\ \notag
&V_{\mathcal{B}}(r)|_{r=r_b}=0\, , \quad V_{\mathcal{B}}'(r)|_{r=r_b}=4\pi/ \beta_{\tau}\,, \quad \text{for bolt}\, ,
\end{align}
where $\beta_{\tau}$ is the period of $\tau$, and the bolt location satifies $r_b>n$. While the first condition determines whether we are considering a NUT or a bolt, the second ensures that the solutions are smooth at $r=n$ and $r=r_b$, respectively and, when possible, it will fix the other integration constant in \req{mastere} for our solutions. 

For concreteness, we shall restrict ourselves to four- and six-dimensional theories. In $D=4$, we will focus on the simplest possible modification to the Einstein-Hilbert action, namely, the ECG term. For this, we will construct solutions with base spaces $\mathcal{B}=\mathbb{S}^2$ and $\mathbb{T}^2$. These are, to the best of our knowledge, the first  higher-curvature generalizations of the Einstein gravity Taub-NUT solutions in four dimensions. Turning to $D=6$,  we find that no non-trivial solutions of the form \req{FFnut} can be constructed  at cubic order in the GQTG family. They will exist, however, when quartic invariants are included, and we will restrict ourselves to that case. For those, we will construct solutions with
$\mathcal{B}=\mathbb{CP}^2,\,\mathbb{S}^2\times  \mathbb{S}^2,\, \mathbb{S}^2\times \mathbb{T}^2$ and $\mathbb{T}^2\times \mathbb{T}^2$.

Although we will not be able to solve \req{mastere} analytically for $V_{\mathcal{B}}(r)$ in general (except for the critical theories), the thermodynamic properties of the solutions will be accesible in a fully analytic fashion, again similar  to what happened for the black hole solutions constructed in \cite{Hennigar:2016gkm,PabloPablo2,Hennigar:2017ego,PabloPablo3,Ahmed:2017jod,PabloPablo4}. The relation between the ADM energy of the solutions, the NUT charge and $r_b$ (when present) will be accessible in each case from the asymptotic and near $r=n$ or $r=r_b$ expansions. On the other hand, in order to compute the free energy of the solutions, we will make use of the method introduced in \cite{HoloECG}. According to this, given some higher-curvature gravity with Lagrangian density $\mathcal{L}(g^{ef},R_{abcd})$ whose linearized equations on pure AdS match those of Einstein gravity (up to a normalization of Newton's constant), the Euclidean on-shell action of any asymptotically AdS solution can be computed using the formula\footnote{The most remarkable aspect of \req{assd} is the fact that, for any theory of the kind exlained above, the usual Gibbons-Hawking-York boundary term of Einstein gravity \cite{York:1972sj,Gibbons:1976ue} only appears modified through an overall factor proportional to $a^*$. This is a considerable simplification with respect to the standard approach of trying to construct the generalized version of $K$ which makes the corresponding gravitational action differentiable \cite{Myers:1987yn, Teitelboim:1987zz,Dehghani:2011hm}.}
\begin{equation}\label{assd}
I_E=-\int_{\mathcal{M}} d^D x \sqrt{g}\mathcal{L}(g^{ef},R_{abcd})-\frac{2a^*}{\Omega_{(D-2)}\tilde{L}^{D-2}}\int_{\partial \mathcal{M}}\sqrt{h}\, \Big[ K+ \text{counterterms} \Big]\, ,
\end{equation}
where $\Omega_{D-2}\equiv 2\pi^{(D-1)/2}/\Gamma((D-1)/2)$ is the area of the unit sphere $\mathbb{S}^{D-2}$, $\tilde{L}$ is the AdS radius, and $a^*$ is the charge appearing in the universal contribution to the entanglement entropy across a spherical entangling surface $\mathbb{S}^{D-3}$ in the dual CFT. This quantity is related, for any higher-curvature theory of gravity, to the on-shell Lagrangian of the theory on pure AdS through \cite{Imbimbo:1999bj,Schwimmer:2008yh,Myers:2010tj,Myers:2010xs,HoloECG}
\begin{equation}
a^*=-\frac{\pi^{(D-1)/2}\tilde{L}^{D}}{(D-1)\Gamma\left[\frac{D-1}{2}\right]}\left. \mathcal{L}\right|_{\text{AdS}}\,.
\end{equation}
 In \cite{HoloECG}, it was also argued that the same counterterms required to produce finite on-shell actions for Einstein gravity solutions can also be used for higher-curvature gravities of this class if we weight them by the same overall coefficient. With minor modifications in the $D=6$ case --- see discussion below Eq. \req{full6} --- associated with the fact that the solutions are only locally asymptotically AdS, Eq. \req{assd}  satisfactorily removes all divergent terms in the corresponding on-shell actions, and yields thermodynamic masses that agree with the ADM ones in all cases.

 The structure of our paper is simple. In sections \ref{ECGs} and \ref{sIx} we construct Taub-NUT/bolt solutions of $D=4$ Einsteinian cubic gravity and $D=6$ Quartic Generalized quasi-topological gravities, respectively. In each case, we compute the relevant thermodynamic quantities of the solutions, with special emphasis on the most standard cases $\mathcal{B}=\mathbb{S}^2$ and $\mathcal{B}=\mathbb{CP}^2$, for which we study the corresponding phase spaces finding interesting new phenomena. Subsection \ref{critic} is somewhat different from the rest. It is devoted to the critical limit of Einsteinian cubic gravity, for which the solutions can be constructed analytically. We conclude in section \ref{conclusions}. In appendix \ref{quarticAp}, we repeat the $D=4$ analysis in section \ref{ECGs} and construct solutions for Einsteinian cubic gravity plus an additional quartic density of the Generalized quasi-topological class. In appendix \ref{freeen}, we present a detailed calculation of the on-shell action for the $D=4$ NUT solution with $\mathcal{B}=\mathbb{S}^2$ which should be illustrative of the method utilized for the other cases. Some details regarding our numerical computations can be found in appendix \ref{methods}.

\section{Four dimensions: Einsteinian cubic gravity} \label{ECGs}

The first theory we will consider is four-dimensional Einsteinian cubic gravity with a negative cosmological constant \cite{PabloPablo}.
 
Its Euclidean action reads 
\begin{equation}\label{ECG}
I_E=-\frac{1}{16\pi G}\int d^4x \sqrt{g}\left[\frac{6}{L^2}+R-\frac{\mu L^4}{8} \mathcal{P} \right]\, ,
\end{equation}
where   the cubic density $\mathcal{P}$ is defined as
\begin{equation}
\mathcal{P}=12 R_{a\ b}^{\ c \ d}R_{c\ d}^{\ e \ f}R_{e\ f}^{\ a \ b}+R_{ab}^{cd}R_{cd}^{ef}R_{ef}^{ab}-12R_{abcd}R^{ac}R^{bd}+8R_{a}^{b}R_{b}^{c}R_{c}^{a}\, .
\end{equation}
The explicit form of the field equations of \req{ECG} can be found \eg in \cite{Hennigar:2016gkm}. 
The theory admits pure AdS$_4$ solutions of radius $\tilde{L}$ related to the action scale $L$ by $\tilde L^2=L^2/f_{\infty}$, where $f_{\infty}$ is determined through
\begin{equation}\label{finn}
1-f_{\infty}+\mu f_{\infty}^3=0\, .
\end{equation}
Throughout the paper we will assume $0\leq \mu \leq 4/27$, for which a unique branch of stable AdS vacua reducing to the Einstein gravity one as $\mu\rightarrow 0$ exists. In general, stable vacua exist for $\mu<0$ as well, but these are eliminated by the requirement that black holes have positive energy \cite{Hennigar:2016gkm,PabloPablo2}. On the other hand, values of $\mu$ larger than $4/27$ always give rise to
unstable vacua. The ``critical'' limit of the theory \cite{Feng:2017tev}, corresponding to $\mu= 4/27$, warrants
special attention.  For that value of the coupling, the effective Newton constant diverges, and a number of simplifications take place, including the existence of analytic black hole solutions --- as well as various exotic results from the point of view of a putative CFT dual \cite{HoloECG}.

Let us consider a metric ansatz with NUT charge $n$ of the form \req{FFnut} where, initially, we choose the base spaces $\mathcal{B}=\mathbb{S}^2$, $\mathbb{T}^2$ and $\mathbb{H}^2$, although we shall only construct explicit solutions for the first two. The base-space metrics and $1$-forms appearing in \req{FFnut} can then be written, respectively, as 
\begin{equation}\label{base}
d\sigma^2_{\mathcal{B}}=\begin{cases}
d\theta^2+\sin^2\theta d\phi^2 \quad {\rm if}\quad {\mathcal{B}}=\mathbb{S}^2\, ,\\
\frac{1}{L^2}(d\eta^2+d\zeta^2)\quad {\rm if}\quad {\mathcal{B}}=\mathbb{T}^2\, ,\\
d\chi^2+\sinh^2\chi d\rho^2 \quad {\rm if}\quad {\mathcal{B}}=\mathbb{H}^2\, ,\\
\end{cases}\, \quad
A_{\mathcal{B}}=\begin{cases}
2\cos\theta d\phi \quad {\rm if}\quad {\mathcal{B}}=\mathbb{S}^2\, ,\\
\frac{2\eta d\zeta}{L^2} \quad {\rm if}\quad {\mathcal{B}}=\mathbb{T}^2\, ,\\
2\cosh \chi d\rho \quad {\rm if}\quad {\mathcal{B}}=\mathbb{H}^2\, .\\
\end{cases}
\end{equation}
We stress again that the most general ansatz for a Taub-NUT metric in a general higher-curvature gravity should involve an additional function --- for example, $g_{\tau\tau}=V_{\mathcal{B}}(r)N_{\mathcal{B}}(r)^2$ instead. It is a remarkable and highly nontrivial property of ECG that, when evaluated on \req{FFnut} with the above choice of base spaces, its field equations reduce to a single differential equation for $V_{\mathcal{B}}(r)$. This is given by (omitting the `$\mathcal{B}$' subscript to reduce the clutter)
\begin{equation}
	\begin{aligned}
		&-2 r V'+\frac{2 V \left(n^2+r^2\right)}{n^2-r^2}+\frac{2 k L^2-6 n^2+6 r^2}{L^2}+\mu L^4\Bigg[\frac{6 V^3 n^2 \left(n^4-16 n^2 r^2-45 r^4\right)}{\left(n^2-r^2\right)^5}\\
		&+\frac{3 V r^2 \left(V''\right)^2}{2 \left(r^2-n^2\right)}+\left(V'\right)^2 \left(\frac{3 V \left(n^4-37 n^2 r^2-2 r^4\right)}{\left(n^2-r^2\right)^3}-\frac{3 k \left(n^2+r^2\right)}{2 \left(n^2-r^2\right)^2}\right)-\frac{3 n^2 r \left(V'\right)^3}{\left(n^2-r^2\right)^2}\\
		&+V'' \left(\frac{6 V^2 \left(2 n^4-15 n^2 r^2-r^4\right)}{\left(n^2-r^2\right)^3}-\frac{6 V r V' \left(5 n^2+r^2\right)}{\left(n^2-r^2\right)^2}+\frac{3 V k \left(n^2-2 r^2\right)}{\left(n^2-r^2\right)^2}\right)\\
		&+V' \left(\frac{6 V k r^3}{\left(r^2-n^2\right)^3}-\frac{6 V^2 r \left(3 n^4+62 n^2 r^2+r^4\right)}{\left(n^2-r^2\right)^4}\right)\\
		&+V^{(3)} \left(-\frac{3 V^2 r \left(4 n^2+r^2\right)}{\left(n^2-r^2\right)^2}+\frac{3 V r^2 V'}{2 \left(r^2-n^2\right)}+\frac{3 V k r}{r^2-n^2}\right)\Bigg]=0\, ,
	\end{aligned}
\end{equation}
where $k=+1,0,-1$ for $\mathbb{S}^2$, $\mathbb{T}^2$  and $\mathbb{H}^2$, respectively.

Despite its challenging appearance, the above equation has two remarkable properties. First, it is of third order, instead of fourth, which is what one would have naively expected. Second, it allows for an integrable factor: after multiplying by $(1-n^2/r^2)$, the equation becomes a total derivative and it can be integrated once. By doing so, we are left with a second-order differential equation of the form \req{mastere}, namely
\begin{equation}
	\label{eqV}
	\begin{aligned}
		&V \left(\frac{2 n^2}{r}-2 r\right)+\frac{2 \left(k L^2 \left(n^2+r^2\right)-3 n^4-6 n^2 r^2+r^4\right)}{L^2 r}+\mu L^4 \Bigg[\frac{6 V^3 n^2 \left(n^2+9 r^2\right)}{r \left(n^2-r^2\right)^3}\\
		&+\left(V'\right)^2 \left(\frac{3 V n^2}{n^2 r-r^3}-\frac{3 k}{2 r}\right)-\frac{\left(V'\right)^3}{2}+V' \left(\frac{3 V^2 \left(17 n^2+r^2\right)}{\left(n^2-r^2\right)^2}+\frac{3 V k}{n^2-r^2}\right)+\\
		&V'' \left(-\frac{3 V^2 \left(4 n^2+r^2\right)}{r^3-n^2 r}+\frac{3 V V'}{2}+\frac{3 V k}{r}\right)\Bigg]=4C\, ,
	\end{aligned}
\end{equation}
where $C$ is an integration constant which will be related to the energy of the solution.  

We now require the metric \req{FFnut} to be locally asymptotically AdS; as a consequence we must demand $V(r)\rightarrow f_{\infty}\frac{r^2}{L^2}+\mathcal{O}(1)$ as $r\rightarrow +\infty$. Performing a $1/r$ expansion   we find
\begin{equation}
	V(r)=f_{\infty}\frac{r^2}{L^2}+k-5f_{\infty}\frac{n^2}{L^2}-\frac{2C}{r(1-3 f_{\infty}^2\mu)}+\mathcal{O}(r^{-2})
	= V_{p}(r)  \, .
	\label{rasymp}
\end{equation}
The effective Newton constant of the theory is given by
\beq
G_{\rm eff}=\frac{G}{1-3 f_{\infty}^2\mu}\, ,
\eeq
so, at least in the spherical case, we can identify the integration constant in \req{eqV} with the ADM mass of the solution as $C=GM$. In an abuse of notation, we will use this definition for all  base spaces. Now, note that since \req{eqV} is a second-order differential equation, it possesses a two-parameter family of solutions, of which \req{rasymp} corresponds to a particular one. In order to find the remaining asymptotic solutions, let us write $V(r)=V_{p}(r)+\frac{r^2}{L^2} g(r)$ and expand linearly in $g$. Taking into account only the leading terms when $r\rightarrow+\infty$, we find that $g$ satisfies the following equation
\begin{equation}\label{heq}
9 L^2 GM\mu f_{\infty} g''(r)-2r(1-3\mu f_{\infty}^2)^2g(r)=0\, .
\end{equation}
Leaving aside the limiting values $\mu=0,4/27$, the general solution is given by 
\begin{equation}\label{hSOL}
g(r)=A \textrm{AiryAi}\left[\left(\frac{2(1-3\mu f_{\infty}^2)^2}{9L^2{G M\mu f_{\infty}}}
\right)^{1/3}r\right]+B \textrm{AiryBi}\left[\left(\frac{2(1-3\mu f_{\infty}^2)^2}{9L^2{G M\mu f_{\infty}}}
\right)^{1/3}r\right]
\end{equation}
where $\textrm{AiryAi}[x]$ and $\textrm{AiryBi}[x]$ are the Airy functions of the first and second kind, respectively.
When $GM\mu>0$, the solution involving $\textrm{AiryBi}$ grows exponentially, while the one with $\textrm{AiryAi}$  decays.
Therefore, we must set $B=0$ in order for the solutions to be locally asymptotically AdS. Hence, we learn that the asymptotic boundary condition is fixing one of the integration constants in \req{eqV}. The remaining one will be fixed by the corresponding regularity conditions in the bulk, as we will show in the following sections. When $GM\mu<0$, the solutions  \req{hSOL} have an oscillatory character and they are all singular at infinity (except the trivial one, $g=0$). To remove this behaviour we would need to set both $A$ and $B$ to zero, which would fully specify the solution. This would leave us with no integration constants to impose regularity in the bulk. This behaviour is very similar to that found for the static black hole solutions of the theory \cite{Hennigar:2016gkm,PabloPablo2} --- see also \cite{PabloPablo4}, and leads us to choose $\mu\ge 0$, so that the solutions with $GM\mu>0$ have positive energy.  

\subsection*{Einstein gravity}
In the following subsections we will consider the base spaces $\mathbb{S}^2$ and $\mathbb{T}^2$ independently, and we will construct new Taub-NUT and bolt solutions for them for general values of $\mu$. It is illustrative however to start analyzing the Einstein gravity case, for which the analysis can be performed at the same time for all base spaces. Indeed, if we set $\mu=0$, \req{eqV} can be easily solved for $V_{\mathcal{B}}(r)$. Imposing the NUT condition $V_{\mathcal{B}}(r=n)=0$ first, one is left with
\begin{equation}\label{NUTE}
V_{\mathcal{B}}(r)=\frac{(r-n)\left[(r-n)(3n+r)+ kL^2\right]}{L^2(n+r)}\, ,
\end{equation}
where we already fixed the integration constant as
\begin{equation}\label{massE}
GM=k n - \frac{4n^3}{L^2}\, .
\end{equation}
The regularity condition \req{smooth} imposes
\begin{equation}\label{betaEin}
\beta_{\tau}=\frac{8\pi n}{k}\, ,
\end{equation}
which means that $\tau$ cannot be a compact coordinate for $\mathcal{B}=\mathbb{T}^2$ or, in other words, the solution is extremal, in the sense that the temperature $T\equiv 1/\beta_{\tau}$ is forced to vanish. Similarly, for $\mathcal{B}=\mathbb{H}^2$, one finds that the period of $\tau$ would need to be negative. This means that $V_{\mathbb{H}^2}(r)$ actually becomes negative for values of $r$ greater than $n$, which is forbidden by assumption. Hence, no regular Taub-NUT solution exists in that case for Einstein gravity. 

If we impose the bolt condition $V_{\mathcal{B}}(r=r_b)=0$ instead, we find
\begin{equation}\label{boltEi}
V_{\mathcal{B}}(r)=\frac{(r-r_b)\left[(6n^2 r r_b-3 n^4+k L^2 (n^2-r r_b)-r r_b (r^2+r r_b+ r_b^2)) \right]}{L^2(n^2-r^2)r_b}\, ,
\end{equation}
where the integration constant was fixed as
\begin{equation}
GM=\frac{k L^2(n^2+r_b^2)-3n^4-6n^2r_b^2+r_b^4}{2 L^2 r_b}\, .
\end{equation}
The regularity condition \req{smooth} fixes now the bolt radius as a function of $n$ and $\beta_{\tau}$, namely
\begin{equation}\label{boltrb}
r_b=\frac{2L^2 \pi}{3\beta_{\tau}} \left[ 1\pm \sqrt{ 1-\frac{3k \beta_{\tau}^2}{4L^2 \pi^2} +\frac{9n^2 \beta_{\tau}^2}{4L^4\pi^2}}\right]\, .
\end{equation}
In order for each solution to be allowed, it must be such that $r_b>n$.  Furthermore,
the quantity inside the square root must be positive, which restricts the allowed values of $n$ for which the corresponding solutions exists.

 On general grounds,	in order to remove the so-called Misner string \cite{Misner:1963fr}, an additional condition must be imposed on $\beta_{\tau}$ both for NUT and bolt solutions when $\mathcal{B}=\mathbb{S}^2$. As we explain in the next subsection, this reads $\beta_{\tau}=8\pi n$. It is a remarkable (and peculiar) fact that in Einstein gravity Eq.~\req{smooth} automatically implements this condition in the case of the NUT solution. In general, both conditions must be imposed separately.

\subsection{$\mathcal{B}=\mathbb{S}^2$}\label{S2cubic4d}
Let us now turn on the Einsteinian cubic gravity coupling. We begin by assuming the base space to be
the one-dimensional complex projective space $\mathbb{CP}^1 $ or, equivalently, the two-dimensional round sphere, $\mathcal{B}=\mathbb{S}^2$. Then, the metric \req{FFnut} reads
\begin{equation}
	ds^2=V_{\mathbb{S}^2}(r)(d\tau+2 n \cos\theta d\phi)^2+\frac{dr^2}{V_{\mathbb{S}^2}(r)}+(r^2-n^2)\left(d\theta^2+\sin^2\theta d\phi^2\right)\, .
	\label{taub}
\end{equation}
This metric has ``wire singularities'' at $\theta=0, \pi$, for which it becomes noninvertible. As shown by Misner \cite{Misner:1963fr}, it can nevertheless be  made regular everywhere using two coordinate patches. The idea is to define new coordinates $\tau^{\pm}= \tau \pm 2n \phi$ covering the $\theta\geq \pi/2$ and $\theta \leq \pi/2$ regions respectively. In the overlap region, $\tau^+=\tau^-+4n\phi$, and since $\beta_{\phi} = 2\pi$, one is forced to impose the periods of $\tau^{\pm}$ to be $\beta_{\tau^{\pm}}=8 \pi n$.
For clarity reasons, in what follows we will work with the metric \req{taub} in a single patch, but taking into account  that the period of $\tau$ is related to the NUT charge through  $\beta_{\tau}=8\pi n$. Observe that this condition is a consequence of choosing $\mathcal{B}=\mathbb{S}^2$ and does not depend on the theory. When combined with the general regularity condition \req{smooth}, this gives rise to the conditions $V_{\mathbb{S}^2}'(r=n)=1/(2n)$ and $V_{\mathbb{S}^2}'(r=r_b)=1/(2n)$ respectively for NUTs and bolts.

The function $V_{\mathbb{S}^2}$ in \req{taub} is determined from \req{eqV} with $k=1$. Using the asymptotic expansion \req{rasymp}, we see that when $r\rightarrow +\infty$ the metric induced on a constant-$r$ hypersurface is given by
\begin{equation}
	\frac{^{(3)}ds^2}{r^2}=\frac{4 f_{\infty} n^2}{L^2}(d\psi+\cos\theta d\phi)^2+d\theta^2+\sin^2\theta d\phi^2+\mathcal{O}(r^{-2})\, ,
\end{equation}
where we have introduced the angle coordinate $\psi= \tau/(2n)$, whose period is $4\pi$. When $4 f_{\infty} n^2=L^2$, the previous metric is the one of a round $\mathbb{S}^3$. For any other value of $n$, it is the metric of a squashed sphere, and it is customary \cite{Chamblin:1998pz,Hartnoll:2005yc,Anninos:2012ft,Bobev:2016sap,Bobev:2017asb} to rewrite the NUT charge in terms of a `squashing parameter' $\alpha$ as $4 f_{\infty} n^2/L^2=1/(1+\alpha)$.
In order to specify the solution, we need to choose a boundary condition at some finite $r=r_b$.
Depending on whether we choose $r_b=n$, or $r_b>n$, we will be considering Taub-NUT or Taub-bolt solutions.

\subsubsection{Taub-NUT solutions}
\begin{figure}[t]
	\centering 
	\includegraphics[scale=0.6]{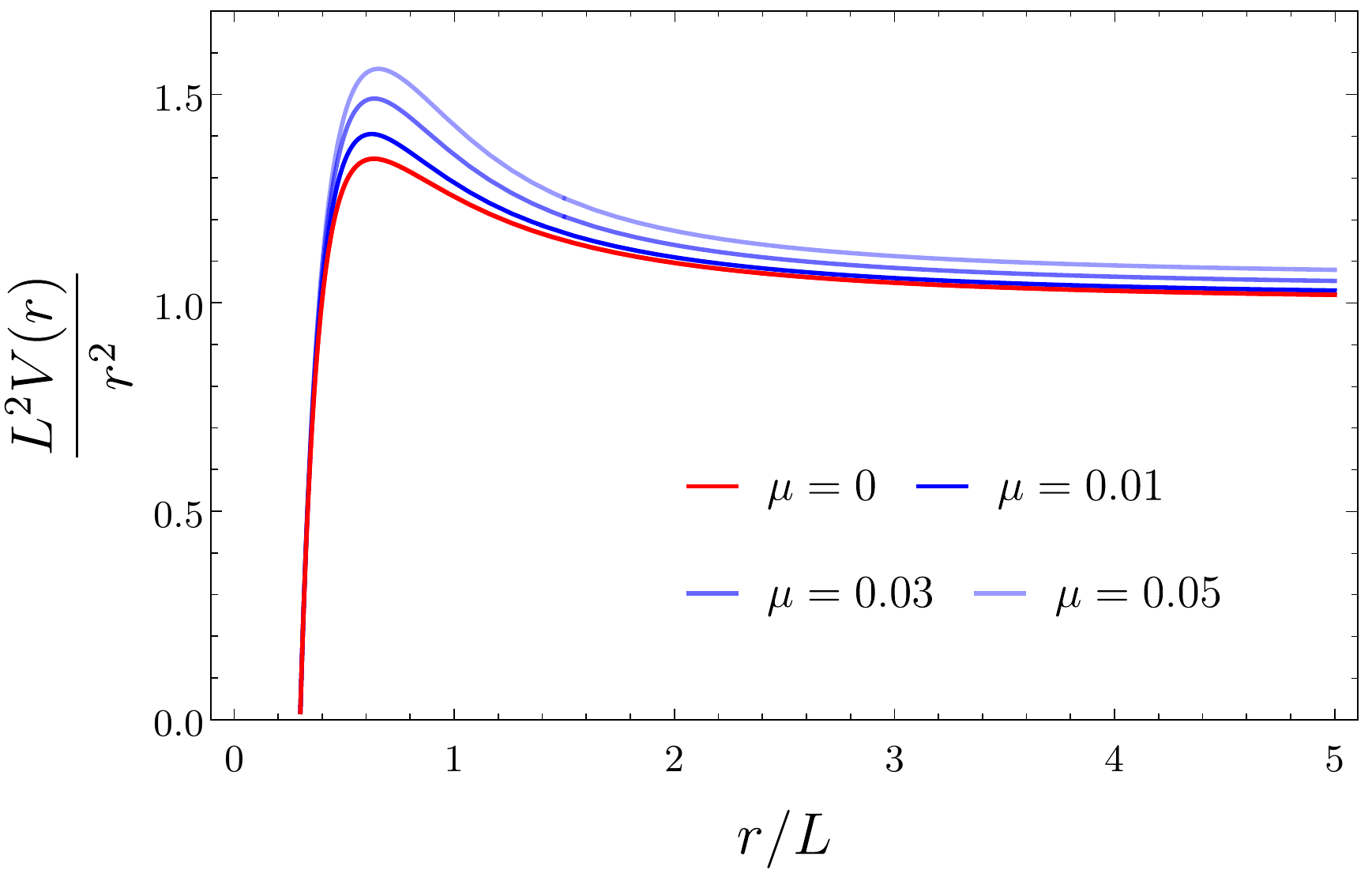}
	\caption{We plot the metric function $ V_{\mathbb{S}^2}(r)\cdot L^2/r^2$ corresponding to Taub-NUT solutions of ECG with $n/L=0.3$ for several values of $\mu$. The examples shown correspond to values of $\mu$ and $n$ which satisfy the positive-mass inequalities \req{condi}.  The red curve corresponds to the metric function for the Einstein gravity Taub-NUT solution given by \req{NUTE} with $k=1$. } 
	\label{NUTS2ECG}
\end{figure}

As we have explained, the Euclidean Taub-NUT metric is characterized by the conditions $V_{\mathbb{S}^2}(r=n)=0$ and $V_{\mathbb{S}^2}'(r=n)=1/(2n)$. Let us then expand $V_{\mathbb{S}^2}(r)$ around $r=n$ as 
\begin{equation}\label{near}
V_{\mathbb{S}^2}(r)=\frac{(r-n)}{2 n}+\sum_{i=2}^\infty (r-n)^i a_i\, ,
\end{equation}
for some $a_i$.
Plugging this expansion into \req{eqV}, we observe that the $\mathcal{O}\left(r-n\right)$ and $\mathcal{O}\left((r-n)^2\right)$ equations are automatically satisfied, whereas the $\mathcal{O}(1)$ one gives rise to the following relation between the mass and the NUT charge,
\begin{equation}\label{masss}
	GM=n-\frac{4 n^3}{L^2}-\frac{\mu L^4}{16 n^3}\, .
\end{equation}
Observe that this reduces to the Einstein gravity expression \req{massE} for $\mu=0$.
The following term in the expansion gives a relation between $a_3$ and $a_2$, which we can use to write the former as a function of the latter,
$a_3(a_2)$. Similarly, the following term allows us to obtain $a_4 (a_2 )$, and so on. Hence, as in the black hole case \cite{Hennigar:2016gkm,PabloPablo2}, the full series is determined
by a single free parameter $a_2$. This parameter must be chosen in a way such that $B = 0$ in \req{hSOL}, which ensures that the solution is locally asymptotically AdS. In practice, the shooting method can be used to identify $a_2$ for each value of $\mu$, so that the near $r=n$ expansion yields a good approximation to the exact solution that connects with the asymptotic expansion \req{rasymp}. There is a unique $a_2$ for each $\mu$ that does the job, corresponding to a unique Taub-NUT solution in each case. We plot the metric function $V_{\mathbb{S}^2}(r)$ for different values of $\mu$ in Fig. \ref{NUTS2ECG}. These solutions generalize the Einstein gravity Taub-NUT solution (the red curve in Fig. \ref{NUTS2ECG}), whose metric function is given by \req{NUTE} with $k=1$. As we can see, the qualitative behaviour of $V_{\mathbb{S}^2}(r)$ is very similar to that of Einstein gravity  for nonvanishing values of the ECG coupling.

One peculiarity of \req{masss} for nonvanishing $\mu$ is that the mass becomes negative for small values of $n$. In particular, the mass is non-negative only when 
\begin{equation}\label{condi}
\frac{L^2}{4f_{\infty}} \geq n^2 \geq \frac{L^2 \mu f_{\infty}^2}{8}\left[1+\sqrt{\frac{4-3\mu f_{\infty}^2}{\mu f_{\infty}^2}}\right]\, .
\end{equation}
The existence of a finite lower bound for $n$
is a new feature, which does not occur for Einstein gravity. Indeed, in that case \req{condi} becomes $L^2/(4f_{\infty})\geq n^2 \geq 0$. 
For general values of the gravitational coupling, we cannot expect the solution to exist whenever $n$ lies outside the interval in \req{condi}, because the asymptotic behaviour for negative masses is pathological.  Indeed, as we explained in the discussion below Eq.~\req{hSOL}, negative mass solutions would be highly oscillating at infinity, and hence they are not asymptotically AdS. 
Solutions with zero mass occur when either the upper or the lower bounds are saturated. In terms of $\mu$, the $M=0$ condition reads
\begin{equation}\label{num0}
\mu=\frac{16 n^4}{L^4}\left[1-\frac{4n^2}{L^2}\right]\, .
\end{equation}
In the case of Einstein gravity, the possibilities are $n^2=L^2/4$ and $n=0$, for which the solution reduces to pure Euclidean AdS$_4$ foliated by round $\mathbb{S}^3$ slices and $\mathbb{S}^1\times \mathbb{S}^2$ slices, respectively\footnote{Observe however that the $n\rightarrow 0$ limit is problematic, in the sense that the period of $\tau$ would vanish in that case. One can of course just set $n=0$ from the beginning, which makes the problem disappear.} \cite{Emparan:1999pm}. For any nonvanishing value of $\mu$, \req{num0} is satisfied identically for $n^2=L^2/(4f_{\infty})$, which can be straightforwardly checked using \req{finn}. In this case, the solution also reduces to pure AdS$_4$, the metric factor being simply given by $V_{\mathbb{S}^2}(r)=f_{\infty}r^2/L^2-1/4$. Besides this solution, there exists another one obtained by choosing $n$ to saturate the lower bound in \req{condi}, and which is analogous to the $n=0$ one in Einstein gravity. Interestingly, this solution no longer reduces to pure Euclidean AdS$_4$ for  $0< \mu < 4/27$ but, rather, it has a nontrivial profile. In the critical limit, $\mu=4/27$, the range allowed by \req{condi} collapses to a single possible value, corresponding to $n^2=L^2/6$. In that case, the solution does correspond to pure AdS$_4$. 
For other values of $n$, the critical solution can also be accessed analytically (see Section \ref{critic}), and the result for the metric function reads
\begin{equation}\label{cric}
V_{\mathbb{S}^2}^{\rm cr}(r)=\frac{3}{2L^2}(r^2-n^2)\, .
\end{equation}
This solution has a vanishing mass parameter $M=0$, namely, it only exists if we fix the integration constant $C$ to zero in \req{eqV}. Note that this solution has ${V_{\mathbb{S}^2}^{\rm cr}}'(n)=3n/L^2$;  it has a conical singularity at $r=n$ in all cases but one, corresponding to the value $n^2=L^2/6$, for which it becomes pure AdS$_4$, as mentioned above.

\begin{figure}[t]
	\centering 
	\includegraphics[scale=0.65]{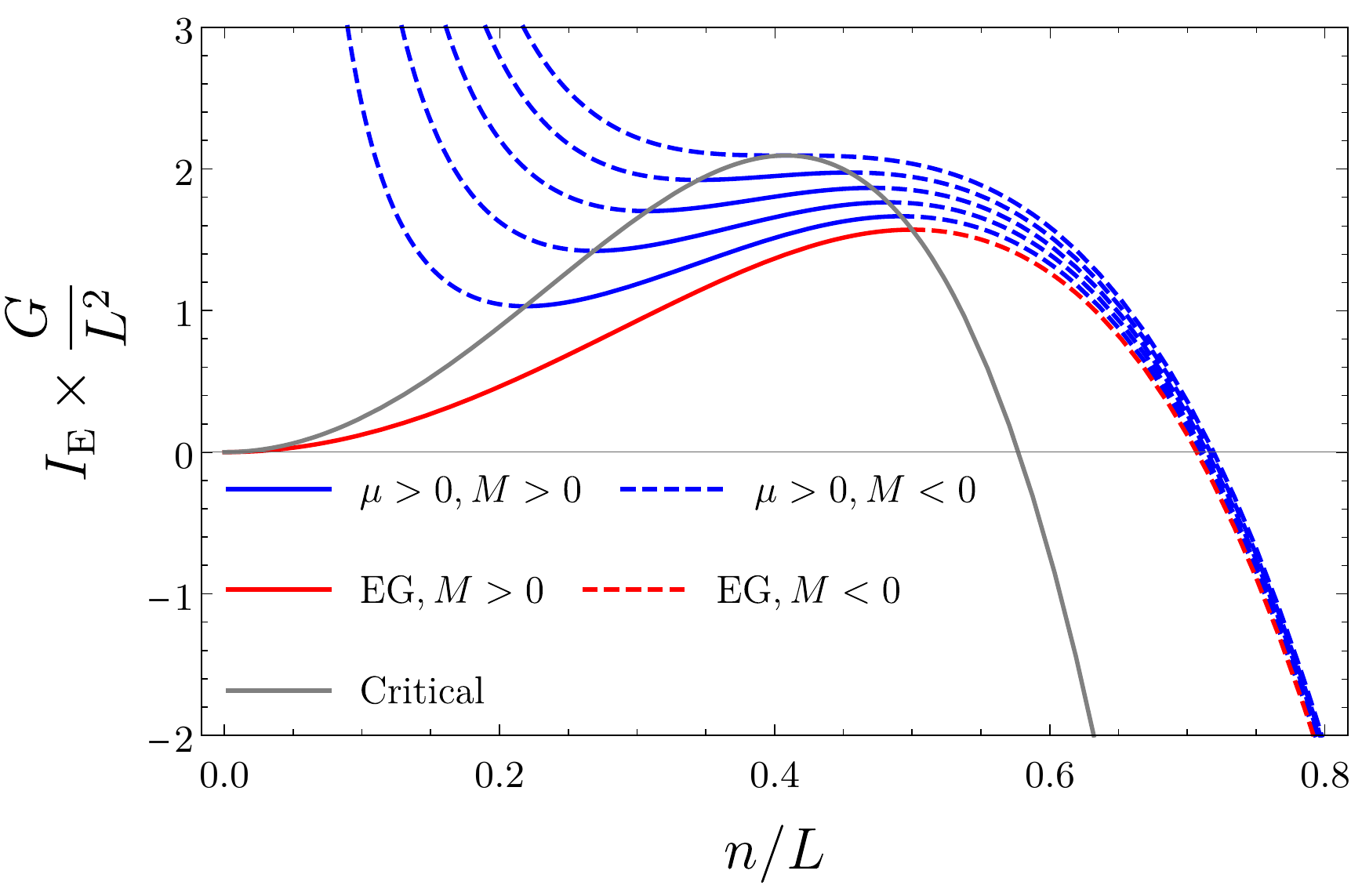}
	\caption{Free energy of the NUT solution with $\mathbb{S}^2$ base. In red we show the Einstein gravity result and in blue the ECG ones for $\mu=8/270,\, 16/270,\, 24/270,\, 32/270,\, 4/27$. The mass $M$ is proportional to the slope of each curve and solid lines represent $M>0$ while dashed lines represent $M<0$. For ECG only the solutions with $M>0$ exist. We also plot the free energy of the special critical solution \req{IEcr}, whose metric function is given in \req{cric}. Remarkably, this curve is the envelope of the free energies with $M>0$. }
	\label{solNUTS2}
\end{figure}

Let us now evaluate the on-shell action of the solutions. In order to do so, we make use of the generalized action \req{assd} where, for ECG, the charge $a^*$ is given by $a^*=(1+3\mu f_{\infty}^2)\tilde{L}^2/(4G)$.
Using this, the full ECG action takes the form
\begin{equation}\label{EuclideanECG}
\begin{aligned}
I_E=-\int  \frac{d^4x \sqrt{g}}{16\pi G}\left[\frac{6}{L^2}+R-\frac{\mu L^4}{8} \mathcal{P} \right]-\frac{(1+3\mu f_{\infty}^2)}{8 \pi G}\int_{\partial \mathcal{M}}d^3x\sqrt{h}\left[K-\frac{2\sqrt{f_{\infty}}}{L}-\frac{L}{2\sqrt{f_{\infty}}}\mathcal{R}\right]\, ,
\end{aligned}
\end{equation}
where $\mathcal{R}$ stands for the Ricci scalar of the induced metric on the boundary. The last two terms in the second line are the standard counterterms in $D=4$ which, as explained in the Introduction, also appear weighted by $a^*$ without further modification according to the prescription in \cite{HoloECG}. A detailed evaluation of all the terms in the above expression for our new Taub-NUT solutions, which can be performed fully analytically using the near $r=n$ and asymptotic expansions, is presented in appendix \ref{freeen}. It yields the following remarkably simple finite answer 
\begin{equation}\label{freeee1}
	I_E=\frac{4\pi}{G}\left[n^2-\frac{2n^4}{L^2}+\frac{\mu L^4}{16 n^2}\right]\, .
\end{equation}
This reduces to the free energy of the corresponding Einstein gravity Taub-NUT solution when $\mu=0$, as it should. The energy and entropy can be easily obtained now from
$
	E=\partial I_E/\partial \beta$ and $S=\beta E-I_E\, .
$
Using this, we find that the energy precisely matches the result for the ADM mass obtained in \req{masss}, $E=M$, which is a highly nontrivial check of the calculation, whereas for the entropy we obtain
\begin{equation}\label{entR}
S=\frac{4\pi}{G}\left[n^2-\frac{6n^4}{L^2}-\frac{3\mu L^4}{16 n^2}\right] \, ,
\end{equation}
 which is not given by a simple area law due to contributions from the Misner string.

It is also possible to consider the thermodynamics of these NUT charged solutions from the perspective of extended phase space thermodynamics. Within this framework, one introduces potentials conjugate to the cosmological constant  --- interpreted as a pressure $P = - \Lambda/(8 \pi G)$ --- and any  higher-curvature couplings that appear in the action~\cite{Kastor:2009wy, Kastor:2010gq}.  These considerations are motivated by scaling arguments, since without these terms the Smarr relation fails to hold. In the case of Taub-NUT solutions in ECG, the extended first law reads
\begin{equation}
dE = TdS + VdP + \Upsilon^{\ssc \rm ECG} d ( \mu L^4 ) \, ,
\end{equation}
where we have restored the dimensions to the ECG coupling constant. The new potentials read
\begin{equation}
V = - \frac{8 \pi n^3}{3} \, , \quad \Upsilon^{\ssc \rm ECG} = \frac{1}{32 G n^3} \, .
\end{equation}
Interestingly, the thermodynamic volume here is precisely the same as for Taub-NUT solutions in Einstein gravity~\cite{Johnson:2014xza}. The same conclusion holds for the thermodynamic volume of black holes in  higher-curvature gravities that belong to the generalized quasi-topological class. With the thermodynamic quantities defined as above, the Smarr relation that follows directly from a scaling argument is found to hold:
\begin{equation}\label{smarrfS2}
E = 2 TS - 2 VP + 4 \mu L^4 \Upsilon^{\ssc \rm ECG} \, .
\end{equation}

In Fig. \ref{solNUTS2} we plot the Euclidean action \req{freeee1} for several values of $\mu$. Dashed lines correspond to negative values of the mass, whereas solid lines correspond to solutions with $M>0$. As we mentioned earlier, in principle we only expect solutions with positive mass $M>0$ to exist. A numerical analysis seems to confirm this, since we were not able to construct any solution with $M<0$.  This also constrains the validity of the thermodynamic expressions \req{freeee1} and \req{entR} to the interval defined by \req{condi}. This interval becomes smaller as $\mu$ grows, and it reduces to a single point, $n^2=L^2/6$, in the critical limit. Interestingly, we observe that the free energy of the critical theory solutions (solid gray curve) acts as an envelope of all possible solutions with positive mass and arbitrary values of $\mu$.  Observe that this free energy cannot be obtained from \req{freeee1} in the $\mu\rightarrow 4/27$ limit; the same applies to the mass, which cannot be obtained from \req{masss}. The correct result for the on-shell action associated to the critical solutions with metric function \req{cric} reads however
\begin{equation}\label{IEcr}
I_{ E}^{\rm cr}=\frac{8\pi n^2}{G}\left[1-\frac{3n^2}{L^2}\right]\, .
\end{equation}
As we mentioned above, all these solutions  except for the one with $n^2=L^2/6$ have conical singularities, so the result must be taken with care --- \eg the mass cannot derived from \req{IEcr} using standard thermodynamic identities. It is a remarkable and somewhat striking fact that the free energy of this singular solution, as given by \req{IEcr}, precisely separates the free energies of negative-mass solutions from those corresponding to completely regular positive-mass solutions for general values of $\mu$. The different nature of the critical solutions can also be seen from the fact that 
whenever $\mu \neq 4/27$, the solutions with $M=0$ correspond to values of $n$ for which $I_E(n)$ is locally extremized, whereas the whole  $\mu = 4/27$ curve has $M=0$.

\subsubsection{Taub-bolt solutions}
\begin{figure}[t]
	\centering 
	\includegraphics[scale=0.62]{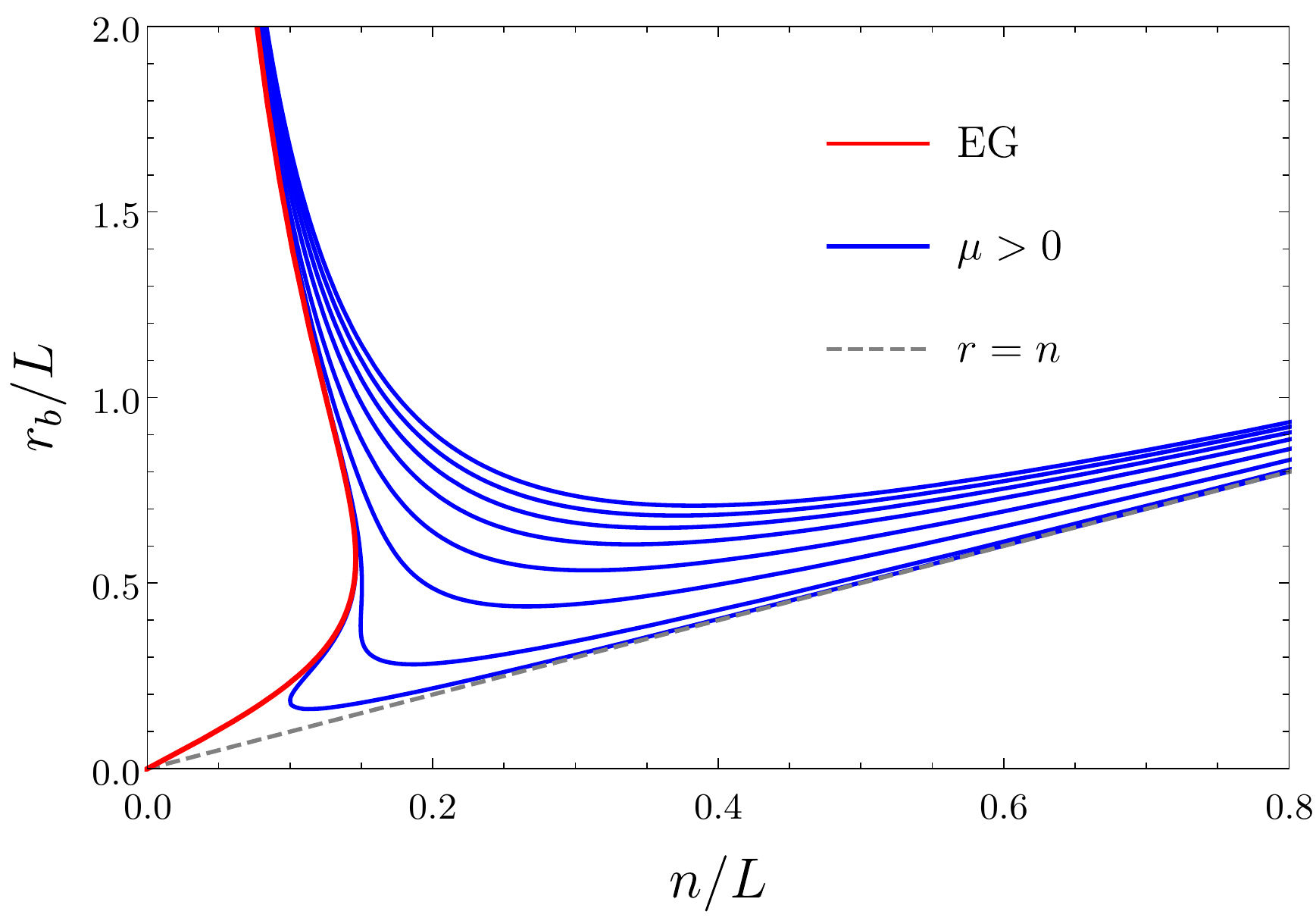}
	\caption{We show the bolt radius $r_b$ for several values of $\mu$. In red we show the Einstein gravity value $(\mu=0)$, and the blue lines correspond to $\mu=0.0001,\, 0.001126,\, 0.01,\, 8/270,\, 16/270,\, 24/270,\, 32/270,\, 4/27$. For any non-vanishing $\mu$ there is at least one solution for every value of $n$. For large $n$ there is a new solution which approaches $r_b\rightarrow n$ asymptotically. The gray dashed line corresponds to NUT solutions.}
	\label{rhbolt}
\end{figure}

Let us now turn to  bolt solutions. These are obtained by imposing $V_{\mathbb{S}^2}(r)$ to vanish for some $r_b>n$, \ie $V_{\mathbb{S}^2}(r_b)=0$, plus the regularity condition $V_{\mathbb{S}^2}'(r_b)=1/(2n)$. If we plug a Taylor expansion for $V_{\mathbb{S}^2}(r)$ around $r=r_b$ including these conditions in \req{eqV}, the equations corresponding to the first nontrivial orders give rise to two equations involving the mass of the solution $M$, the bolt radius $r_b$, and the NUT charge $n$. These read
\begin{eqnarray}\label{gmm3}
GM&=&\frac{n^2+r_b^2}{2r_b}+\frac{1}{L^2}\left[\frac{r_b^3}{2}-\frac{3 n^4}{2r_b}-3 r_b n^2\right]-\frac{\mu L^4}{64n^2}\frac{(6n+r_b)}{n r_b}\, ,\\ \label{rbb}
0&=&\frac{6}{L^2}(r_b^2-n^2)^2+(2-r_b/n)(r_b^2-n^2)-\frac{3\mu L^4}{8n^2}\frac{(r_b^2+n r_b+n^2)}{(r_b^2-n^2)}\, .
\label{Eqr}
\end{eqnarray}
The relation $r_b(n)$ has several remarkable differences with respect to the Einstein gravity case. Indeed, for $\mu=0$, \req{rbb} has two nontrivial roots, given by \req{boltrb}, namely
\begin{equation}\label{Einss}
r_b(\mu=0)=\frac{L^2}{12 n}\left[1\pm \sqrt{1-\frac{48 n^2}{L^2}+ \frac{144n^4}{L^4}}\right]\, .
\end{equation}
Since we want $r_b$ to be real and larger than $n$, this implies that $n/L<\left[(2-\sqrt{3})/12\right]^{1/2}\simeq 0.1494$. In particular, there is no bolt solution near the undeformed $\mathbb{S}^3$ case, corresponding to $n/L=1/2$. The situation is very different in ECG. Indeed, for any nonvanishing value of $\mu$ and for any value of $n$, there always exists at least one solution satisfying $r_b>n$. For small and large $n/L$, there is a unique solution in each case, while intermediate values of $n/L$ give rise to one or three possible solutions, depending on the value of $\mu$. For $\mu<0.001126$, there is a region of values of $n/L$ for which three solutions with $r_b>n$ exist. If $\mu$ is greater than this quantity, there is a two-to-one relation between $n$ and $r_b$ for all $n$. All this is shown in Fig. \ref{rhbolt}.

For the set of parameters for which a unique bolt solution exists, the profile of $V_{\mathbb{S}^2}(r)$ can be accessed numerically following exactly the same logic as for the NUT solutions. We plot the resulting metric functions for some values of $\mu$ in Fig. \ref{rhbolt}.
\begin{figure}[t]
	\centering 
	\includegraphics[scale=0.6]{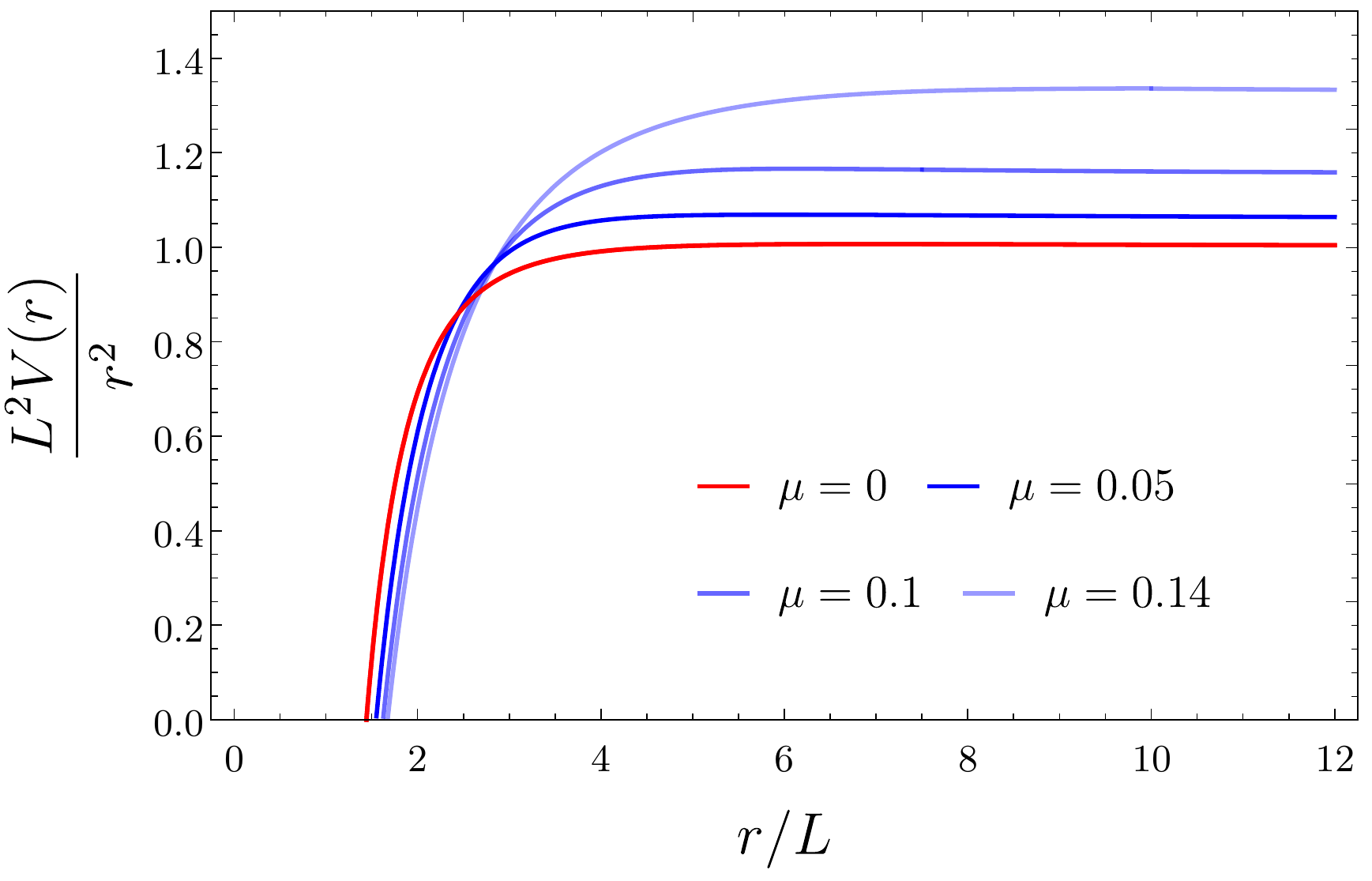}
	\caption{We plot the metric function $ V_{\mathbb{S}^2}(r)\cdot L^2/r^2$ corresponding to Taub-bolt solutions of ECG with $n/L=0.1$ for several values of $\mu$. We choose the largest bolt radius $r_b$ when there are several solutions for the fixed $n$. 
	 } 
	\label{boltS2ECG}
\end{figure}
We can also compute the on-shell action for the bolt solutions analogously to the NUT case.
The final result can be written as
\begin{equation}
I_E=\frac{\pi}{G}\left[n^2-r_b^2+4 n r_b+\frac{4n r_b}{L^2}(r_b^2-3n^2)+\mu L^4\frac{5 n^2+12 n r_b+r_b^2}{16n^2(r_b^2-n^2)}\right]\, .
\end{equation}
Using the chain rule and the relation \req{Eqr}, one can show again that $E\equiv \partial_{\beta} I_E= M$ as given in \req{gmm3}, which is a consistency check of the calculation.  In addition, the entropy, given by $S=\beta E-I_{\rm E}$, reads
\begin{equation}
S=\frac{\pi}{G r_b}\left[ -\frac{12 n^3}{L^2} \left(n^2+r_b^2\right)+4 n^3-n^2 r_b+r_b^3 +\frac{3 \mu  L^4 \left(4 n^3-n^2 r_b-8 n r_b^2-r_b^3\right)}{16 n^2(r_b^2-n^2) }\right]\,.
\end{equation}

 Just as in the case of the NUT solutions, we can also study the thermodynamics of the bolts in extended phase space. The extended first law has the same form as in the NUT case, but now the thermodynamic volume and coupling potential read
\begin{equation}
V = \frac{4 \pi \rh}{3}(\rh^2 - 3 n^2) \, , \quad \Upsilon^{\ssc \rm ECG} = \frac{\rh^2 +12 \rh n + 5 n^2}{128 G n^3 (\rh^2 - n^2)}\, ,
\end{equation}
and satisfy the Smarr formula that follows from scaling, which is of the same form as in the NUT case --- see \eqref{smarrfS2}. Note that, once again, the basic formula for the thermodynamic volume of the bolts is unaltered by the  higher-curvature terms~\cite{Johnson:2014xza}. However the thermodynamic volume is implicitly sensitive to the ECG coupling since the ECG term is important in determining the value of $\rh$ for a given $n$.

\begin{figure}[t]
	\centering 
	\includegraphics[scale=0.485]{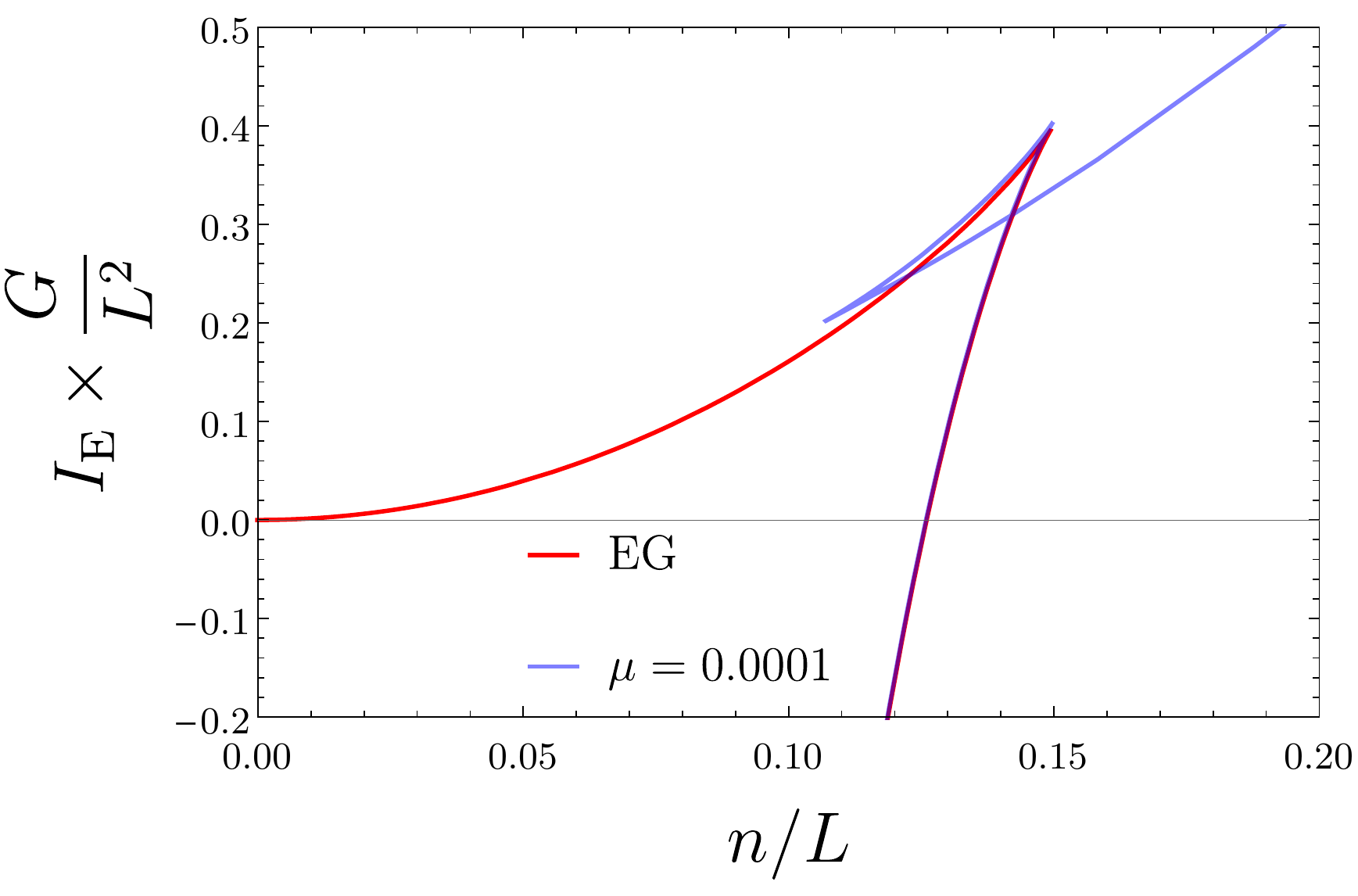}
	\includegraphics[scale=0.48]{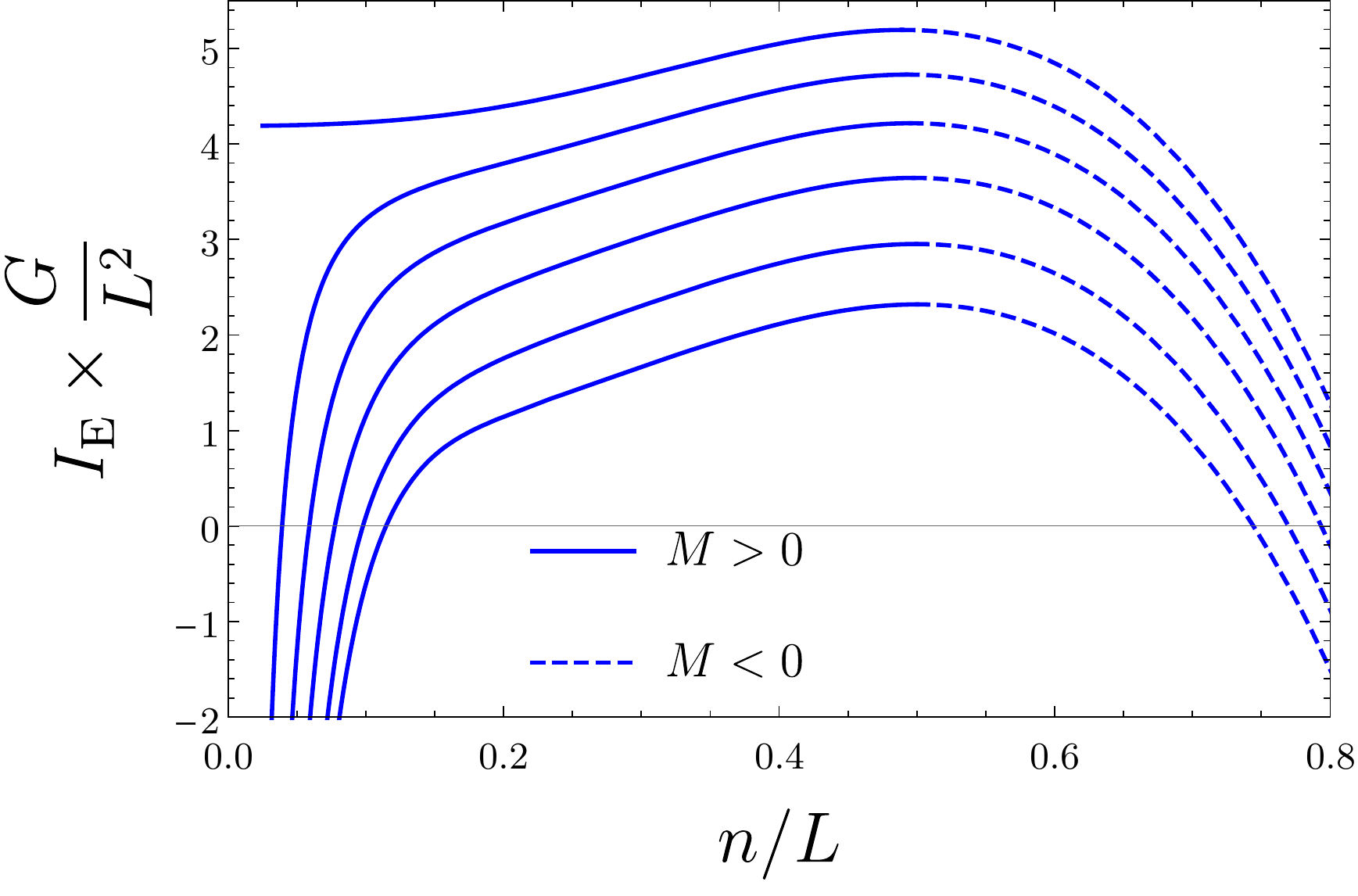}
	\caption{Euclidean on-shell action for $\mathcal{B}=\mathbb{S}^2$ bolt solutions in ECG. In the left panel we compare the Einstein gravity result with the ECG one with  $\mu=0.0001$, which contains three branches. For $\mu>0.001126$ there is only one branch, which is shown in the right panel for $\mu=0.01,\, 8/270,\, 16/270,\, 24/270,\, 32/270,\, 4/27$. The dashed lines correspond to $M<0$, so they should be excluded. As we can see, when $n\rightarrow0$ the free energy diverges to $-\infty$, except for the critical case $\mu=4/27$, which corresponds to the upper line.} 
	\label{IEboltS2ECG}
\end{figure}

Although we cannot solve \req{Eqr} exactly, we can study its behaviour in several limits. For example, let us consider the new branch of solutions for which $r_b$ is close to $n$ in the limit $\mu\ll 1$. We can expand $r_b$ in powers of $\mu^{1/2}$. To second order, we get
\begin{equation}
r_b=n+\frac{L^2}{n}\sqrt{\frac{3\mu}{8}}+\frac{3 \mu L^2}{16 n^3}\left(L^2-12 n^2\right)+\mathcal{O}(\mu^{3/2})\, .
\end{equation}
For Einstein gravity, we get $r_b=n$, and the solution reduces to the NUT one. However, for any given nonvanishing $\mu$, we have two inequivalent solutions:  the NUT constructed in the previous subsection, and this one. In particular, as opposed to the Einstein gravity case, a bolt solution does exist for $n^2=L^2/(4 f_{\infty})$, which corresponds to a nonsquashed spherical boundary geometry. 
Expansions for the free energy and the mass of this branch of solutions can be easily obtained in the $\mu\ll 1$ limit, the results being
\begin{align}
I_E&=\frac{\pi}{G}\left[4 n^2-\frac{8 n^4}{L^2}+\frac{3L^2}{\sqrt{2}}\mu^{1/2}+\left(\frac{27 L^2}{8}-\frac{L^4}{8 n^2}\right)\mu\right]+\mathcal{O}(\mu^{3/2})\, ,\\
GM&=n-\frac{4 n^3}{L^2}+\frac{\mu L^4}{32 n^3}+\mathcal{O}(\mu^{3/2})\, .
\end{align}
Note that the mass is nonvanishing when the boundary geometry is that of a round $\mathbb{S}^3$, namely, $GM(n^2=L^2/(4f_{\infty}))=3\mu L/4+\mathcal{O}(\mu^{3/2})$, so the free energy is not extremized in that case. Instead, the maximum is reached for $n^2/L^2=1/4[1+\mu/2+\mathcal{O}(\mu^{3/2})]$, which is also the $M=0$ value. Greater values of $n$  would give rise to negative mass solutions, as illustrated in Fig. \ref{IEboltS2ECG}. Note also that $I_E(n^2=L^2/(4f_{\infty}))$  is greater than the one for the NUT solution, so that in the region near $n^2=L^2/(4f_{\infty})$, the NUT would dominate the corresponding holographic partition function.

We can also study the behaviour near $n= 0$, for which there is also a single bolt solution for each nonvanishing $\mu$. We get approximately
\begin{equation}
r_b=\frac{L^2}{6n}-2\left(1-\frac{27}{4}\mu\right)n+\mathcal{O}(n^2)\, ,
\end{equation}
and for the free energy,
\begin{equation}
I_E=\frac{\pi}{G}\left[-\frac{L^4}{108 n^2}\left(1-\frac{27}{4}\mu\right)+\frac{2L^2}{3}\left(1+\frac{27}{4}\mu\right)\right]+\mathcal{O}(n^2)\, .
\end{equation}
If we set $\mu=0$ in these expressions, we recover the small $n$ expansions for Einstein gravity bolts corresponding to the $(+)$ root in \req{Einss}. Observe that in the critical limit, $\mu=4/27$, the leading term disappears, and the on-shell action is finite for $n=0$.

We try to summarize the different possibilities in Fig. \ref{IEboltS2ECG}, where we plot $I_{\rm E}$ for ECG bolt solutions for several values of $\mu$. As we can see, the result is very different from that of Einstein gravity. There are two cases that we can distinguish: if $0<\mu<0.001126$, the diagram contains three branches, since there are three different bolt solutions; for $\mu>0.001126$ there is a single (elephant-shaped) branch. At $\mu\simeq 0.001126$, we expect to have a critical point which would represent a second-order phase transition if the bolt solution were dominant.  
 In all cases, the solutions exist for much larger values of $n$ than in Einstein gravity. However, there is an additional upper bound on $n$ coming from imposing $M>0$.

\subsubsection*{Free energy comparison}
\begin{figure}[t]
	\centering 
	\includegraphics[scale=0.485]{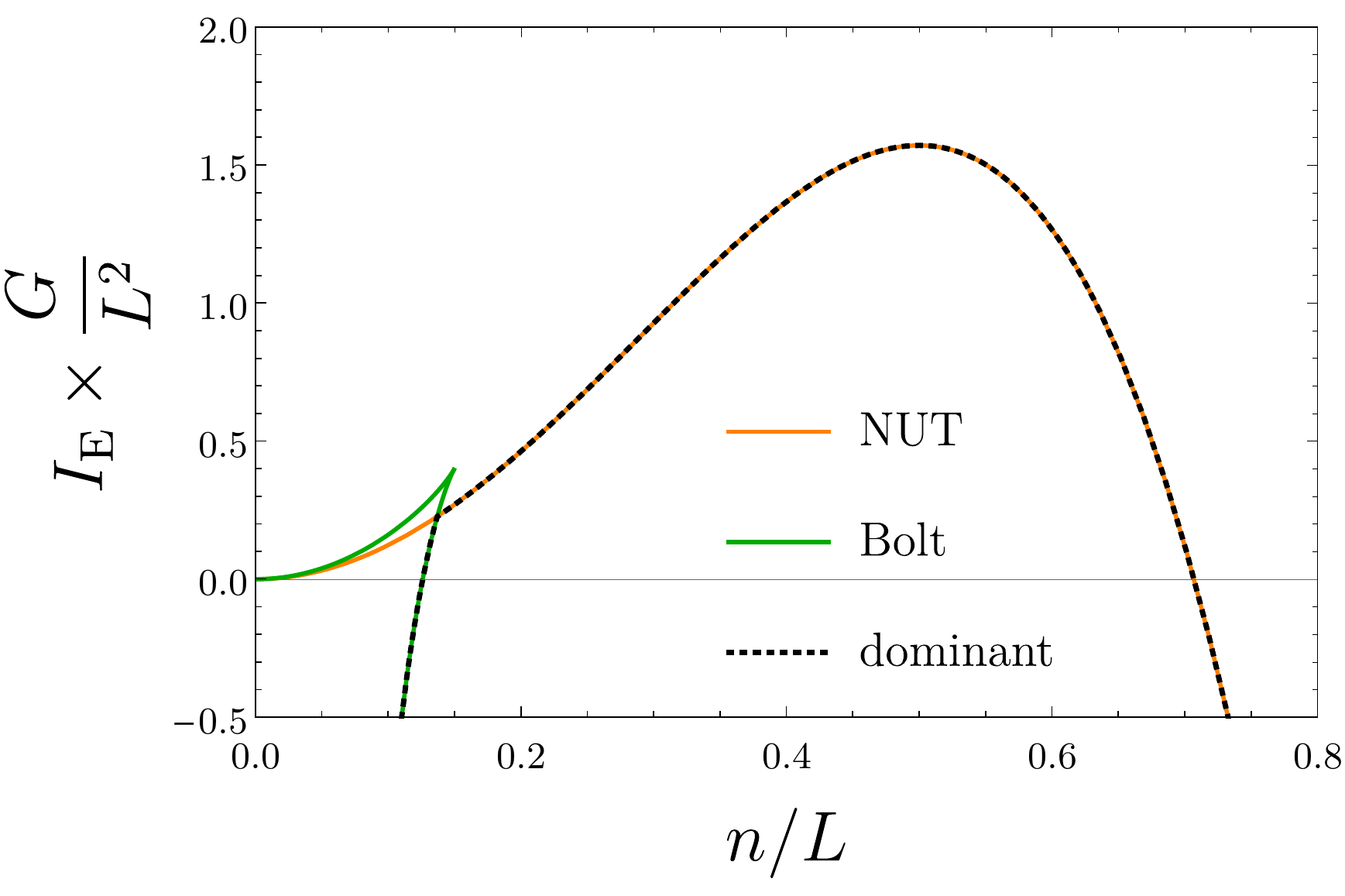}
	\includegraphics[scale=0.475]{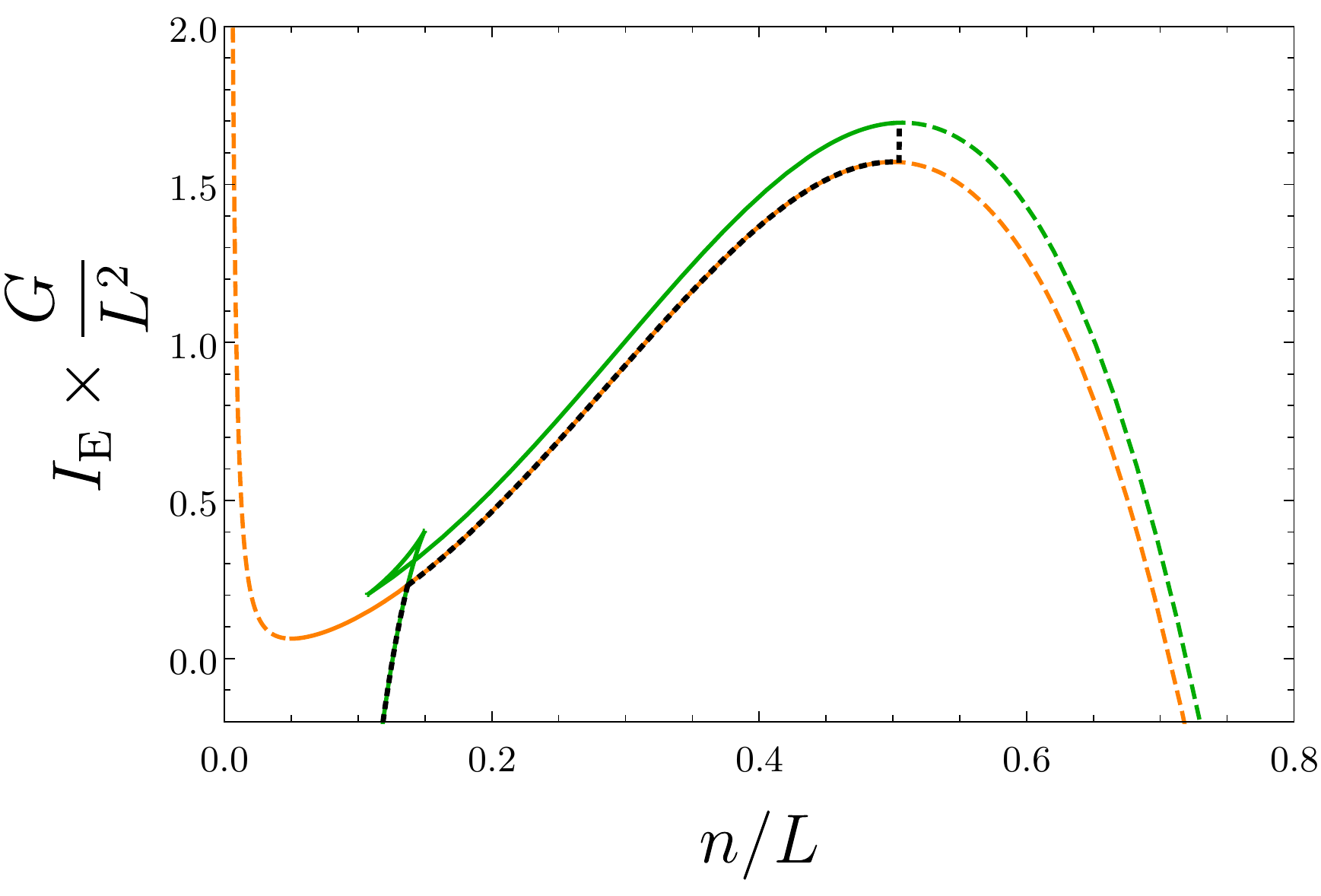}
	\includegraphics[scale=0.485]{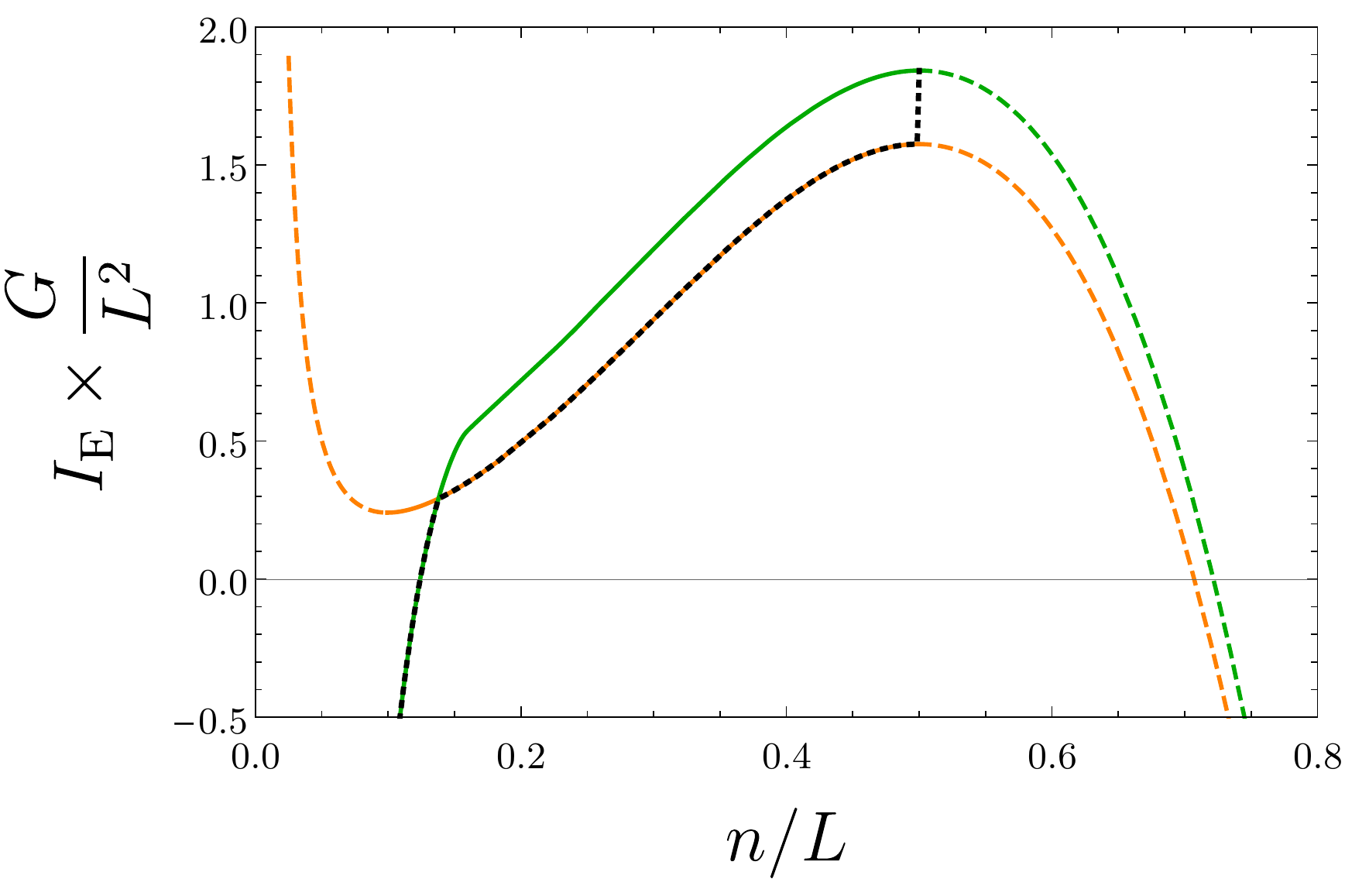}
	\includegraphics[scale=0.47]{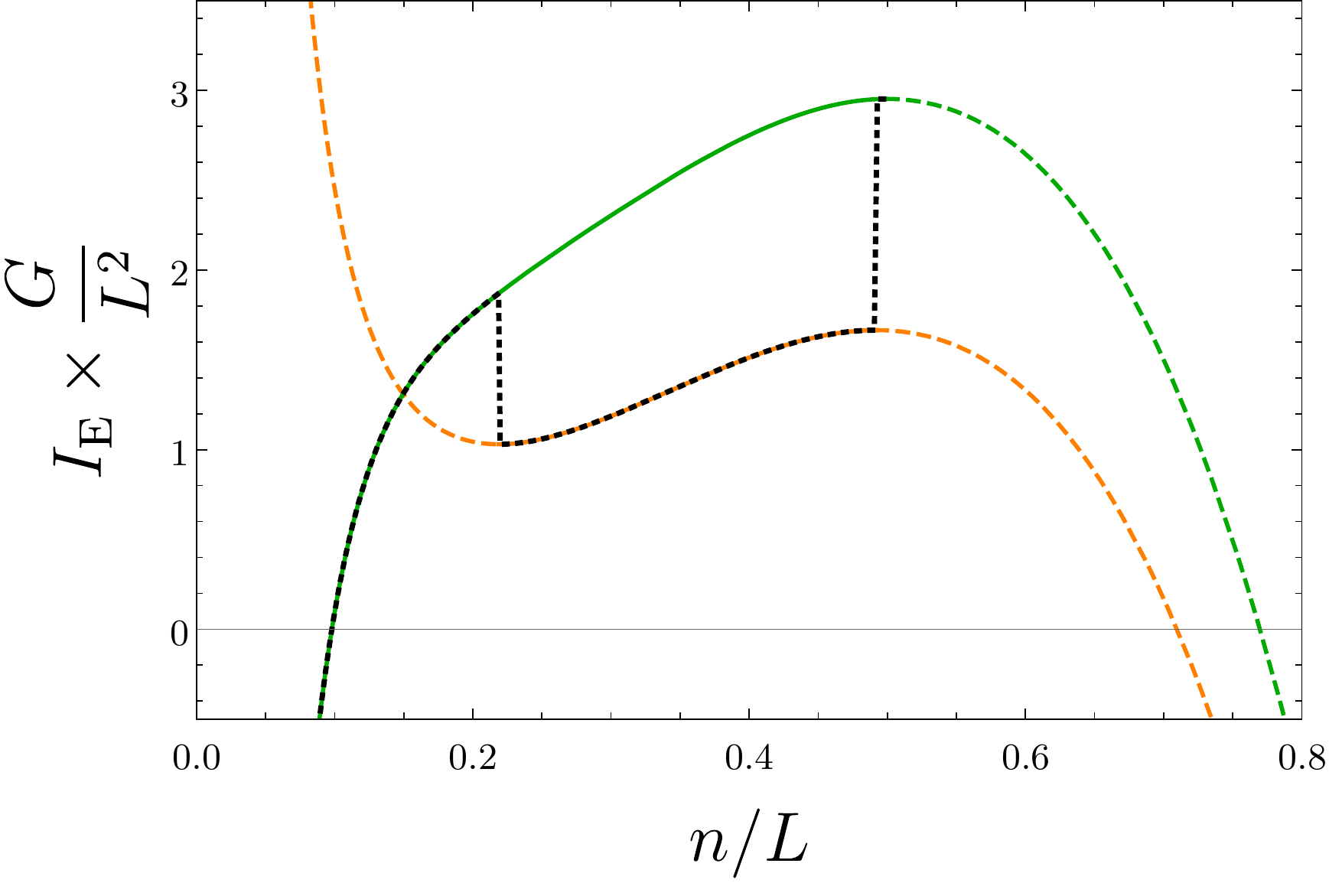}
	\caption{Comparison of Euclidean on-shell actions for NUT and bolt solutions for $\mathcal{B}=\mathbb{S}^2$ in ECG. Orange lines correspond to NUT solutions,  green ones correspond to bolt solutions, and the black dotted line represents the dominant contribution. Dashed lines represent configurations with $\mu M<0$, so that such solutions do not exist.
	Top, left: Einstein gravity result $(\mu=0)$. Top, right: $\mu=0.0001$. Bottom, left: $\mu=0.0015$. Bottom, right: $\mu=8/270\approx0.0296$. The vertical black dotted lines correspond to zeroth-order phase transitions in the points where the NUT solutions cease to exist.  } 
	\label{IEcomp4D}
\end{figure}

Finally, let us compare the Euclidean action of NUT and bolt solutions in order to determine which one dominates the partition function.  In Fig.~\ref{IEcomp4D} we compare the Euclidean actions for several values of $\mu$. The case for Einstein gravity is shown in the top-left panel, and we can see that the NUT solution dominates in all the region $n>L/6\sqrt{7-2\sqrt{10}}$, where a first order phase transition NUT/bolt takes place. In particular, there are no bolt solutions near the undeformed 3-sphere $n^2=L^2/4$. When we switch $\mu$ on, there are some drastic changes. Specially, we recall that for positive values of $\mu$ there are no solutions with negative mass. We plot with dashed lines the would-be Euclidean action of these solutions, but they do not actually exist. This has the effect of inducing zeroth-order phase transitions in the points where some solution ceases to exist. Another new feature is the existence of bolt solutions near the round 3-sphere $n^2=L^2/(4 f_{\infty})\equiv n_0^2$ for all values of $\mu>0$. In all the cases, we observe that for $n=n_0$ the NUT solution (corresponding to pure AdS) dominates, but for $n>n_0$ the NUT solution does not exist because it would have negative mass. However, for values of $n$ slightly larger than $n_0$, there is still a bolt solution of positive mass, and a zeroth-order phase transition from NUT to bolt must take place at $n_0$.  For larger values of $n$, the bolt solution also acquires a negative mass and there are no solutions. The behaviour is more interesting in the region $n<n_0$. In all the cases the NUT solution dominates until certain value $n=n_{\rm min}$, where there is a transition to a bolt solution. When $\mu<0.00569$, the transition is of first-order, as shown in top-right and bottom-left panels in Fig.~\ref{IEcomp4D}. When $\mu>0.00569$, the mass of the NUT solution vanishes before the value of the Euclidean action crosses that of the bolt solution, and a zeroth-order phase transition takes place, as shown in the bottom-right panel. After that phase transition the bolt solution exists and dominates for $0<n<n_{\rm min}$.

The appearance of zeroth-order transitions in Taub-NUT solutions is a new feature whose interpretation is not clear to us.  This seems to be a problem that only appears in four dimensions, since, as we will see, in six dimensions there is no restriction on the mass of the solutions.

\subsection{$\mathcal{B}=\mathbb{T}^2$}\label{SecT2}
Let us now consider a toroidal base space, so that the metric ansatz \req{FFnut} reads
\begin{equation}\label{k0metric}
ds^2=V_{\mathbb{T}^2}(r) \left(d\tau+ \frac{2n}{L^2} \eta d\zeta \right)^2+\frac{dr^2}{V_{\mathbb{T}^2}(r)}+\frac{(r^2-n^2)}{L^2}(d\eta^2+d\zeta^2)\, ,
\end{equation}
Here, the coordinates $(\eta,\zeta)$ parametrize a $\mathbb{T}^2$ with periods which we choose to be equal, $\beta_{\eta,\zeta}=l$.
We note that, unlike the spherical case, the periodicity of the variable $\tau$, which we denote $\beta_{\tau}\equiv 1/T$, is not a priori fixed in terms of $n$  \cite{Dehghani:2005zm,Astefanesei:2004ji}. 
The function $V_{\mathbb{T}^2}(r)$ satisfies \req{eqV} with $k=0$. From the general asymptotic expansion \req{rasymp}, we can obtain the metric of constant-$r$ hypersurfaces for $r\gg n$. This reads
\begin{equation}\label{bdryplanar}
\frac{^{(3)}ds^2_{\infty}}{r^2}= \left[\frac{f_{\infty}}{L^2}\left(d\tau+ \frac{2n}{L^2} \eta d\zeta \right)^2+\frac{(d\eta^2+d\zeta^2)}{L^2}\right]\, .
\end{equation}
If we define $z=\tau/(2n)$, $\eta/L=-x$, $\zeta/L=y$, this can be rewritten as
\begin{equation}\label{nil1}
\frac{^{(3)}ds^2_{\infty}}{r^2}= \left[\frac{4n^2 f_{\infty}}{L^2}\left(dz- x dy\right)^2+dx^2+dy^2\right]\, .
\end{equation}
Remarkably, when $n^2=L^2/(4f_{\infty})$ --- \ie for the same value $n$ for which in the $\mathcal{B}=\mathbb{S}^2$ case the corresponding boundary metric becomes that of a round $\mathbb{S}^3$ --- this reduces to the so-called Nil geometry\footnote{In fact, an additional change of variables can be used to rewrite \req{nil1} in the Nil form for any value of $n$, up to an overall factor. However, such coordinate change would involve making the periods of $\eta$ and $\zeta$ depend on $n$.} \cite{Thurston}. 
The appearance of this kind of geometry should not come as a surprise, as $\mathbb{T}^m$-bundles over tori $\mathbb{T}^n$ are always compact 2-step nilmanifolds (and vice versa) \cite{Palais} --- in our case above, $m=1$ and $n=2$. 


On the other hand, it is also natural to define $\hat \tau=\sqrt{f_{\infty}}\tau$, whose periodicity is $\beta_{\hat \tau}=\sqrt{f_{\infty}}\beta_{ \tau}\equiv 1/\hat T$. Then, \req{bdryplanar} reduces to the standard metric on $\mathbb{T}^3$ for $n=0$, 
\begin{equation}\label{hatT}
\frac{L^2}{r^2}\, \left. ^{(3)}ds^2_{\infty}\right|_{n=0}=d\hat \tau^2+d\eta^2+d\zeta^2\,,
\end{equation}
so we can also understand \req{bdryplanar} as a sort of twisted three-torus metric.


\subsubsection{Taub-NUT solutions}
Let us start with the NUT solutions. Just like in the previous section, we assume that $V_{\mathbb{T}^2}(r=n)=0$, and we impose $V_{\mathbb{T}^2}'(r=n)=4\pi T$ in order to avoid a conical singularity at the NUT. Then, we can consider a Taylor expansion around $r=n$ of the form
\begin{equation}
V(r)=4\pi T (r-n)+\sum_{i=2}^{\infty}(r-n)^i a_ i \, .
\end{equation}
Plugging it into \req{eqV} and solving order by order in $(r-n)$, we obtain the following relations for the first terms
\begin{eqnarray}
GM&=&-\frac{4 n^3}{L^2}+\mu L^4 (4\pi  T)^3\, ,\\
0&=&\mu L^4 (4\pi T)^2\left(a_2-\frac{2\pi T}{n}\right)\, ,\\
0&=&8\pi T\left[-2+3\mu  L^4 \left(a_2^2 -\frac{3 \pi  a_2 T}{n}+\frac{2 \pi   T \left(5 \pi   T-2 a_3 n^2\right)}{n^2}\right)\right]\, .
\end{eqnarray}
The first equation fixes the ``mass'' $M$ in terms of $n$, $L$, $\mu$ and $ T$, while the rest give us relations between the coefficients of the expansion and the temperature. We can try to solve these relations in two inequivalent ways. 

The first possibility, which is the only one available for Einstein gravity, 
consists in setting $ T=0$ --- see equations \req{NUTE}, \req{massE} and \req{betaEin} with $k=0$. This solves the last two equations and, in fact, completely determines the series expansion for any value of $\mu$, \ie we can obtain $a_2$, $a_3$, etc., from the subsequent equations. The series is convergent in a vicinity of $r=n$. However, note that in that case, $GM=-4 n^3/L^2 <0$. Hence, according to the general discussion about the asymptotic behaviour, we expect this solution to be pathological at infinity unless some miraculous fine-tuning occurs. Unfortunately, this is not the case, and when we solve \req{eqV} starting from the near-horizon expansion with $ T=0$, the oscillatory character appears asymptotically. We are then led to conclude that regular extremal NUT solutions do not exist for any allowed value of $\mu$.

The second possibility is setting $a_2=2\pi T/n$, which solves the second equation. The following equations can be used to determine the remaining coefficients, which turn out to have a nonperturbative dependence on $\mu$, \eg 
\begin{equation}
a_3=\frac{2 \pi  T}{n^2}-\frac{1}{6 \pi \mu L^4  T}\, .
\end{equation}
Observe that these do not possess a finite limit when $\mu\rightarrow 0$. As a matter of fact, we have failed to construct these solutions for any nonvanishing value of $\mu$ different from the critical limit value $\mu=4/27$ --- see Section \ref{critic} --- so we strongly suspect that no regular NUT solution exists for $\mathcal{B}=\mathbb{T}^2$ for any $0<\mu<4/27$. 


\subsubsection{Taub-bolt solutions}
\begin{figure}[t]
	\centering 
	\includegraphics[scale=0.6]{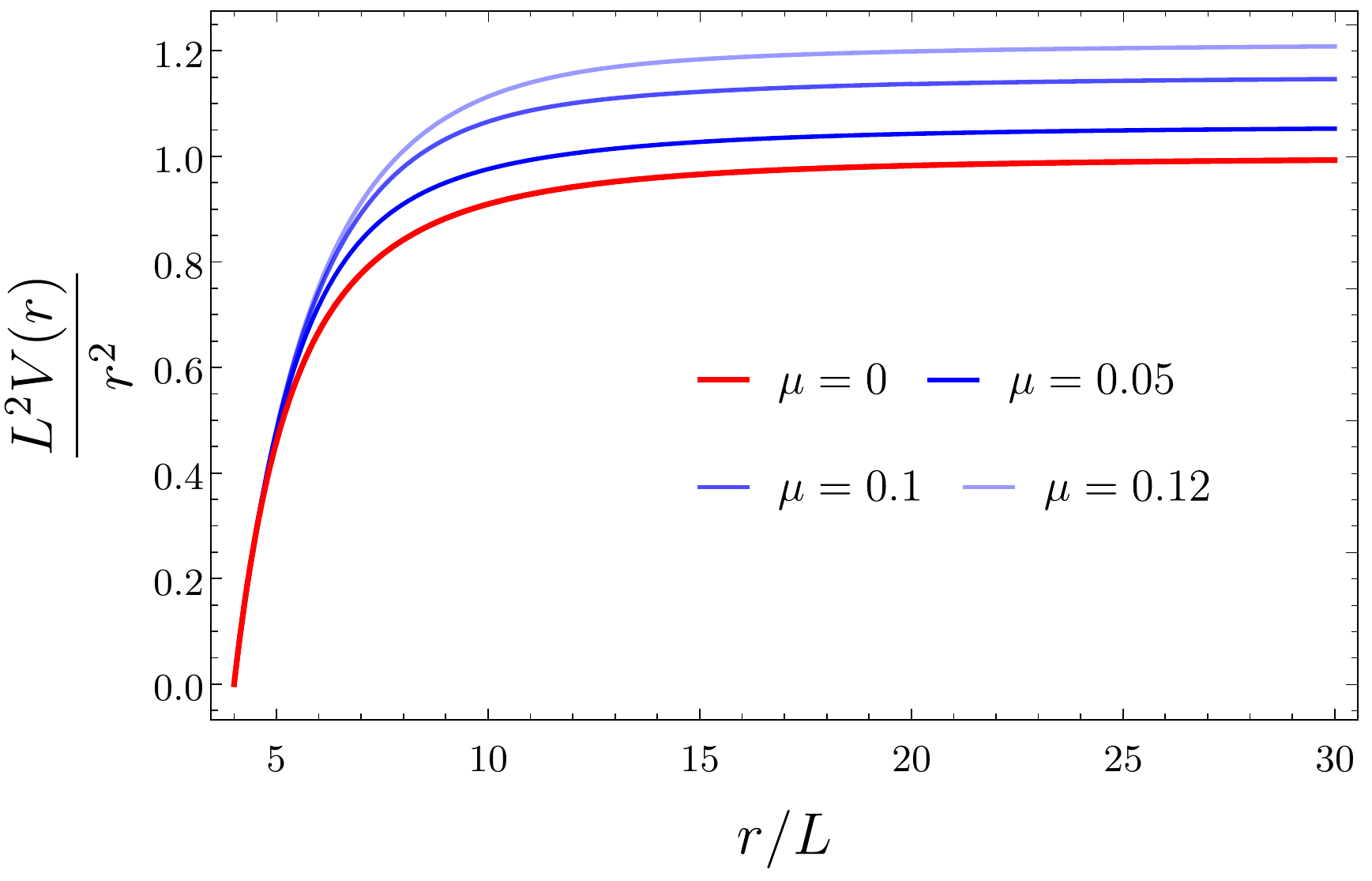}
	\caption{We plot the metric function $ V_{\mathbb{T}^2}(r)\cdot L^2/r^2$ corresponding to Taub-bolt solutions of ECG with $n/L=1$ and $r_b/L=4$ for several values of $\mu$.  	} 
	\label{Planarbolt}
\end{figure}

Fortunately, the situation is different for bolt solutions. In that case, we impose the existence of some $r_b>n$ such that near $r=r_b$,
\begin{equation}\label{PBexp}
V_{\mathbb{T}^2}(r)=4\pi  T (r-r_b)+\sum_{i=2}^{\infty}(r-r_b)^i   a_ i\, .
\end{equation}
This fixes $V_{\mathbb{T}^2}(r_b)=0$ and $V_{\mathbb{T}^2}'(r_b)=4\pi  T$. Again, we plug this expansion in \req{eqV}, and from the first two terms we get 
\begin{eqnarray}\label{MTplanar}
GM&=&\frac{ \left(-3 n^4-6 n^2 r_b^2+r_b^4\right)}{2L^2 r_b}-\frac{1}{8}\mu  L^4 (4\pi T)^3\, ,\\ \label{MTplanar2}
0&=&\frac{6 \left(r_b^2-n^2\right)^2}{L^2 r_b^2}-\frac{8 \pi   T \left(r_b^2-n^2\right)}{r_b}-\frac{3\mu L^4 n^2 (4\pi T)^3}{r_b(r_b^2-n^2)}\, .
\end{eqnarray}
As usual, the first equation fixes $M$, while the second relates $r_b$ to $n$ and $ T$. It turns out that for $\mu\ge 0$, there is a unique solution for $ T$ for every $n$ and $r_b>n$. The solution can be written explicitly as
\begin{equation}\label{Tplanar}
 T=\frac{\left(r_b^2-n^2\right) \left(2 r_b^{2/3}-\left(\sqrt{729 \mu  n^2+8 r_b^2}-27n \sqrt{\mu } \right)^{2/3}\right)}{12 \pi  L^2 n r_b^{1/3} \sqrt{\mu} \left(\sqrt{729 \mu  n^2+8 r_b^2}-27 n\sqrt{\mu}\right)^{1/3}}
\end{equation}
For a given $n$, this is a one-to-one relation between every $r_b>n$ and $ T>0$. In particular, we have $\lim_{r_b \rightarrow n} T=0$. However, in order to keep $GM$ positive, $r_b$ is bounded from below, $r_b\ge n \gamma(\mu)$, for some constant $\gamma(\mu)$. In particular, for the limiting cases $\mu=0$ and $\mu=4/27$, we have, respectively, $\gamma(0)=\sqrt{3+2\sqrt{3}}\simeq 2.5425$, and $\gamma(4/27)\simeq 4.0171$. In each case, for a given $n$, the radius $r_b$ and the ``mass'' $M$ are fixed by the periodicity of the coordinate $ \tau$. The remaining coefficients in the expansion \req{PBexp} are fully determined once we choose $a_2$, which is the only free parameter.  Once again, this  is fixed by demanding the solution to have the correct asymptotic behaviour. In all cases we find that there is one and only one value of $a_2$ for which this happens, and so the solutions are completely determined by $n$ and $ T$.  In Fig. \ref{Planarbolt} we show some of the metric functions corresponding to these solutions computed numerically.

Let us now study the thermodynamic properties of the solutions. The Euclidean action can be once again evaluated  using the generalized action \req{assd}, following the same steps as in the previous subsections.  The result is
\begin{equation}\label{FEP4d}
I_E=
-\frac{l^2}{8 \pi GT }\left[\frac{r_b^4+3n^4}{2r_b L^4}-\mu L^2\frac{(11n^2+r_b^2)(2\pi T)^3}{(r_b^2-n^2)}\right]\, ,
\end{equation}
where  we used \req{MTplanar2} to simplify the result. This on-shell action should be understood as a function of $T$ and $n$, which appear implicitly through $r_b$. In the case of Einstein gravity, for which the metric function can be obtained analytically --- see \req{boltEi} with $k=0$ --- the result for the on-shell action can be written explicitly as a function of $T$ and $n$. Using \req{boltrb} with $k=0$, one finds
\begin{equation}
I_{ E}=-\frac{l^2}{108 \pi G T L^4}\left[8 \pi^3 L^6  T^3+\left(4\pi^2 L^4 T^2+9 n^2\right)^{3/2}\right]\, .
\end{equation}
Now we must account for the fact that we have an extended thermodynamic phase space, since $n$ is in this case a free variable. However, $n$ cannot be the appropriate thermodynamic variable as it has units of length instead of energy. Hence, let us define $\theta\equiv 1/n$, which has the right units. Then, associated with $T$ and $\theta$, we have two potentials: the usual entropy $S$, and a new potential $\Psi$. In terms of the free energy $F\equiv T I_E$, these are given by
\begin{equation}\label{SPsi}
S=-\left(\frac{\partial F}{\partial T}\right)_{\theta}\, ,\quad \Psi=-\left(\frac{\partial F}{\partial \theta}\right)_{T}\, ,
\end{equation}
which explicitly read
\begin{eqnarray}
S&=&\frac{\pi l^2 T}{9 G} \left(2 \pi  L^2 T+\sqrt{4 \pi ^2 L^4 T^2+\frac{9}{\theta ^2}}\right)\, ,\quad \Psi=-\frac{l^2\sqrt{4 \pi ^2 L^4 T^2+\frac{9}{\theta ^2}}}{4 \pi  G L^4 \theta ^3 }\, .
\end{eqnarray}
Finally, the energy is defined as $E=F+TS+\theta\Psi$, so that, by construction it satisfies the first law
\begin{equation}
dE=T dS+\theta d\Psi\, .
\end{equation}
The energy is given by
\begin{equation}
E_{}=\frac{l^2}{27\pi G}\left[4 \pi^3 L^2 T^3+\left(2\pi^2T^2-\frac{9}{\theta^2 L^4}\right)\sqrt{4 \pi ^2 L^4 T^2+\frac{9}{\theta ^2}}\right]\, .
\end{equation}
Now, this is a thermodynamic energy, but the energy of the solution should be computed using the ADM formula, which in this case tells us that $E_{\rm ADM}=M l^2/(4\pi L^2)$. Using the expression for $M$ given in \req{MTplanar}, we have checked that both energies actually coincide $E_{\rm ADM}=E$. Hence, the introduction of the variable  $\theta$ is crucial for the first law of black hole mechanics to hold in this case.

\begin{figure}[t]
	\centering 
	\includegraphics[scale=0.45]{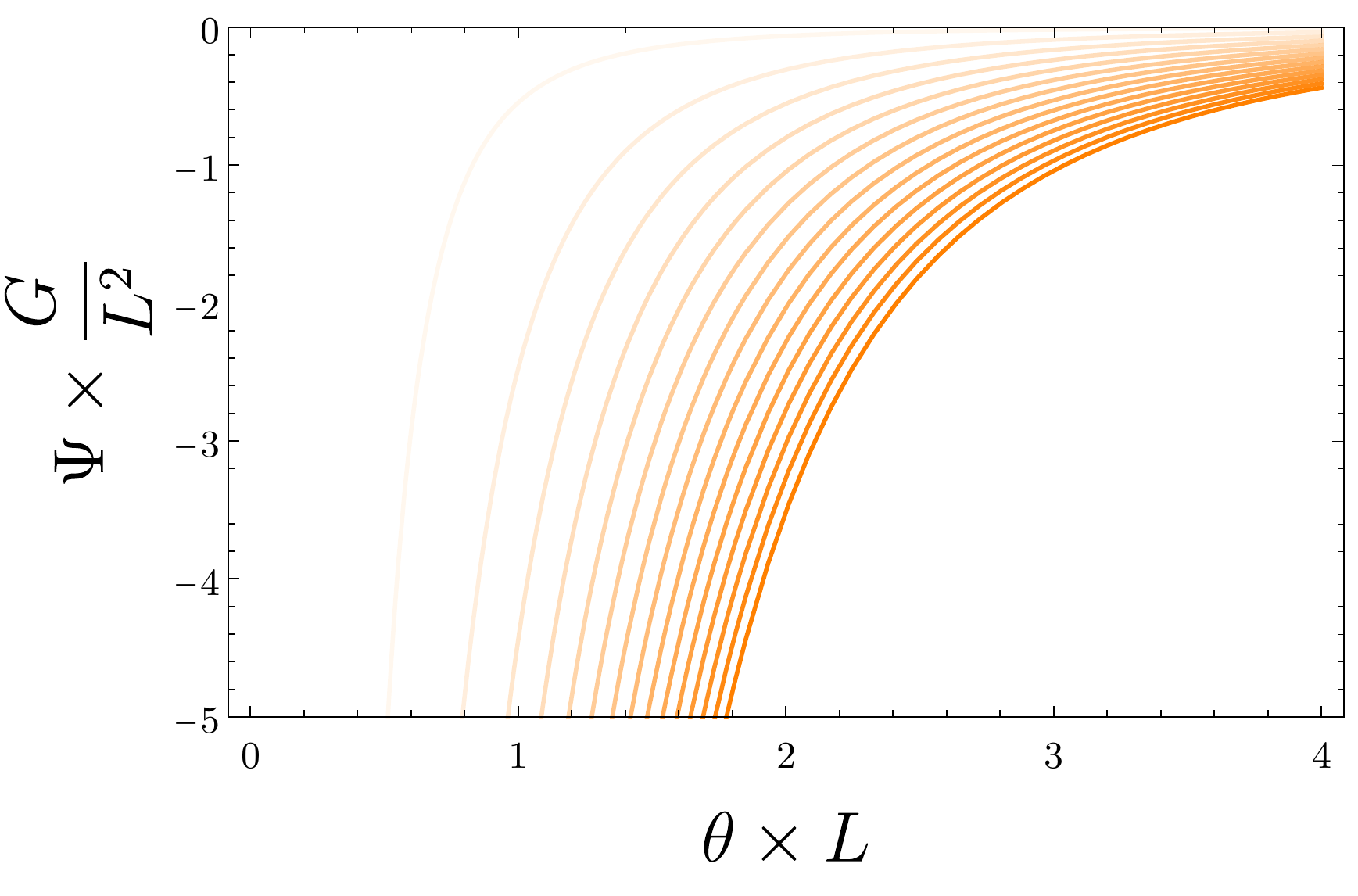}
	\includegraphics[scale=0.45]{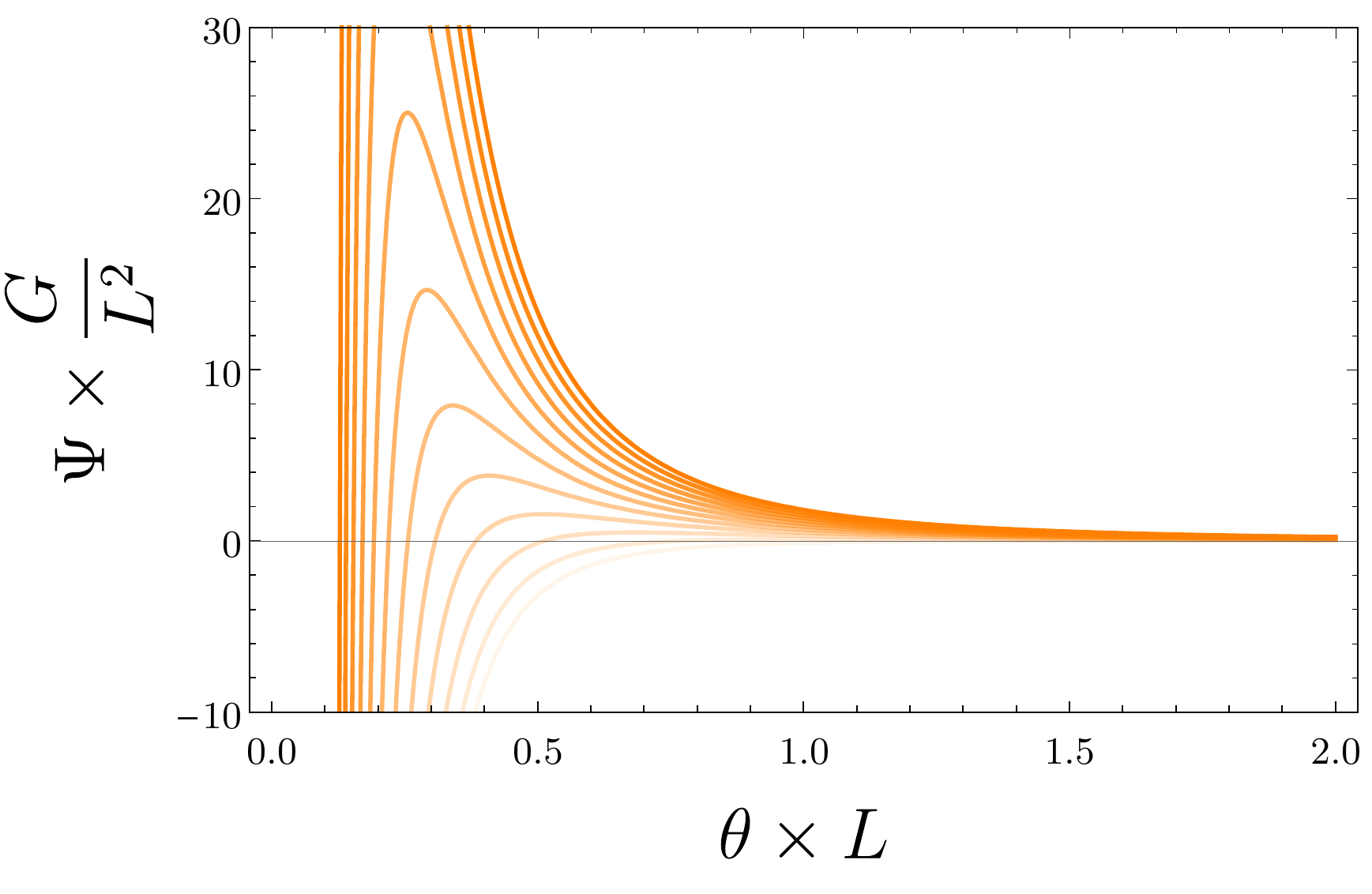}
	\caption{Isotherms in the $\Psi$-$\theta$ plane. Left: $\mu=0$. Right: $\mu=0.12$.} 
	\label{Psidiagram}
\end{figure}

This picture goes through nicely  when the ECG term is turned on. In that case, it is convenient to express the thermodynamic quantities in terms of the rescaled temperature $\hat T=T/\sqrt{f_{\infty}} $ introduced above equation \req{hatT}. In terms of this, we have the free energy $F(\hat T,\theta; \mu)=\sqrt{f_{\infty}}\hat T I_E$, which can be obtained from \req{FEP4d}, and the thermodynamic potentials $S(\hat T, \theta; \mu)$ and $\Psi(\hat T,\theta; \mu)$ defined as in \req{SPsi} (but with respect to $\hat T$ instead of $T$). We find that
\begin{equation}
E_{\rm ADM}=F+\hat TS+\theta\Psi\, ,
\end{equation}
where the ADM energy is now given by $E_{\rm ADM}=M l^2/(4\pi L^2\sqrt{f_{\infty}})$ and
\begin{eqnarray}
S&=&\frac{l^2}{4GL^2}\left[\frac{r_b^2\theta^2-1}{\theta^2}-12\mu L^4\frac{\pi^2f_{\infty} \hat T^2(5+r_b^2\theta^2)}{\rh^2\theta^2-1}\right]\, ,\\
\Psi&=&\frac{l^2}{8\pi G L^4\sqrt{f_{\infty}}\theta^3}\left[3r_b(\theta^2r_b^2-3)+4\pi L^2 \hat T \sqrt{f_{\infty}}(1-\theta^2r_b^2)\right]\,  .
\end{eqnarray}
It is interesting to study the isotherms on the diagram of $\Psi$-$\theta$. These are shown in Fig. \ref{Psidiagram} for $\mu=0$ and $\mu=0.12$. In the case of Einstein gravity --- the same happens for small values of $\mu$ --- the isotherms are monotonous. However, when $\mu$ is large enough, the diagram changes drastically. In that case, the isotherms develop a maximum, and the limit $\Psi \rightarrow 0$ corresponding to $\theta \rightarrow + \infty$ is approached from above instead of from below. However, the phase space seems to be free of critical points.

We have seen that to satisfy the quantum-statistical relation and first law, it is necessary to treat $\theta = 1/n$ as a thermodynamic variable. If we also wish to satisfy the Smarr formula, then once again we must consider both $\Lambda$ and $\mu$ to be thermodynamic parameters. The basic construction is identical to the $\mathcal{B} = \mathbb{S}^2$ case, but now we include $\theta$ as well. A simple computation then reveals the following for the thermodynamic volume and coupling potential
\begin{equation}
V = \frac{l^2 \rh}{3 L^2} \left(\rh^2 - \frac{3}{\theta^2} \right) \, , \quad \Upsilon^{\ssc \rm ECG} = \frac{\pi^2 l^2 T^3}{G L^2 } \left(\frac{\rh^2\theta^2 + 5}{\rh^2 \theta^2-1} \right)\, ,
\end{equation}
where in the second equation above we note that it is the un-normalized temperature that appears (i.e. $T$ rather than $\hat{T}$).  With these definitions, the extended first law and Smarr formula hold, with the latter being identical to Eq.~\eqref{smarrfS2} with the additional term $2\theta\Psi$ added.

Let us close this section by mentioning the possibility that the solutions considered in this subsection can be relevant holographically. In that context, and in analogy to the Taub-bolt solutions with $\mathcal{B}=\mathbb{S}^2$, we expect them to represent saddle points in the semiclassical partition function for boundary theories living on deformed tori with metric \req{bdryplanar}.
%
 While in the spherical case the boundary is only characterized by $n$ --- or, equivalently, the squashing parameter $\alpha$ ---, the $\mathcal{B}=\mathbb{T}^2$ case is richer, given that  $n$ and $\hat T$ are independent parameters in that case. 

\subsection{Exact Taub-NUT solutions in the critical limit}\label{critic}
In this subsection we study the Taub-NUT solutions of critical ECG, which can be constructed analytically. As we mentioned earlier, when $\mu=4/27$, the only AdS vacuum has a length scale $\tilde L^2=2 L^2/3$, and the linearized equations on that background vanish identically \cite{Feng:2017tev}. 
 The field equations simplify considerably, which has allowed for the construction of analytic black hole solutions \cite{Feng:2017tev}.  In the case of NUT-charged metrics, a similar simplification takes place, and we find the following family of exact Taub-NUT solutions,
\begin{equation}\label{tauc}
ds^2=(r^2-n^2)\left[\frac{3}{2L^2}\left(d\tau+n A_{\mathcal{B}}\right)^2+d\sigma_{\mathcal{B}}^2\right]+\frac{2 L^2dr^2 }{3(r^2-n^2)}\, .
\end{equation}
As we mentioned before, in the case of a spherical base space, $\mathcal{B}=\mathbb{S}^2$, this solution has a conical singularity at $r=n$, except for $n^2=L^2/6$, in whose case the solution is simply globally Euclidean AdS$_4$ --- also known as  $\mathbb{H}^4$. Hence, only the cases $\mathcal{B}=\mathbb{T}^2$, $\mathbb{H}^2$ are of interest in Euclidean signature. 

The solutions \req{tauc} can be analytically continued to Lorentzian signature in different ways, giving rise to very interesting metrics. For example if we make the replacement $n\rightarrow i n$ and $\tau=i t$ we get the following metric
\begin{equation}
ds^2=(r^2+n^2)\left[-\frac{3}{2L^2}\left(dt+n A_{\mathcal{B}}\right)^2+d\sigma_{\mathcal{B}}^2\right]+\frac{2 L^2dr^2 }{3(r^2+n^2)}\, .
\end{equation}
This metric is regular everywhere and, in fact, we can allow $r$ to take values in the whole real line. Hence, this solution usually represents a wormhole or wormbrane, depending on the topology, connecting two asymptotically AdS$_4$ regions. The cases $k=n=0$ and $k=-6n^2/L^2=-1$ are special as they correspond to pure AdS$_4$. Let us introduce a new radial coordinate $r=n \cosh\left(\rho/(\sqrt{2/3}L)\right)$, so that the metric reads
\begin{equation}\label{Lcrit}
ds^2=n^2\cosh^2\left(\frac{\rho}{\sqrt{2/3}L}\right)\left[-\frac{3}{2L^2}\left(dt+n A_{\mathcal{B}}\right)^2+d\sigma_{\mathcal{B}}^2\right]+d\rho^2\, ,
\end{equation}
which has an explicit wormhole character. In the spherical case $\mathcal{B}=\mathbb{S}^2$ the solution reads 
\begin{equation}
ds^2=n^2\cosh^2\left(\frac{\rho}{\sqrt{2/3}L}\right)\left[-\frac{3}{2L^2}\left(dt+2n \cos\theta d\phi\right)^2+d\theta^2+\sin^2\theta d\phi^2\right]+d\rho^2\, .
\end{equation}
This solution has the problem that it suffers from closed time-like curves, because the time coordinate must be periodic $t\rightarrow t+8\pi n$. The $\mathcal{B}=\mathbb{T}^2$, $\mathbb{H}^2$ cases are free of them, because there is no periodicity condition on the time coordinate.
In particular, after some rescalings we can write the $\mathbb{T}^2$ solution as
\begin{equation}\label{whT2-1}
ds^2=\cosh^2\left(\frac{\rho}{\sqrt{2/3}L}\right)\left[-\left(dt+\frac{\sqrt{6}}{L}x dy \right)^2+dx^2+dy^2\right]+d\rho^2\, ,
\end{equation}
where the NUT charge has been absorbed in the period of the coordinates $x$, $y$. However, we can also allow $x$ and $y$ to be noncompact.
Interestingly, there is an inequivalent Lorentzian solution that can be obtained by rotating the coordinates as $(t,y)\rightarrow (i y, i t)$. This reads

\begin{equation}
ds^2=\cosh^2\left(\frac{\rho}{\sqrt{2/3}L}\right)\left[-dt^2+dx^2+\left(dy+\frac{\sqrt{6}}{L}x dt \right)^2\right]+d\rho^2\, .
\end{equation}

Going back to the general solution \req{Lcrit}, we can consider the following transformation: $t\rightarrow z$, $\rho \rightarrow i t$, $L\rightarrow i L$. Here we are changing the sign of $L^2$, which amounts to changing the sign of the cosmological constant in the ECG action \req{ECG}. Hence, the corresponding metric is a solution of the critical theory with a positive cosmological constant. The general solution reads
\begin{equation}
ds^2=-dt^2+n^2\cosh^2\left(\frac{t}{\sqrt{2/3}L}\right)\left[\frac{3}{2L^2}\left(dz+ n A_{\mathcal{B}}\right)^2+d\sigma_{\mathcal{B}}^2\right]\, .
\end{equation}
These represent bouncing cosmologies with different topologies for the spatial sections connecting two asymptotically (NUT charged) de Sitter spaces for $t\rightarrow\pm \infty$. The only exception is the case $k=1$, $n^2=L^2/6$, which is actually de Sitter space foliated by $\mathbb{S}^3$ spheres. Particularly relevant for cosmology is the flat case $k=0$, which after rescaling of the coordinates can be written as
\begin{equation}\label{bouncing}
ds^2=-dt^2+\cosh^2\left(\frac{t}{\sqrt{2/3}L}\right)\left[\left(dz+ \frac{\sqrt{6}}{L}x dy \right)^2+dx^2+dy^2\right]\, .
\end{equation}
The transverse geometry is again a Nil space. Interestingly, this solution represents a homogeneous but nonisotropic bouncing cosmology. Homogeneity follows from the fact that Nil space is a coset space and it possesses the isometries $(x,y,z)\rightarrow (x+a,y+b, z+c-a\sqrt{6}y/L)$, for arbitrary $(a, b, c)$. Let us also mention in passing that this solution seems to be disconnected from the isotropic and homogeneous bouncing solution found in \cite{Feng:2017tev}, since we do not recover it in any limit.

\section{Six dimensions: Quartic generalized quasi-topological gravities}\label{sIx}


We now move on to consider theories in six dimensions. In this case the generalized quasi-topological gravity class \cite{Hennigar:2017ego,PabloPablo3,Ahmed:2017jod} includes additional densities beyond those in   $D=4$. In particular, the Gauss-Bonnet term $\mathcal{X}_4$ is no longer topological. Analytic generalizations of the Einstein gravity static black-hole \cite{Boulware:1985wk,Cai:2001dz} and Taub-NUT/bolt solutions \cite{Dehghani:2005zm,Dehghani:2006aa} have been constructed in the presence of this contribution. In particular, the Taub-NUT solutions of Gauss-Bonnet \cite{Dehghani:2005zm,Dehghani:2006aa}  
and 3rd-order Lovelock gravity \cite{Hendi:2008wq}   were, prior to this paper, and to the best of our knowledge, the only known examples of solutions of that class for any  higher-curvature gravity theory.
 
 In principle, the six-dimensional GQTG family includes two nontrivial terms at cubic order, corresponding to the usual quasi-topological gravity density, plus an additional one. As observed in \cite{Quasi,Quasi2,Oliva:2010zd}, including the quasi-topological gravity density in $D\geq 6$ is equivalent, from the point of view of static black-hole solutions, to including the cubic Lovelock interaction. In $D=6$, this is a topological term, and therefore no new nontrivial black holes exist in that case for quasi-topological gravity. They do exist, however, when the additional GQTG term is included \cite{Hennigar:2017ego}. Hence, following the same reasoning as for ECG in $D=4$, one would have expected that new Taub-NUT solutions of the form \req{FFnut} should also exist for GQTG. Remarkably, we find that this is not the case, and that no cubic theory admits nontrivial generalizations of the Einstein gravity or Gauss-Bonnet Taub-NUT solutions characterized by a single function in six dimensions (independent of the base manifold considered).
While there do exist cubic theories that satisfy the necessary constraints to admit solutions of the form~\eqref{FFnut}, it turns out that for any such theory the field equations on these spaces vanish identically.  This is analogous to the ${\cal C}^{(i)}_{D}$ terms discussed in~\cite{Ahmed:2017jod} for the case of static, spherically symmetric metrics.

Happily, at quartic order in curvature there exist non-trivial options of both quasi-topological and generalized quasi-topological type.\footnote{Recall that the distinction between both classes comes from considering the theories for static spherically symmetric spacetimes. While both admit solutions of the form \req{FFbh}, the field equations for the quasi-topological theories reduce to algebraic polynomial equations for $f(r)$, while for those  of the generalized quasi-topological type, the metric function satisfies a non-linear second order differential equation in each case. } We will not present a detailed classification of such theories here --- see \cite{Dehghani:2011vu} and \cite{Ahmed:2017jod} --- but limit ourselves to some brief remarks. Beginning from a general action containing all 26 possible quartic invariants \cite{Aspects}, we constrain the action by imposing the conditions listed in appendix A of~\cite{Ahmed:2017jod}. This selects theories admitting black hole solutions of the form \req{FFbh} and which, as a consequence \cite{PabloPablo3}, do not propagate ghosts on maximally symmetric backgrounds. Next, for each of the possible four dimensional base manifolds listed below, we generate additional constraints to ensure the corresponding theory admits solutions of the form~\eqref{FFnut}. Surprisingly, demanding the constraints to be simultaneously satisfied for all four base manifolds $\mathcal{B} = \mathbb{CP}^2, \mathbb{S}^2 \times \mathbb{S}^2$, $\mathbb{S}^2 \times \mathbb{T}^2$ and $\mathbb{T}^2 \times \mathbb{T}^2$ results in a family of theories that yield trivial field equations. When one relaxes this condition, considering only a subset of the base manifolds, then nontrivial options exist.    

The non-trivial theories can be classified into two groups: quasi-topological and generalized quasi-topological.
For the base manifold $\mathbb{S}^2 \times \mathbb{S}^2$, the only nontrivial theories are of the generalized quasi-topological type. The constraints can be satisfied simultaneously for $\mathcal{B} = \mathbb{CP}^2, \mathbb{S}^2 \times \mathbb{S}^2$ and $\mathbb{T}^2 \times \mathbb{T}^2$, resulting in a three parameter family of non-trivial theories, each making the same contribution to the field equations for a given base manifold.  We can deduce all of the relevant physics by considering only one member of this class, which we denote as $\mathcal{S}$ below.     


Excluding the base manifold $\mathbb{S}^2 \times \mathbb{S}^2$, then quasi-topological options exist for all remaining base manifolds. The constraints can be satisfied simultaneously, resulting in a three parameter family of non-trivial quasi-topological theories. Again, each theory makes the same contribution to the field equations for a given base manifold, and we need only consider a single member of this family, which we denote as $\mathcal{Z}$ below. Thus, in six dimensions, we will consider the following two quartic Lagrangian densities:  
\begin{align}
{\cal S} &=  992 R_{a}{}^{c} R^{ab} R_{b}{}^{d} R_{cd} + 28 R_{ab} R^{ab} R_{cd} R^{cd} - 192 R_{a}{}^{c} R^{ab} R_{bc} R - 108 R_{ab} R^{ab} R^2 
\nonumber\\
&+ 1008 R^{ab} R^{cd} R R_{acbd} + 36 R^2 R_{abcd} R^{abcd} - 2752 R_{a}{}^{c} R^{ab} R^{de} R_{bdce} + 336 R R_{a}{}^{e}{}_{c}{}^{f} R^{abcd} R_{bedf} 
\nonumber\\
&- 168 R R_{ab}{}^{ef} R^{abcd} R_{cdef} - 1920 R^{ab} R_{a}{}^{cde} R_{b}{}^{f}{}_{d}{}^{h} R_{cfeh} + 152 R_{ab} R^{ab} R_{cdef} R^{cdef} 
\nonumber\\
&+ 960 R^{ab} R_{a}{}^{cde} R_{bc}{}^{fh} R_{defh} - 1504 R^{ab} R_{a}{}^{c}{}_{b}{}^{d} R_{c}{}^{efh} R_{defh} + 352 R_{ab}{}^{ef} R^{abcd} R_{ce}{}^{hi} R_{dfhi} 
\nonumber\\
&- 2384 R_{a}{}^{e}{}_{c}{}^{f} R^{abcd} R_{b}{}^{h}{}_{e}{}^{i} R_{dhfi} + 4336 R_{ab}{}^{ef} R^{abcd} R_{c}{}^{h}{}_{e}{}^{i} R_{dhfi} - 143 R_{ab}{}^{ef} R^{abcd} R_{cd}{}^{hi} R_{efhi} 
\nonumber\\
&- 436 R_{abc}{}^{e} R^{abcd} R_{d}{}^{fhi} R_{efhi} + 2216 R_{a}{}^{e}{}_{c}{}^{f} R^{abcd} R_{b}{}^{h}{}_{d}{}^{i} R_{ehfi} - 56 R_{abcd} R^{abcd} R_{efhi} R^{efhi} \, ,
\\
{\cal Z} &=  -112 R_{a}{}^{c} R^{ab} R_{b}{}^{d} R_{cd} - 36 R_{ab} R^{ab} R_{cd} R^{cd} + 18 R_{ab} R^{ab} R^2 - 144 R^{ab} R^{cd} R R_{acbd} 
\nonumber\\
&- 9 R^2 R_{abcd} R^{abcd} + 72 R^{ab} R R_{a}{}^{cde} R_{bcde} + 576 R_{a}{}^{c} R^{ab} R^{de} R_{bdce} - 400 R^{ab} R^{cd} R_{ac}{}^{ef} R_{bdef} 
\nonumber\\
&+ 48 R R_{a}{}^{e}{}_{c}{}^{f} R^{abcd} R_{bedf} + 160 R_{a}{}^{c} R^{ab} R_{b}{}^{def} R_{cdef} - 992 R^{ab} R_{a}{}^{cde} R_{b}{}^{f}{}_{d}{}^{h} R_{cfeh} 
\nonumber\\
&+ 18 R_{ab} R^{ab} R_{cdef} R^{cdef} - 8 R^{ab} R_{a}{}^{cde} R_{bc}{}^{fh} R_{defh} + 238 R_{ab}{}^{ef} R^{abcd} R_{ce}{}^{hi} R_{dfhi} 
\nonumber\\
&- 376 R_{a}{}^{e}{}_{c}{}^{f} R^{abcd} R_{b}{}^{h}{}_{e}{}^{i} R_{dhfi} + 1792 R_{ab}{}^{ef} R^{abcd} R_{c}{}^{h}{}_{e}{}^{i} R_{dhfi} - 4 R_{ab}{}^{ef} R^{abcd} R_{cd}{}^{hi} R_{efhi} 
\nonumber\\
&- 284 R_{abc}{}^{e} R^{abcd} R_{d}{}^{fhi} R_{efhi} + 320 R_{a}{}^{e}{}_{c}{}^{f} R^{abcd} R_{b}{}^{h}{}_{d}{}^{i} R_{ehfi} \, .
\end{align}
The generalized quasi-topological  term ${\cal S}$ is an appropriate choice for all base manifolds besides $\mathbb{T}^2 \times \mathbb{S}^2$, while the quasi-topological term ${\cal Z}$ is an appropriate choice for all base manifolds besides $\mathbb{S}^2 \times \mathbb{S}^2$.

The complete action we consider is then
\be\label{eqn:quart_6d_act}
I_{\rm E} = - \frac{1}{16 \pi G} \int d^6x \sqrt{g} \left[\frac{20}{L^2} + R + \frac{\lambda_{\rm \ssc GB} L^2 }{6} {\cal X}_4 -\frac{ \xi L^6 }{216} \mathcal{S} - \frac{ \zeta L ^6}{144} \mathcal{Z}  \right] \, ,
\ee
where we have allowed for the possible contribution of the Gauss-Bonnet term. In this case, the AdS$_6$ vacua of the theory are characterized by being solutions to $h(f_\infty)=0$, where 
\be  \label{hfinf}
h(f_\infty) \equiv 1 - f_\infty + \lambda_{\rm \ssc GB} f_\infty^2 + \zeta f_\infty^4 + \xi f_\infty^4 \, ,
\ee
a definition that will turn out to be useful later on.

As anticipated, when we insert the single-function Taub-NUT ansatz \req{FFnut} in the equations of motion of this theory, we are left with a single independent equation for $V_{\mathcal{B}}$, which can be integrated once to leave it in the form \req{mastere},
where the function ${\cal E}_\mathcal{B}$ receives contributions from all terms in \req{eqn:quart_6d_act}, namely,
\be \label{eq6d}
 {\cal E}_\mathcal{B}^{\rm E} + \lambda_{\rm \ssc GB} L^2  {\cal E}_\mathcal{B}^{({\rm  GB})} + \xi L^6   {\cal E}_\mathcal{B}^{({\cal S})} + \zeta L^6  {\cal E}_\mathcal{B}^{(\mathcal{Z})}=
 C_{\mathcal{B}}\, ,
\ee 
where $ C_{\mathcal{B}}$ is an integration constant.
The explicit form of the various terms appearing in the field equation is the following.
The Einstein gravity contributions to the field equation can be expressed in the form
\begin{align}
{\cal E}_\mathcal{B}^{\rm E} =& \frac{6 L^2(n-r)^2(n+r)^2 V - 6 r^6 + (30n^2 - 2 L^2)r^4 + (-90 n^4 + 12 L^2 n^2)r^2 - 30n^6 + 6 L^2 n^4}{3 L^2 r} \notag
\\
&-\frac{(3 n^4 + 6 n^2 r^2 - r^4)(1+\kappa)}{3r}\, ,
\end{align}
where $\kappa$ is defined by 
\be \label{kappa6}
\kappa = \begin{cases} 
	-1 \quad &\text{for } \mathcal{B} = \mathbb{CP}^2 \text{ and }\mathbb{S}^2 \times \mathbb{S}^2\, ,
	\\
	0 \quad &\text{for } \mathcal{B} = \mathbb{S}^2 \times \mathbb{T}^2\, ,
	\\
	+1 \quad & \text{for } \mathcal{B} = \mathbb{T}^2 \times \mathbb{T}^2\, .
\end{cases}
\ee
Next are the Gauss-Bonnet contributions, which for the various base spaces read
\begin{align}
{\cal E}_{\mathbb{CP}^2}^{\rm  GB} =& -\frac{2n^2}{9 r} \left(9 V^2 + 6 V + 2 \right) -\frac{2r}{9 } \left(9 V^2 - 6 V + 2 \right) \, ,
\\
{\cal E}_{\mathbb{S}^2 \times \mathbb{S}^2}^{\rm  GB} =& -\frac{2n^2}{3 r} \left(3 V^2 + 2 V + 1 \right) -\frac{2r}{3 } \left(3 V^2 - 2 V + 1 \right) \, ,
\\
{\cal E}_{\mathbb{S}^2 \times \mathbb{T}^2}^{\rm GB} =& -\frac{2n^2}{3 r} V \left(3 V  + 1 \right) -\frac{2r}{3 } V \left(3 V - 1 \right) \, ,
\\
{\cal E}_{\mathbb{T}^2 \times \mathbb{T}^2}^{\rm GB} =& - 2 V^2 \left(\frac{n^2}{r} + r \right) \, .
\end{align}
The quartic contributions to the field equations are, of course, more complicated. The ones due to the generalized quasi-topological term read
\begin{align}
{\cal E}_\mathcal{B}^{(\mathcal{S})} &= - \frac{16}{3} \Bigg[  \bigg(\frac{18n^4 + 37n^2 r^2 + 9r^4}{(n-r)^2(n+r)^2r} V^3  + \frac{19n^2 + 9 r^2}{(n-r)(n+r)} V^2 V' + \frac{n^2 + 9 r^2}{4 r} V (V')^2\bigg)V''
\nonumber\\
&- \frac{n^2 + 9 r^2}{16 r} (V')^4 + \frac{5n^2 - 3 r^2}{4(n-r)(n+r)} V (V')^3 + \frac{31n^4 + 98 n^2 r^2 + 9 r^4}{2(n-r)^2(n+r)^2r} V^2(V')^2   
\nonumber\\
&+ \frac{152n^4+143n^2r^2+9r^4}{(n-r)^3(n+r)^3}V^3V' +\frac{375n^6+1693n^4r^2+817n^2r^4+27r^6}{8 (n-r)^4(n+r)^4r} V^4 + \mathfrak{E}_\mathcal{B}^{(\mathcal{S})} \Bigg]  \, ,
\end{align}
where $\mathfrak{E}_\mathcal{B}$ is a base-dependent contribution, which takes the explicit form
\begin{align}
\mathfrak{E}_{\mathbb{CP}^2}^{(\mathcal{S})} &= \bigg(\frac{6(n^2+r^2)}{r(n-r)(n+r)} V + 3 V' + \frac{1}{2 r} \bigg) V V'' - (V')^3 + \bigg(\frac{n^2}{r(n-r)(n+r)}V - \frac{1}{4 r}\bigg)(V')^2 
\nonumber\\
&+ \bigg(\frac{2(14 n^2 + 3 r^2)}{(n-r)^2(n+r)^2} V^2 + \frac{V}{2(n-r)(n+r)} \bigg) V' + \frac{33n^4+86n^2r^2+9r^4}{2 (n-r)^3(n+r)^3r} V^3 
\nonumber\\
&+ \frac{3(n^2+r^2)}{2(n-r)^2(n+r)^2 r} V^2 \, ,
\\
\mathfrak{E}_{\mathbb{S}^2 \times \mathbb{S}^2}^{(\mathcal{S})} &= \bigg(\frac{6(n^2+r^2)}{(n-r)(n+r)r} V + 3 V' \bigg)V V'' - (V')^3 + \frac{n^2}{(n-r)(n+r)r} V (V')^2 
\nonumber\\	
&+ \frac{2(14 n^2 + 3 r^2)}{(n-r)^2(n+r)^2}V^2 V' 
+ \frac{33n^4+86n^2r^2+9r^4}{2 (n-r)^3(n+r)^3r} V^3 
- \frac{3(3n^2-r^2)}{4(n-r)^2(n+r)^2r} V^2 
\nonumber\\
&- \frac{V}{2(n-r)(n+r)r} - \frac{-r\log\left[\frac{r+n}{r-n} \right]+ 2 n}{16 n^3 r}\, ,
\\
\mathfrak{E}_{\mathbb{T}^2 \times \mathbb{T}^2}^{(\mathcal{S})} &=	0 \, .
\end{align}
On the other hand, the quartic quasi-topological contributions yield
\begin{align}
{\cal E}_\mathcal{B}^{(\mathcal{Z})} =& \frac{2}{9} \bigg[ 40 \bigg(\frac{4 n^2 r}{(n-r)^2(n+r)^2} V^3 + 4 \frac{n^2}{(n+r)(n-r)} V' V^2 + \frac{n^2}{r} (V')^2 V  \bigg) V'' - \frac{10 n^2}{r} (V')^4 
\nonumber\\
&+ \frac{20 n^2}{(n-r)(n+r)} V (V')^3 + \frac{140n^2(n^2 + 2 r^2)}{r(n-r)^2(n+r)^2} V^2 (V')^2  + \frac{560n^2(n^2+r^2)}{(n-r)^3(n+r)^3} V^3 V'
\nonumber\\
&- \frac{(405 n^6 - 425 n^4 r^2 - 293 n^2 r^4 + 9 r^6) }{ r (n-r)^4(n+r)^4} V^4 + \mathfrak{E}_\mathcal{B}^{(\mathcal{Z})} \bigg]  \, ,
\end{align}
where now the base-dependent factors $\mathfrak{E}_\mathcal{B}^{(\mathcal{Z})}$ are
\begin{align}
\mathfrak{E}_{\mathbb{CP}^2}^{(\mathcal{Z})} =& \frac{10 n^2 }{r(n-r)(n+r)} V (V')^2 + \frac{40 n^2}{(n-r)^2(n+r)^2} V^2 V'  - \frac{4(45 n^4 + 8 n^2 r^2 + 3 r^4)}{r(n-r)^3(n+r)^3} V^3 
\nonumber\\
&- \frac{12(2n^2 + r^2)}{r(n-r)^2(n+r)^2} V^2 - \frac{2 V}{3r(n-r)(n+r)} \, ,
\\
\mathfrak{E}_{\mathbb{T}^2 \times \mathbb{S}^2}^{(\mathcal{Z})} &=	0  \, ,
\\
\mathfrak{E}_{\mathbb{T}^2 \times \mathbb{T}^2}^{(\mathcal{Z})} &=	0  \, .
\end{align}
It should be emphasized that while for static and spherically symmetric solutions the quartic quasi-topological term yields algebraic field equations \cite{Dehghani:2011vu}, these become non-linear second-order differential equations for Taub-NUT metrics. This is an interesting difference with respect to the Gauss-Bonnet case, for which the equations determining the metric function are algebraic for both kinds of solutions.

\subsection*{Einstein gravity}
Just like in the four-dimensional case, in the following subsections we will be studying the different base spaces independently. As before, it is illuminating  to start with a quick study of the situation for Einstein gravity, for which the analysis can be performed at the same time for all base spaces. Indeed, if we set $\lambda_{\rm \ssc GB}=\xi=\zeta=0$, \req{eq6d} can be easily solved for $V_{\mathcal{B}}(r)$. Imposing the NUT condition $V_{\mathcal{B}}(r=n)=0$ first, one is left with
\begin{equation}\label{NUTE2}
V_{\mathcal{B}}(r)=\frac{(r-n)\left[6(r^3+3nr^2+n^2 r-5n^3)+(\kappa+3)(3n+r)L^2\right]}{6L^2(n+r)^2}\, ,
\end{equation}
 where we set the integration constant  
\begin{equation}\label{massE2}
C_{\mathcal{B}}=-\frac{8n^3}{3}\left[\frac{12n^2}{L^2}+\kappa-1 \right]\, .
\end{equation}
The regularity condition \req{smooth} imposes
\begin{equation}\label{betaEin2}
\beta_{\tau}=\frac{24\pi n}{(1-\kappa)}\, .
\end{equation}
Hence, we find $\beta_{\tau}=12\pi n$ for $\mathcal{B}=\mathbb{CP}^2$ and $\mathcal{B}=\mathbb{S}^2\times\mathbb{S}^2$, $\beta_{\tau}=24\pi n$ for $\mathcal{B}=\mathbb{S}^2\times \mathbb{T}^2$, and $\beta_{\tau}=\infty$ for $\mathcal{B}=\mathbb{T}^2\times\mathbb{T}^2$, which forbids the existence of regular solutions with a compact $\mathbb{S}^1$ in that case. 


If we impose the bolt condition $V_{\mathcal{B}}(r=r_b)=0$ instead, we find
\begin{align}\label{boltEi6d}
V_{\mathcal{B}}(r)=\frac{1}{6L^2(n^2-r^2)^2 r_b}&\left[6\left(r^6 r_b+15n^4r (r-r_b)r_b-r r_b^6-5n^6(r-r_b)-5n^2r r_b (r^3-r_b^3) \right) \right. \\ \notag & \left.-(\kappa-1)L^2 (r-r_b)\left(3n^4-6n^2 r r_b +r r_b (r^2+r r_b +r_b^2)\right)\right]\, ,
\end{align}
where in this case we related  $C_{\mathcal{B}}$  to $n$ and $r_b$ through
\begin{equation}
C_{\mathcal{B}}=-\frac{1}{3r_b}\left[\frac{6(5n^6+15n^4r_b^2-5n^2r_b^4+r_b^6)}{L^2}+(\kappa-1)(3n^4+6n^2r_b^2-r_b^4)\right]\, .
\end{equation}
Finally, the regularity condition \req{smooth} produces the following relation between $r_b$, $n$ and the period of $\tau$,
\begin{equation}\label{rb6}
r_b=\frac{4L^2\pi}{10\beta_{\tau}}\left[1\pm \sqrt{1+\frac{5(\kappa-1)\beta_{\tau}^2}{8L^2\pi^2}+\frac{25n^2\beta_{\tau}^2}{4L^4\pi^2} }\right]\, .
\end{equation}
Just like in $D=4$, we must require the quantity inside the square root to be positive and, of course, $r_b>0$, which in each case restricts the values of $n$ for which solutions exist.

 Besides the regularity condition \req{smooth}, additional constraints on $\beta_{\tau}$ arise both for NUTs and bolts when demanding the absence of Misner string singularities --- see \eg discussion in \cite{Mann:2005ra}. For example, for $\mathcal{B}=\mathbb{CP}^2$, we must demand $\beta_{\tau}=12\pi n$. Just like in $D=4$, for the Einstein gravity NUT this condition is automatically implemented by \req{smooth}. This is not the case in general, and the conditions must be imposed separately. 

\subsection{$\mathcal{B}=\mathbb{CP}^2$}
Let us now turn on again the  higher-curvature couplings in \req{eqn:quart_6d_act}. The first base space we consider is $\mathbb{CP}^2$. For this, we can write
\begin{align}
A_{\mathbb{CP}^2}&=6\sin^2 \xi_2 (d\psi_2+\sin^2\xi_1 d\psi_1)\, , \\
d\sigma_{\mathbb{CP}^2}^2&=6\left\{ d\xi_2^2+\sin^2\xi_2 \cos^2\xi_2(d\psi_2+\sin^2\xi_1 d\psi_1)^2 +\sin^2\xi_2 (d\xi_1^2+\sin^2\xi_1 \cos^2\xi_1 d\psi_1^2)\right\} \, ,
\end{align}
where the coordinate ranges\footnote{See, for example,~\cite{CPkbook} for a detailed discussion of $\mathbb{CP}^k$ in these octant coordinates.} are $0 \le \xi_{1,2} \le \pi/2$ and $0 \le \psi_{1,2} \le 2 \pi$. Now, we consider the metric asymptotically, making the rescalings $\tau \to 6  n\psi$ and $r \to r/\sqrt{6}$. This gives 
\be 
\frac{^{(5)}ds^2_{\rm bdry}}{r^2} = \frac{6 n^2 f_\infty}{L^2} \left(d\psi + \frac{A_{\mathbb{CP}^2}}{6} \right)^2 + \frac{1}{6} d\sigma^2_{\mathbb{CP}^2} 
\ee
at large $r$.
For the specific case of 
$
6 n^2 f_\infty/L^2 = 1 \, , 
$
this boundary metric is just that of a round $\mathbb{S}^5$ provided that the coordinate $\psi$ has period $2\pi$ to ensure regularity.  In all other cases, it is the metric of a squashed sphere \cite{Bobev:2017asb} and, in analogy with the $D=4$ case, it is customary to parametrize such squashing with the parameter $\alpha$, defined in terms of $n$ through
$
6 n^2 f_\infty/L^2= 1/(1 + \alpha)\, .
$

We begin our study of Taub-NUT/bolt solutions with this base space by considering the asymptotic behaviour of the metric. 
%
%
The asymptotic solution for $V_{\mathbb{CP}^2}(r)$ consists of a particular and homogeneous solution. The particular solution is found by  expanding $V_{\mathbb{CP}^2}(r)$ in a $1/r$ series and solving the field equations to determine the constants order by order. The result is
\begin{align}\label{CP_asymp} 
V_p(r) 
=&f_\infty \frac{r^2}{L^2} + \frac{1}{3} - \frac{3 \fin n^2}{L^2} - \frac{6 L^2 + 12 f_\infty n^2 - 6 L^2 f_\infty + 5 L^2 f_\infty^2 \lambda_{\rm \ssc GB} - 6 f_\infty^3 n^2 \lambda_{\rm \ssc GB}}{9 r^2 f_\infty (2 f_\infty^2 \lambda_{\rm \ssc GB} - 3 f_\infty + 4)} \left(1-\frac{6f_\infty n^2}{L^2} \right)	
\nonumber\\
& - \frac{C_{\mathbb{CP}^2}}{2 h'(f_\infty) r^3} + {\cal O} (r^{-4}) \, ,
\end{align}
where $h'(f_\infty)$ denotes the derivative of $h(f_\infty)$ --- see \req{hfinf} --- with respect to $f_\infty$. To obtain the form of the homogeneous equation, we again write $V(r) = V_p(r) + g(r)$, and work to linear order in $g(r)$. While both $\mathcal{S}$ and $\mathcal{Z}$ contribute in the same way to the particular solution, the contributions differ in the homogeneous equation.  The resulting equation, in the limit of large $r$, takes the form:
\be 
a(r) g''(r) + b(r) g'(r) + c(r) g(r) = 0\, ,
\ee 
where $a(r)$ and $b(r)$ are the leading terms in this expansion, taking the explicit forms
\begin{align}
a(r) =& \frac{8 f_\infty \xi L^2 r}{3 } \left(1-\frac{6f_\infty n^2}{L^2} \right)^2 + (1-\hat\xi)\frac{320 f_\infty \zeta n^2 L^4}{81 r} \left(1-\frac{6f_\infty n^2}{L^2} \right)^2 \, , 
\nonumber\\
b(r) =& -8 f_\infty \xi L^4 \left(1-\frac{6f_\infty n^2}{L^2} \right)^2 - (1-\hat\xi) \frac{1520 f_\infty \zeta n^2 L^4}{81 r^2} \left(1-\frac{6f_\infty n^2}{L^2} \right)^2 \, ,
\nonumber\\
c(r) = & - 2 r^3 h'(f_\infty) \, ,	
\end{align}
and we have defined 
\be
\hat\xi = \begin{cases}
	1 &\text{for } \xi \neq 0\, ,
	\\
	0 &\text{for } \xi = 0  \, ,
\end{cases}
\ee
to simplify the presentation of the terms above. We recognize that the contributions in parenthesis in $a(r)$ and $b(r)$ are directly related to the squashing parameter and vanish when the base is a round sphere; in that case, the solution reduces to just pure AdS. 

In the limit of large $r$, the homogeneous equation can be solved  in terms of special functions. First, when $\xi \neq 0$ the homogeneous solution reads
\be 
g(r) = C_1 r^2 I_{1} \left(r\sqrt{-\frac{c(r)}{4 a(r)}} \right)  + C_2 r^{2}  K_{1} \left(r\sqrt{-\frac{c(r)}{4 a(r)}} \right)\, ,
\ee
while if $\xi = 0$ it takes the form:
\be 
g(r) = C_1 r^{23/8}  I_{\frac{23}{24}} \left(r \sqrt{-\frac{c(r)}{9 a(r)}} \right)  + C_2 r^{23/8}  K_{\frac{23}{24}} \left(r\sqrt{-\frac{c(r)}{9 a(r)}} \right)\, ,
\ee
where $I_\nu(x)$ and $K_\nu(x)$ are the modified Bessel functions of the first and second kinds, respectively. The explicit form of these solutions is not as important as what their asymptotic behaviour tells us: because $c(r) > 0$ by virtue of demanding the graviton is not a ghost, when $a(r) < 0$, in each case the homogeneous solution consists of a super-exponentially growing and super-exponentially decaying part. When $a(r) > 0$, both of the above  solutions oscillate more and more rapidly near infinity and are ultimately pathological. Therefore, to ensure that  the solutions are physically reasonable, we must demand that $a(r) < 0$,  while also requiring $h'(\fin) < 0$. The general solutions to these constraints with $\lambda_{\rm \ssc GB} \neq 0$ are a bit messy, and so we quote the result explicitly only in the case $\lambda_{\rm \ssc GB} = 0$. In that case it is a straight-forward matter to show that these conditions are satisfied --- independent of $n$ --- provided that
\begin{align} 
\label{eqn:good_vac_6dCP}
\xi &< {\rm min} \left\{0, \frac{27}{256} - \zeta \right\} \quad \text{if $\xi \neq 0$ or} \, ,
\nonumber\\
\zeta &< 0 \quad \text{ if $\xi = 0$ } .
\end{align}
Interestingly, in contrast to the four dimensional case, here the mass parameter $M$ does not enter into the constraints, with the result that there is no pathology associated with the negative mass solutions (see below). Furthermore, note that for $\zeta$ non-zero, simply demanding $\xi < 0$ is not enough since one must also require that $h'(\fin) < 0$ --- this is the origin of the more complicated constraint in that case. It can be shown that, for $\lambda_{\rm \ssc GB} = 0$, $\xi = 27/256 - \zeta$ corresponds to the critical limit of the theory, which has $f_\infty = 4/3$. 

A consequence of these bounds on the coupling is that, when one considers a theory that contains only a single one of the quartic terms, then it is not possible to reach the critical limit of the theory at physical coupling. This situation is similar to what happens to cubic GQTG for spherically symmetric black hole solutions in $D \ge 6$.

\subsubsection{Taub-NUT solutions}
\begin{figure}[t]
	\centering
	\includegraphics[scale=0.45]{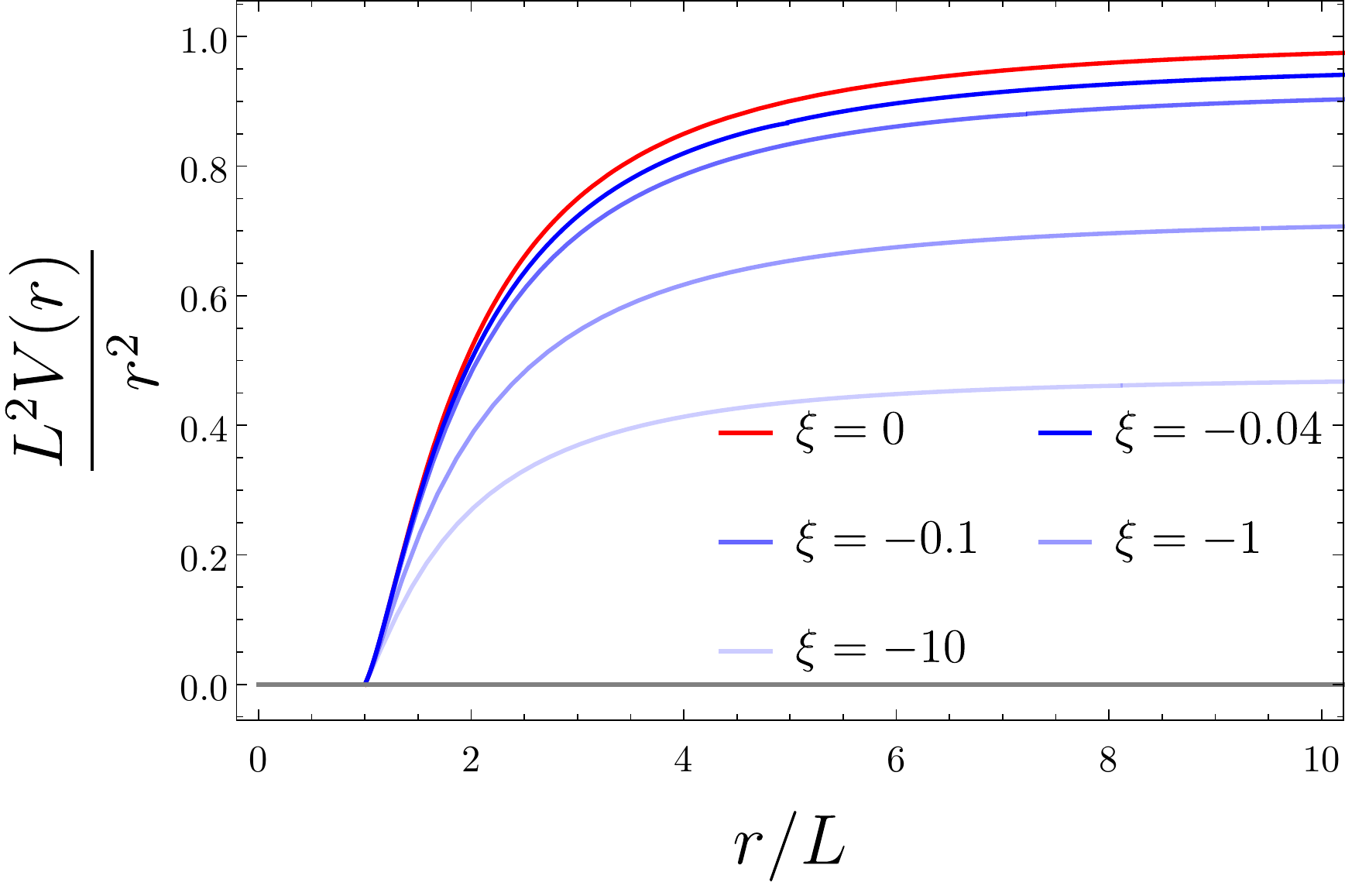}
	\quad
	\includegraphics[scale=0.45]{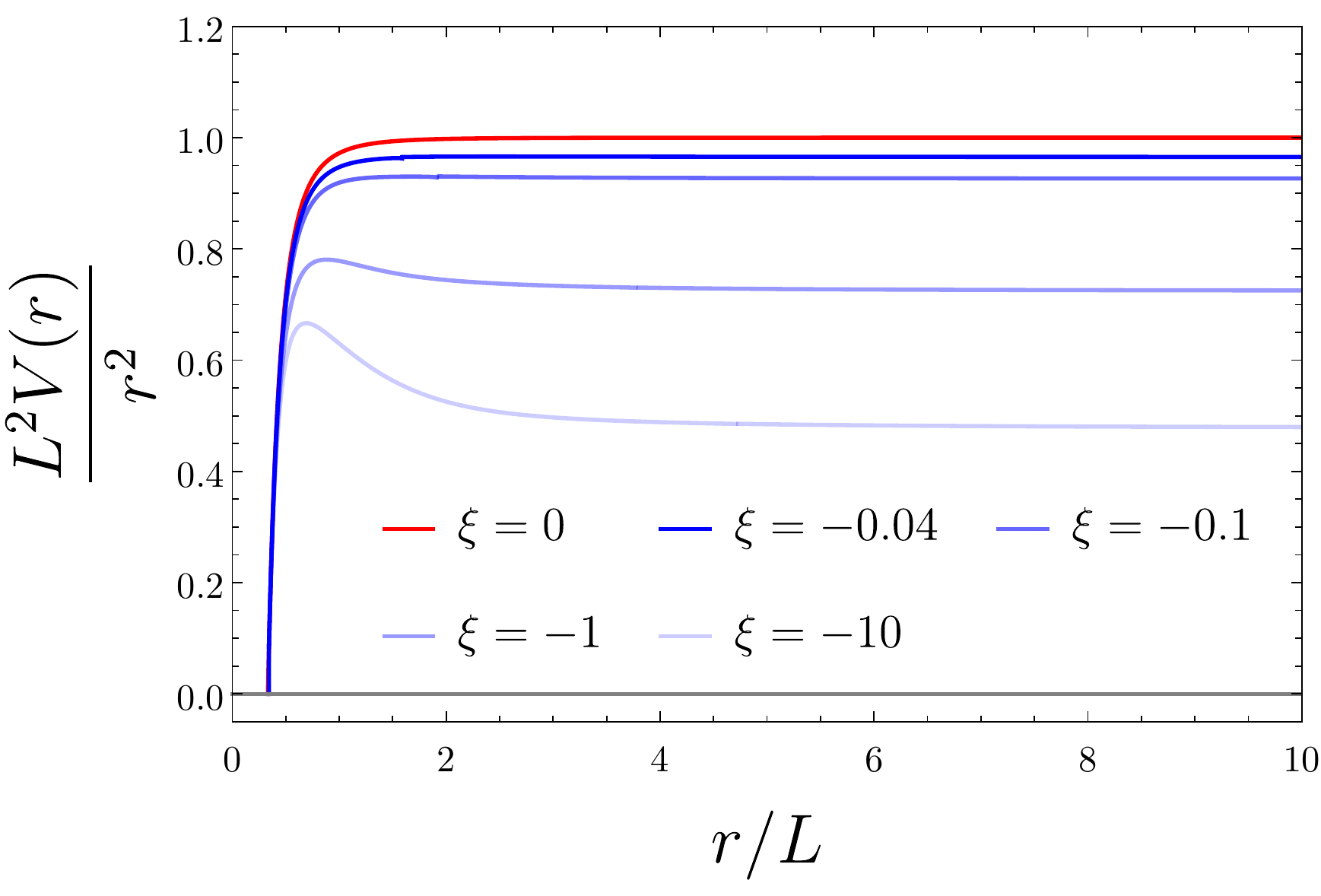}
	\includegraphics[scale=0.45]{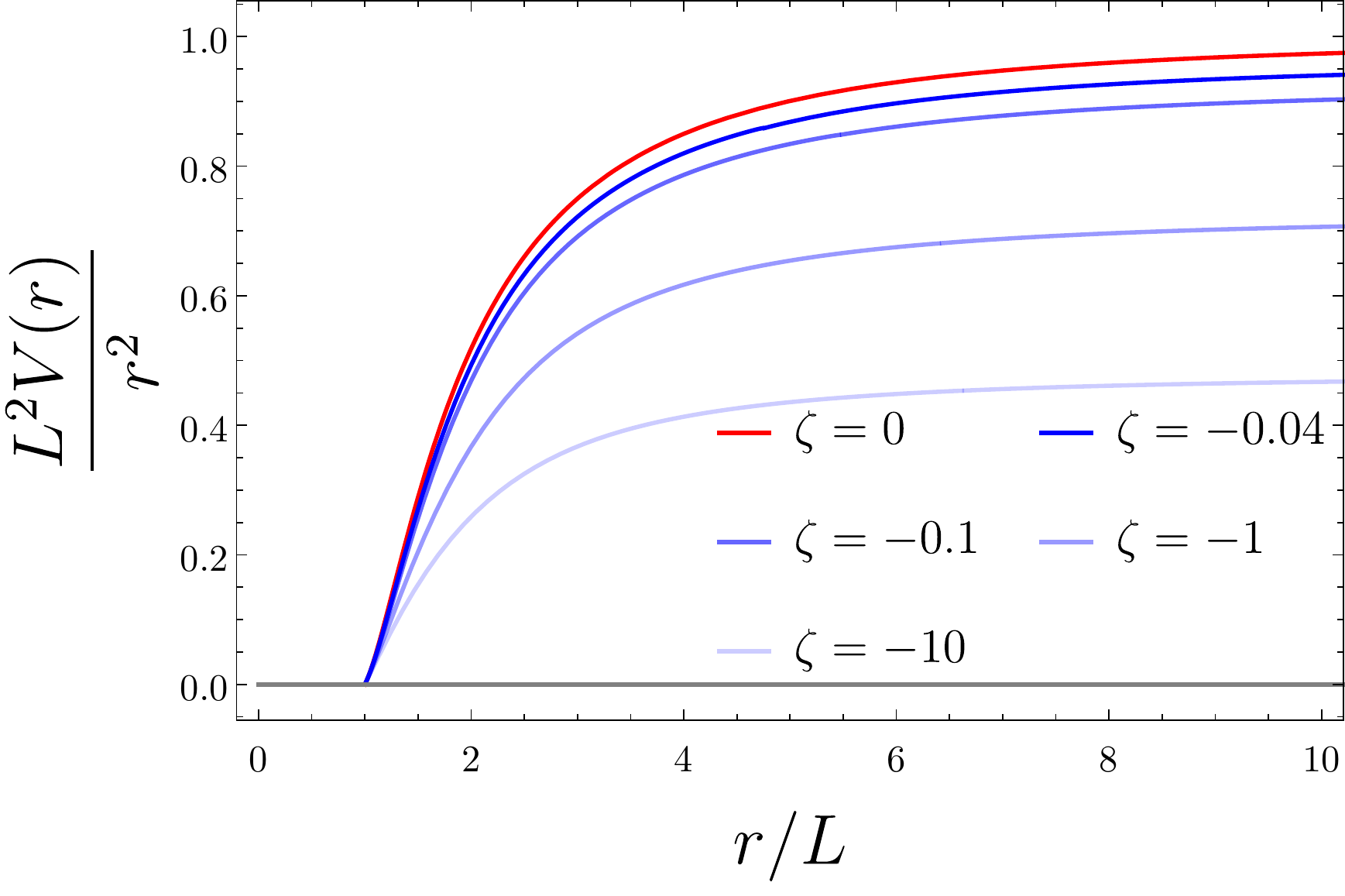}
	\quad
	\includegraphics[scale=0.45]{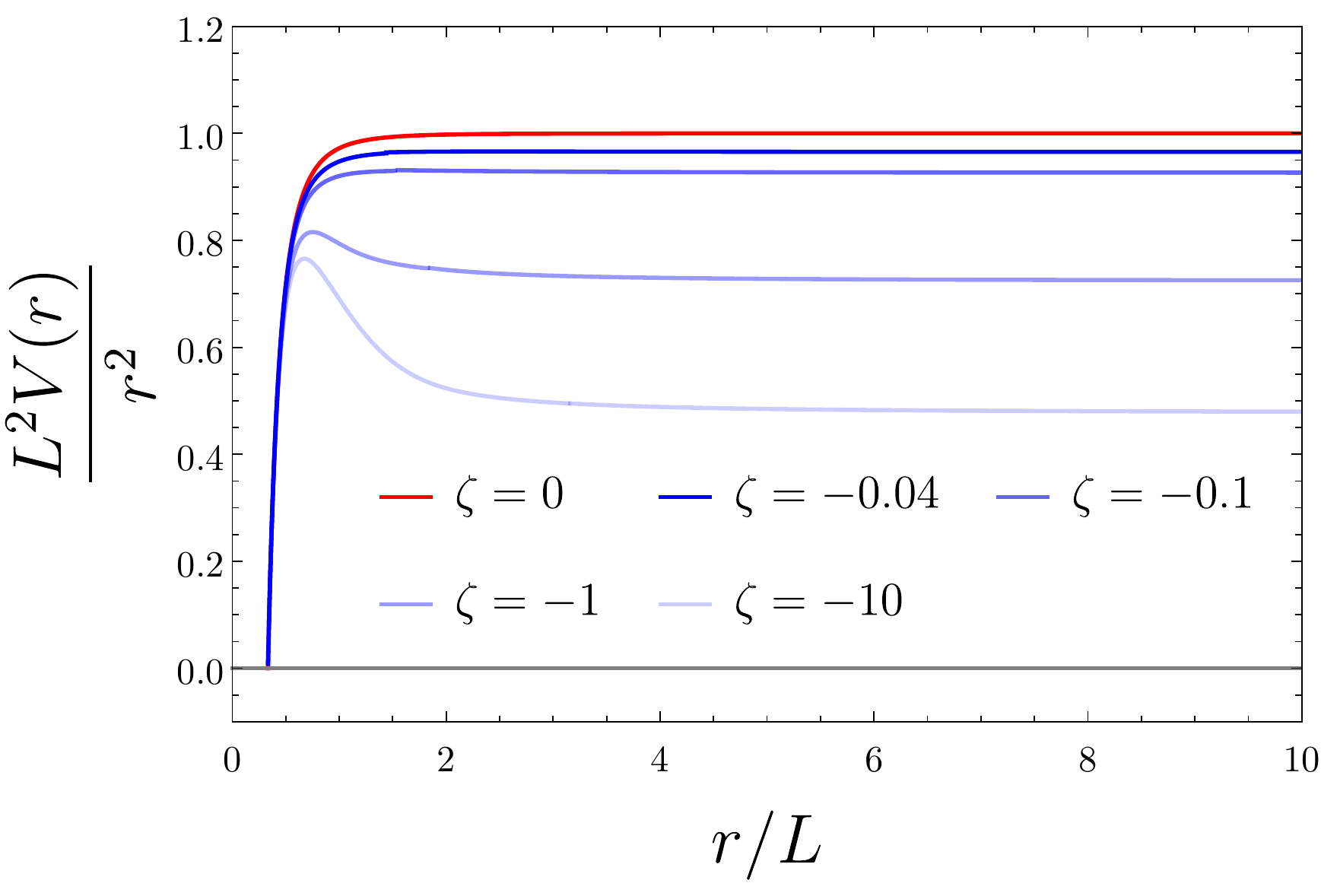}
	\caption{The metric function $L^2V_{\mathbb{CP}^2}(r)/r^2$ is plotted for NUT solutions of the quartic theories.  The top row depicts solutions of the quartic generalized quasi-topological theory with $n/L= 1$ (left) and $n/L = 1/3$ (right). The solutions with $n/L = 1$ all have positive mass, while those with $n/L = 1/3$ have negative mass. The bottom row depicts solutions of the quartic quasi-topological theory  with $n/L= 1$ (left) and $n/L = 1/3$ (right). The solutions with $n/L = 1$ all have positive mass, while those with $n/L = 1/3$ have negative mass.} 
	\label{CP2-numeric_NUT}
\end{figure}

We now consider NUT solutions where $V_{\mathbb{CP}^2}(r=n) = 0$. Further restrictions on $V_{\mathbb{CP}^2}(r)$ arise due to regularity of the metric. Recall from the discussion above that the boundary is a squashed $\mathbb{S}^5$. Regularity of this boundary metric requires that $\psi := \tau/(6n)$ has period $2 \pi$, which in turn means $\tau \sim \tau + 12 \pi n$. A further constraint is imposed on the derivative of $V_{\mathbb{CP}^2}(r)$ near the NUT where the absence of conical singularities at a zero of $V_{\mathbb{CP}^2}$ requires that $\tau$ is periodic with period $\beta_\tau$ given by $\beta_\tau = 4 \pi /V_{\mathbb{CP}^2}'(r=n)$. Consistency of these two regularity conditions fixes $\beta_\tau = 12 \pi n$ and so we therefore have the following series expansion near the NUT:
\be 
V(r) =  \frac{(r-n)}{3n} + \sum_i^\infty (r-n)^i a_i  \, .
\ee 
Substituting this expression into the field equations, and expanding in $(r-n)$, we find
\begin{align}
\frac{16}{3} \frac{n^3(L^2 - 6 n^2)}{L^2} - \frac{8  n L^2 \lambda_{\rm \ssc GB}}{9}  - \frac{2 L^6 \left(\xi + \zeta \right)}{81 n^3} + \frac{4 G M }{9 \pi } + {\cal O} \left((r-n)^3 \right) = 0 \, ,
\end{align}
where we have conveniently redefined the integration constant $C_{\mathbb{CP}^2}=-4GM/(9\pi)$, where $M$ will correspond to the ADM mass of the solution.
The first condition (shown above explicitly) determines $M$ in terms of the couplings and the NUT charge, and the next two relations are automatically satisfied. The next non-trivial relation is linear in $a_3$, allowing one to solve for $a_3$ as a function of the free parameter $a_2$. This trend continues to higher order in the field equations, and thus there is a single free parameter that is left unfixed by the field equations and regularity conditions. This is  fully analogous to the $D=4$ case.

The near horizon solution can be joined to the asymptotic solution that was presented above by numerically integrating the field equations. The near horizon expansion is used as initial data, with the shooting method employed to determine the free parameter $a_2$. A careful choice of this parameter is required to ensure the growing modes present in the asymptotic solution are not excited. Ensuring this, we find a unique $a_2$ for which the solution can be integrated, with the result for several values of the coupling shown in Fig.~\ref{CP2-numeric_NUT}. For comparison, the Einstein gravity solution is shown in red and we see that the solutions to the higher curvature theories are qualitatively the same with the main difference being that they approach $\fin r^2/L^2$ with $\fin$ depending on the value of the couplings.  We also note that, while the top left and bottom left plots depict solutions with positive mass, the top right and bottom right plots depict solutions with negative mass. The fact that the negative mass solutions can be constructed and possess no inherent pathology is in contrast with the four dimensional case, where the negative mass solutions possessed pathological asymptotic structure.

Let us now turn to the free energy of the NUTs and compute the regularized on-shell action for these solutions. 
With minor modifications, the prescription \req{assd} introduced in \cite{HoloECG} can be used to eliminate the divergent terms in the on-shell action. The Euclidean action, completed with the generalized boundary term and counterterms is given by
\begin{align}\label{full6}
I_E &= - \int \frac{d^6 x \sqrt{g}}{16 \pi G} \left[ \frac{20}{L^2} + R + \frac{\lambda_{\rm \ssc GB} L^2 }{6} {\cal X}_4 -\frac{ \xi L^6 }{216} \mathcal{S} - \frac{ \zeta L ^6}{144} \mathcal{Z} \right] 
	\nonumber\\
&- \frac{1 - 4 \lambda_{\rm \ssc GB} \fin  + 8 (\xi + \zeta) \fin^3}{8 \pi G} \int d^5 x \sqrt{h} \left[K  - \frac{4 \sqrt{\fin}}{L} - \frac{L}{6 \sqrt{\fin}} {\cal R} - \frac{L^3}{18 \fin^{3/2}} \left( {\cal R}_{ab}{\cal R}^{ab} - \frac{5}{16} {\cal R}^2 \right)\right]
	\nonumber\\
&+ \frac{ \lambda_{\rm \ssc GB} \fin - 6 (\xi + \zeta) \fin^3}{8 \pi G} \frac{L^3}{18 \fin^{3/2}} \int d^5 x \sqrt{h} \left( 4 {\cal R}_{ab}{\cal R}^{ab} - \frac{5}{4} {\cal R}^2 + \frac{3}{2} {\cal X}^{(h)}_4 \right) \, .	
\end{align}
The evaluation is facilitated via the asymptotic expansion presented above and the expansions near $r = n$ in the NUT case or $r = \rh$ for the bolts. Near the boundary, the bulk action has several divergent components that are precisely canceled by the generalized boundary and counterterms. Note the addition of a new counterterm, nonproportional to $a^*$, on the last line above. This appears because, strictly speaking, the spacetime is not asymptotically AdS --- the boundary is not maximally symmetric except for the choice of NUT parameter that yields the undeformed five sphere. The additional counterterm was chosen since it vanishes identically when the boundary is maximally symmetric (and so could be dropped in those cases) but allows for the cancellation of the linear divergence in the case of $\mathcal{B} = \mathbb{CP}^2$ considered here.

Just like  the four-dimensional case discussed in detail in appendix \ref{freeen}, eliminating the divergent terms also removes all possible constant terms coming from boundary contributions, leaving us with the bulk action evaluated at $r=n$, and nothing else. The final result is
\be
\label{eqn:CP_nut_IE} 
I_E = \frac{36\pi^2}{ G}\left[n^4\left(\frac{4n^2}{L^2}-1\right) + \frac{L^2 n^2 \lambda_{\rm \ssc GB}}{3} -\frac{L^6(\xi + \zeta)}{108n^2} \right]  \, ,
\ee
from which the total energy and entropy can be found to be,
\begin{align} 
E &= \frac{12\pi}{G} \left[n^3\left(\frac{6n^2}{L^2}-1\right) + \frac{ n L^2 \lambda_{\rm \ssc GB}}{6} + \frac{L^6(\xi+\zeta)}{216 n^3} \right] = M \, , 
\nonumber\\
S &= \frac{36\pi^2}{G} \left[ n^4\left(\frac{20n^2}{L^2}-3\right)+  \frac{ n^2 L^2 \lambda_{\rm \ssc GB} }{3}+ \frac{L^6(\xi+\zeta)}{36n^2} \right]  \, ,
\end{align}
and the first law $dE = TdS$ is verified to hold.  

 Similar to the discussion for the $\mathbb{S}^2$ base in the case of ECG, here we can also enlarge the thermodynamic phase space and construct the extended first law. The expression is slightly more complicated reading
\begin{equation}\label{eflawCP}
dE = TdS + VdP + \Upsilon^{\ssc \rm GB} d(L^2\lambda_{\ssc \rm GB}) + \Upsilon^{\cal S} d (L^6\xi) + \Upsilon^{\cal Z} d(L^6\zeta) \, ,
\end{equation}
where we have again restored the dimensionality to the coupling constants. The potentials appearing in the extended first law read
\begin{equation}
V = \frac{48 \pi^3}{5} n^5 \, , \quad \Upsilon^{\ssc \rm GB} = \frac{1}{n G} \, ,\quad \Upsilon^{\cal S} = \Upsilon^{\cal Z} = -\frac{\pi}{36 G n^3} \, .
\end{equation}
The expression above for the thermodynamic volume holds also in Einstein gravity though there appear to be no previous computations of this quantity in the literature for higher dimensional Taub-NUT solutions. It is noteworthy that the thermodynamic volume here is positive, while the thermodynamic volume is negative in the $D = 4$ case. Of course, we find that the Smarr relation consistent with scaling is satisfied by the thermodynamic quantities defined above:
\begin{equation}
\label{smarrCP}
3 E = 4 TS - 2 PV + 2 \Upsilon^{\ssc \rm GB} L^2 \lambda_{\ssc \rm GB} + 6 \Upsilon^{\cal Z} (L^6 \zeta) + 6 \Upsilon^{\cal S} (L^6 \xi)   \, .
\end{equation}

Note that if we turn off the quartic couplings, then the result for the free energy reduces to that previously calculated for Einstein gravity and Gauss-Bonnet gravity in~\cite{Clarkson:2002uj, KhodamMohammadi:2008fh} up to an overall factor of $8/9$, the same discrepancy noted in~\cite{Bobev:2017asb}. We have carefully revisited the calculations in~\cite{Clarkson:2002uj, KhodamMohammadi:2008fh} and have traced the discrepancy to the ratio of volumes of  $\mathbb{S}^2 \times \mathbb{S}^2$ to $\mathbb{CP}^2$. In~\cite{Clarkson:2002uj} it is claimed that the thermodynamic quantities for both base spaces are identical. However, we have found this to be true only up to an overall ratio of the volumes of the base spaces. For $\mathbb{CP}^2$ normalized so that $R_{ab} = g_{ab}$ the volume is ${\rm Vol}\left(\mathbb{CP}^2 \right) = 18 \pi^2$, while the volume of $\mathbb{S}^2 \times \mathbb{S}^2$ is given by ${\rm Vol} \left(\mathbb{S}^2 \times \mathbb{S}^2 \right) = (4 \pi )^2$.  The ratio of these volumes is precisely $8/9$, which accounts for the observed discrepancy.\footnote{In~\cite{Clarkson:2002uj} a different set of coordinates is used, but the metric of $\mathbb{CP}^2$ is still normalized so that $R_{ab} = g_{ab}$. Using the fact that the coordinates in~\cite{Clarkson:2002uj} have ranges $0 \le u \le \infty$, $0 \le \theta \le \pi$, $0 \le \phi \le 2 \pi$ and $0 \le \psi \le 4 \pi$ we obtain the result presented here with the correct overall factor. In higher dimensions, ${\rm Vol}\left(\mathbb{CP}^k \right) = 2^k(k+1)^k \pi^k/k!$ and the volume of $k$ 2-spheres is ${\rm Vol} \left(\mathbb{S}^2 \times \cdots \times \mathbb{S}^2 \right) = (4 \pi )^k$. In higher dimensions, we find the ratio between thermodynamic quantities for the two bases is $2^k k!/(k+1)^k$.}

\begin{figure}[t]
	\centering 
	\includegraphics[scale=0.57]{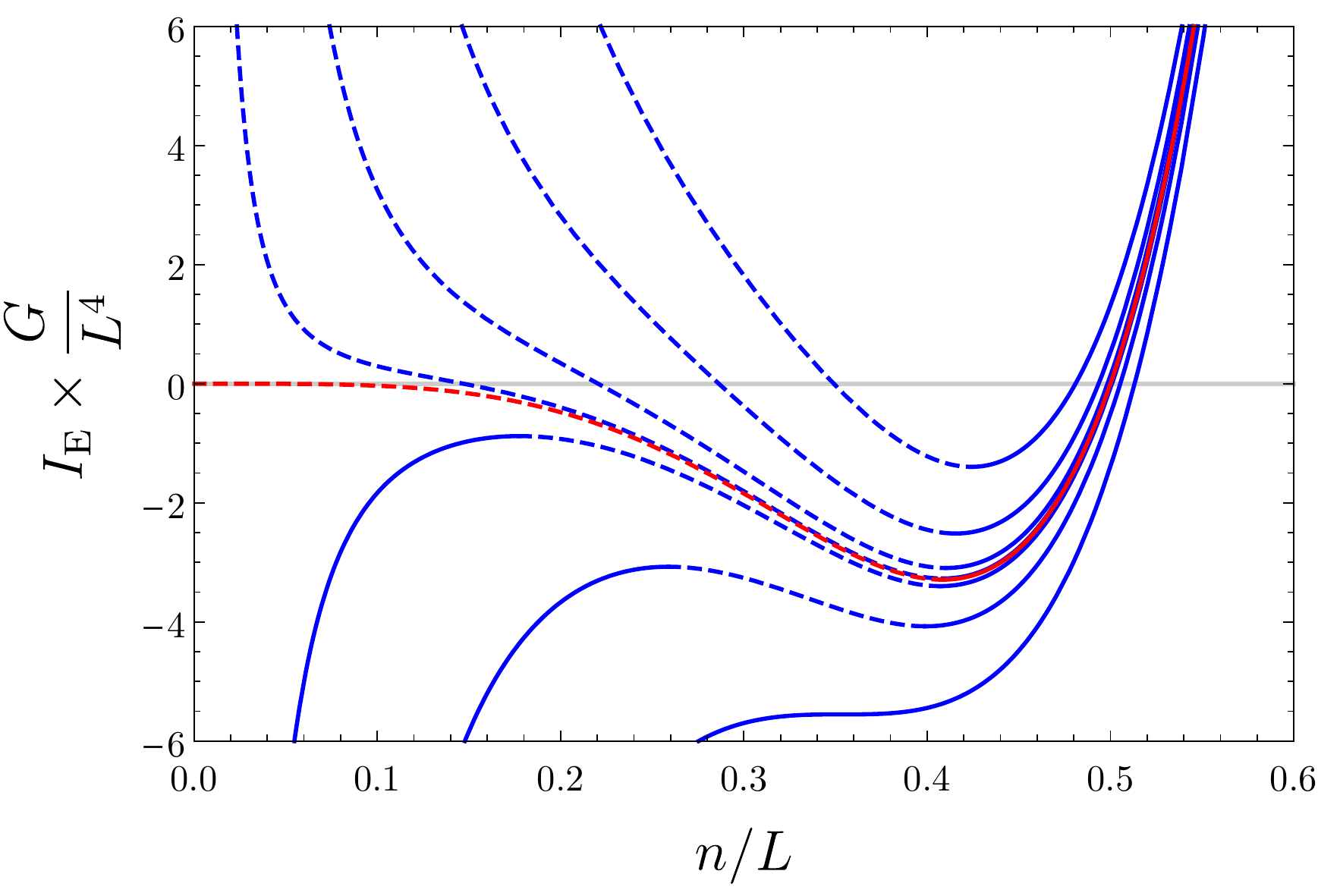}
	\caption{Euclidean on-shell action for $\mathcal{B} = \mathbb{CP}^2$ NUT solutions. The red curve corresponds to the Einstein gravity result, while the blue curves correspond to $\xi + \zeta = 27/256, 27/256 - 1/10, 27/256 - 10/95,  - 10^{-3}, - 10^{-2}, - 4  \times 10^{-2}$ and $-10^{-1}$ (bottom to top through a vertical slice). The dashed portions of the curves indicate solutions with negative mass, though there is no pathology associated with these solutions in this case.} 
	\label{CPNUTIE}
\end{figure}

In Fig.~\ref{CPNUTIE} we show plots of the Euclidean on-shell action for the NUT solutions with $\lambda_{\rm \ssc GB} = 0$. As is clear from Eq.~\eqref{eqn:CP_nut_IE}, this depends on the higher curvature couplings only through the combination $\xi + \zeta$. 
 In each case, there is only a single branch and, from the figure, we see that its qualitative structure depends on whether $\xi + \zeta$ is positive or negative. For consistency with the plots presented earlier in the document, we have indicated regions of negative mass with dashed curves. However, unlike the four dimensional case, there is no pathology associated with the negative mass solutions for the quartic theories in six dimensions. The region of negative mass solutions shrinks and eventually vanishes as $\xi + \zeta \to 27/256$, which corresponds to the critical limit of the theory.

\subsubsection{Taub-bolt solutions}

We now consider Taub-bolt solutions which satisfy $V_{\mathbb{CP}^2}(\rh) = 0$ for $\rh > n$.  In this section, we turn off the Gauss-Bonnet coupling to limit the size of the parameter space. Regularity demands that $V_{\mathbb{CP}^2}'(\rh) = 1/(3n)$, and therefore we write the near horizon expansion as
\be 
V(r) = \frac{(r-\rh)}{3n} + \sum_{i=2}^\infty(r-\rh)^i  a_i  \, .
\ee
Substituting this expansion into the field equations and solving order by order in $(r-\rh)$, we find the first two relations fix the integration constant $C_{\mathbb{CP}^2}$ and the relationship between $\rh$ and $n$:
\begin{align}
\label{eqn:bolt_nh_cp2}
0 =&  \frac{4 G M}{9 \pi} + \frac{\xi L^6 ( 9 \rh^2 + 48 \rh n + 37 n^2)}{243n^4 \rh} - \frac{20 \zeta L^6}{729 n^2 \rh} 
\nonumber\\
&- \frac{2\left( L^2 \rh^4 - 6 L^2 \rh^2 n^2 - 3 L^2 n^4 + 3 \rh^6 - 15 \rh^4 n^2 + 45 \rh^2 n^4 + 15 n^6 \right)}{3 L^2 \rh} \, , 
\\ \label{eqn:bolt_nh_cp22}
0 =& \frac{2 (\rh^2 - n^2)^2(L^2 \rh - 3 L^2 n - 15 \rh^2 n + 15 n^3)}{3 n L^2 \rh^2} - \frac{\xi L^6 (3 \rh^4 - 46 \rh^2 n^2 - 48 \rh n^3 - 37 n^4)}{243 n^4\rh^2(\rh^2 - n^2)} 
\nonumber\\
&- \frac{20 \zeta L^6\left(\rh^2 + \frac{6}{5} n \rh + n^2 \right)}{729 n^2 \rh^2 (\rh^2 - n^2)} \, ,	
\end{align} 
where, just as in the NUT case, we have set $C_{\mathbb{CP}^2}=-4GM/(9\pi)$.

Let us now examine the second relation above in more detail. When the higher curvature terms are turned off, the bolt radius is given by \req{rb6} with $\kappa=-1$, \ie
\be 
\rh(\zeta, \xi=0) = \frac{L^2}{30 n} \left[1 \pm \sqrt{1 - 180 \frac{ n^2}{L^2} + 900 \frac{n^4}{L^4}} \right] \, .
\ee
Since $\rh$ must be real and larger than $n$, we must then have $n < \sqrt{15} (2-\sqrt{2})L/30$. As in the four-dimensional case, there is a maximum value of $n$ for bolts in Einstein gravity. In particular, this means that there does not exist a bolt solution near the undeformed five sphere, for which $n= L/\sqrt{6}$. Of course, the behaviour is different with higher curvature corrections, but there are some notable differences from what was observed in the four dimensional case. 

Depending on the relative size of $\xi$ and $\zeta$, the behaviour of $\rh$ as a function of $n$ can either resemble that of Einstein gravity (namely, there is a largest value of $n$ for which a bolt exists) or resemble that observed in the four dimensional cubic case discussed earlier in this paper (bolts exist for arbitrarily large $n$). In fact, this classification is completely determined by the sign of the quantity $\zeta - 6\xi$. If this quantity is positive, there exists a maximum value of $n$; if this quantity is negative, bolts exist for arbitrarily large $n$.\footnote{This can be deduced in the following way. Take the numerator of Eq. \eqref{eqn:bolt_nh_cp22} and set $\rh = n + x$. Next, apply Descartes' rule of signs, treating $x$ as the independent variable, and notice that in the limit of large $n$ there will be a single sign flip provided that $\zeta - 6\xi < 0$. This guarantees a single positive root for $x$, which in turn guarantees the existence of a bolt with $\rh > n$. }

\begin{figure}[t]
	\centering
	\includegraphics[scale=0.65]{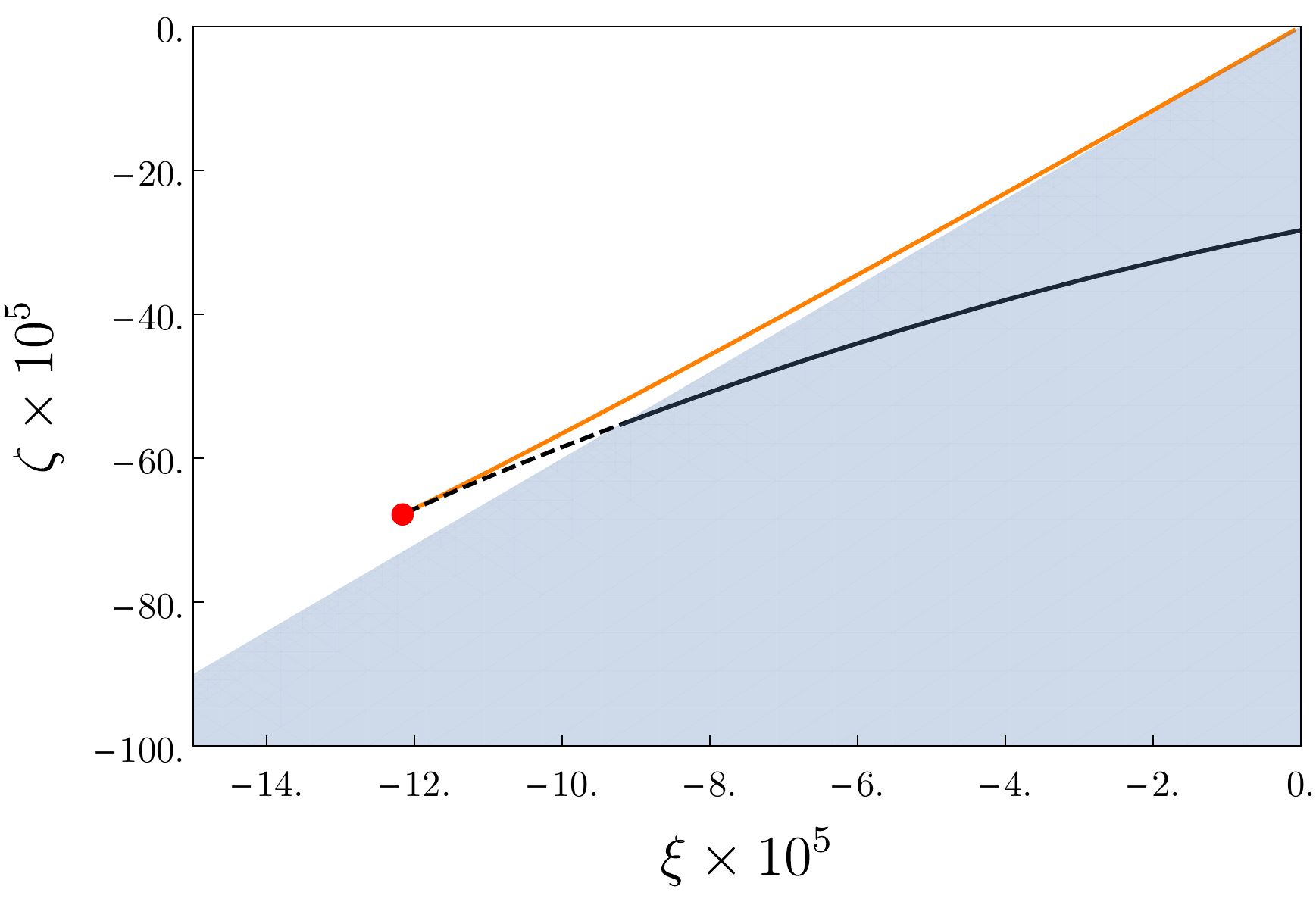}
	\caption{A breakdown of the coupling parameter space into useful regions for bolt solutions. Here the orange and black curves denote lines of `critical points', \ie for a given NUT charge, three solutions for $\rh$ coalesce. The red dot represents the single point in the physical parameter space where there is a coalescence of four roots. Within the blue shaded region, $\zeta < 6 \xi$ and there are bolts for arbitrarily large $n$. In the complement, $\zeta > 6 \xi$ and there is a largest value of $n$ for which bolts exist.  On the black locus of critical points, the critical point is always physical (i.e. of lowest free energy) within the blue region, otherwise (for the dashed portion of the curve) the situation can depend on which branch of the cusp minimizes the free energy, and also on whether or not there are re-entrant phase transitions as described in the text.  }
	\label{fig:bolt_params_CP2}
\end{figure}

From the perspective of the phase structure of the bolts, the most interesting scenario occurs when there are three values of $\rh$ for a particular $n$ --- one would expect these cases could yield swallowtail type behaviour and critical phenomena. We can constrain the regions of parameter space where three bolts exist by searching for `critical points'. More specifically, such a critical point would occur when $\partial n/\partial \rh = \partial^2 n /\partial \rh^2 = 0$, while respecting  Eq. \eqref{eqn:bolt_nh_cp22}. These points will mark transitions in the maximum number of bolts for given couplings. We were unable to solve the resulting constraints analytically, but it is straightforward to do so numerically. This results in the breakdown of parameter space shown in Fig.~\ref{fig:bolt_params_CP2}.  

\begin{figure}[htp]
	\centering
	\includegraphics[scale=0.45]{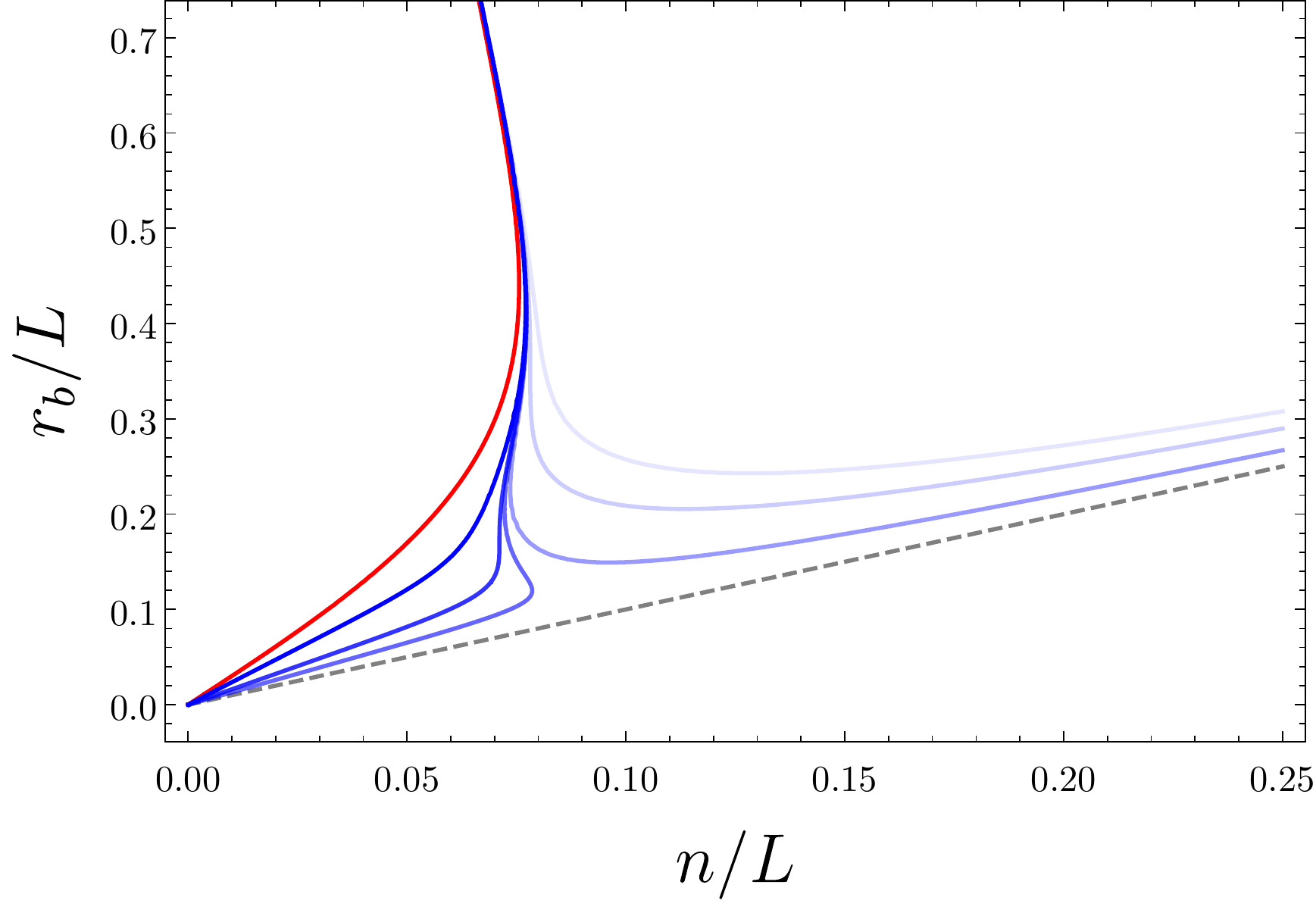}
	\includegraphics[scale=0.45]{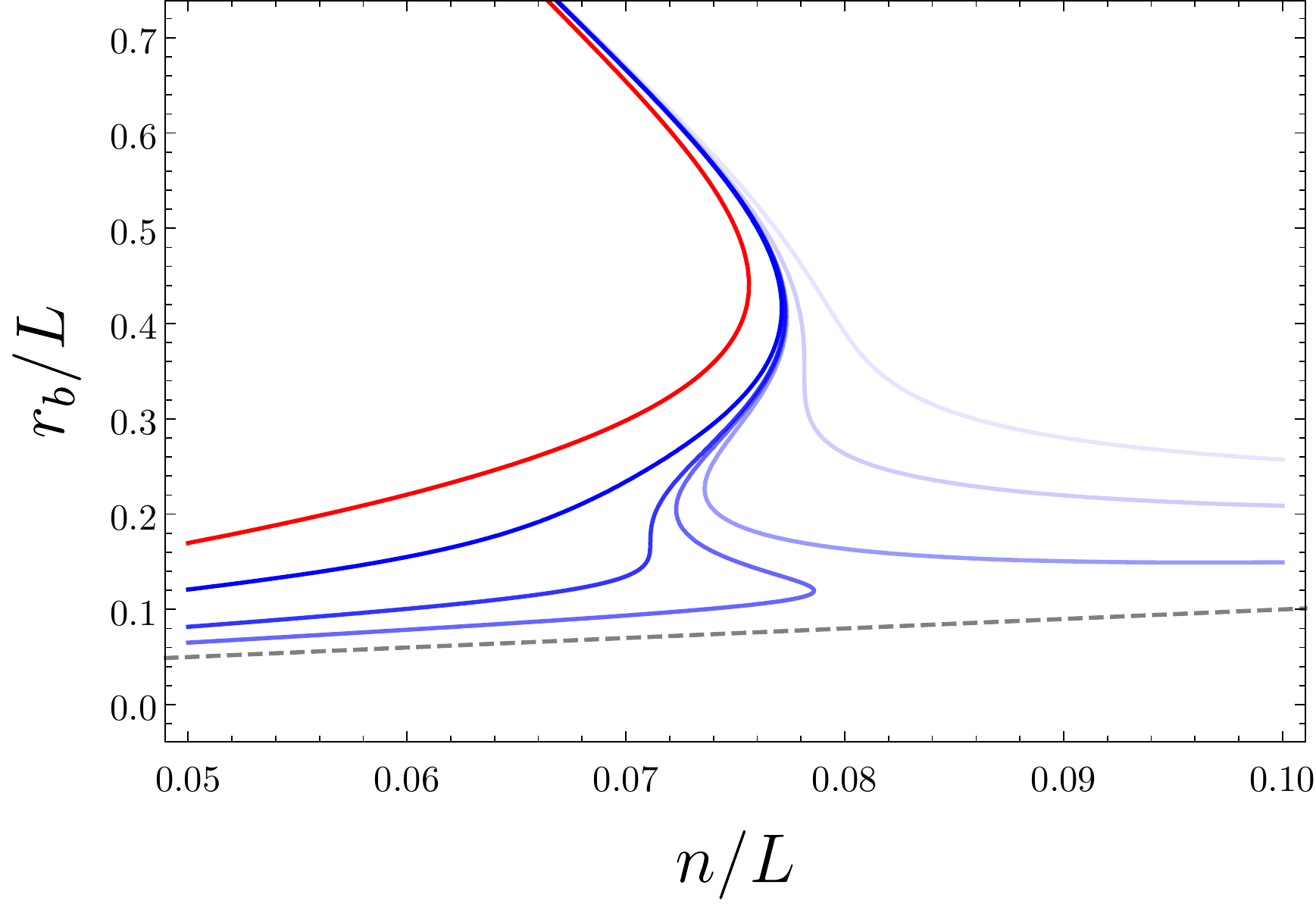}
	\includegraphics[scale=0.45]{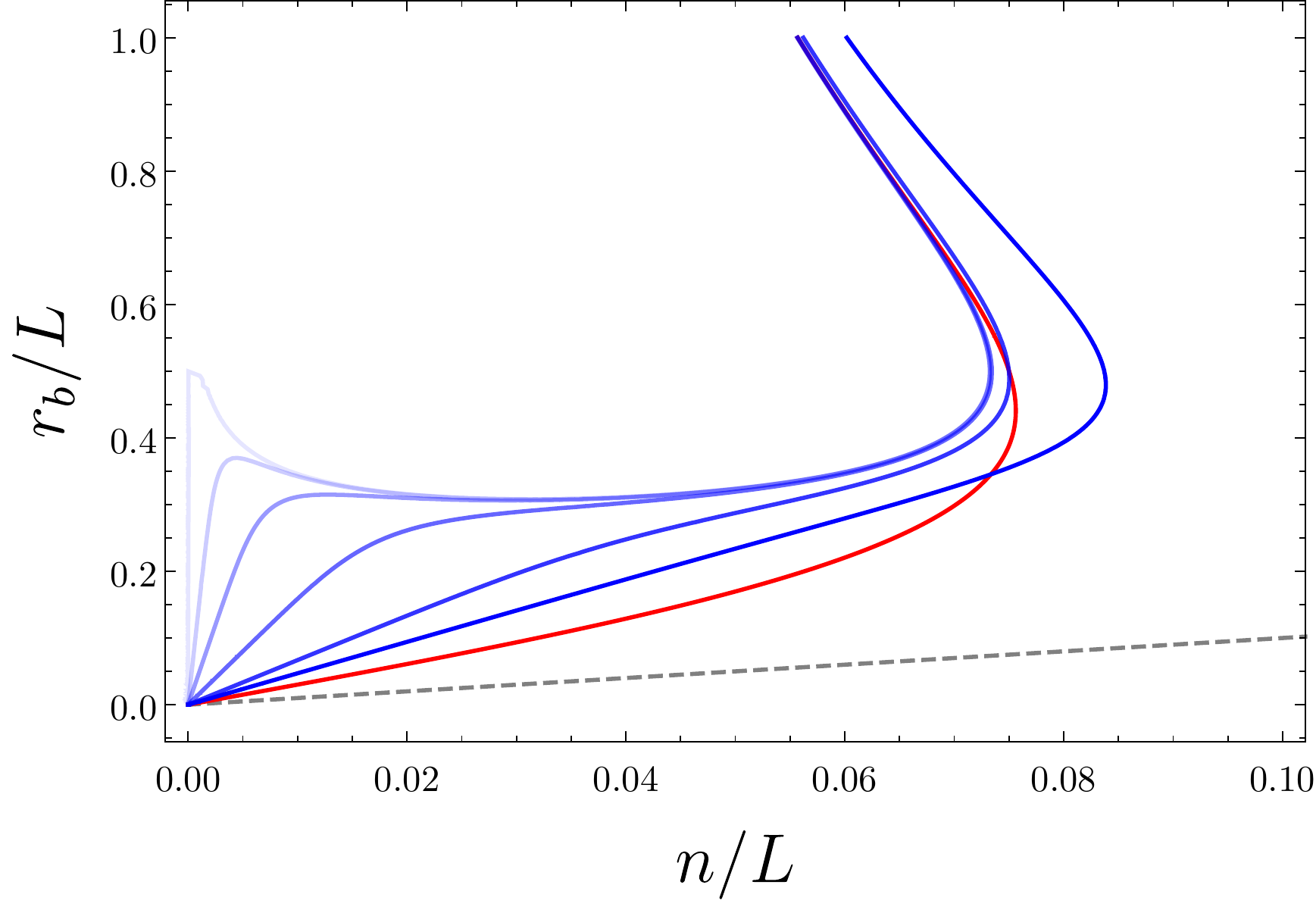}
	\includegraphics[scale=0.45]{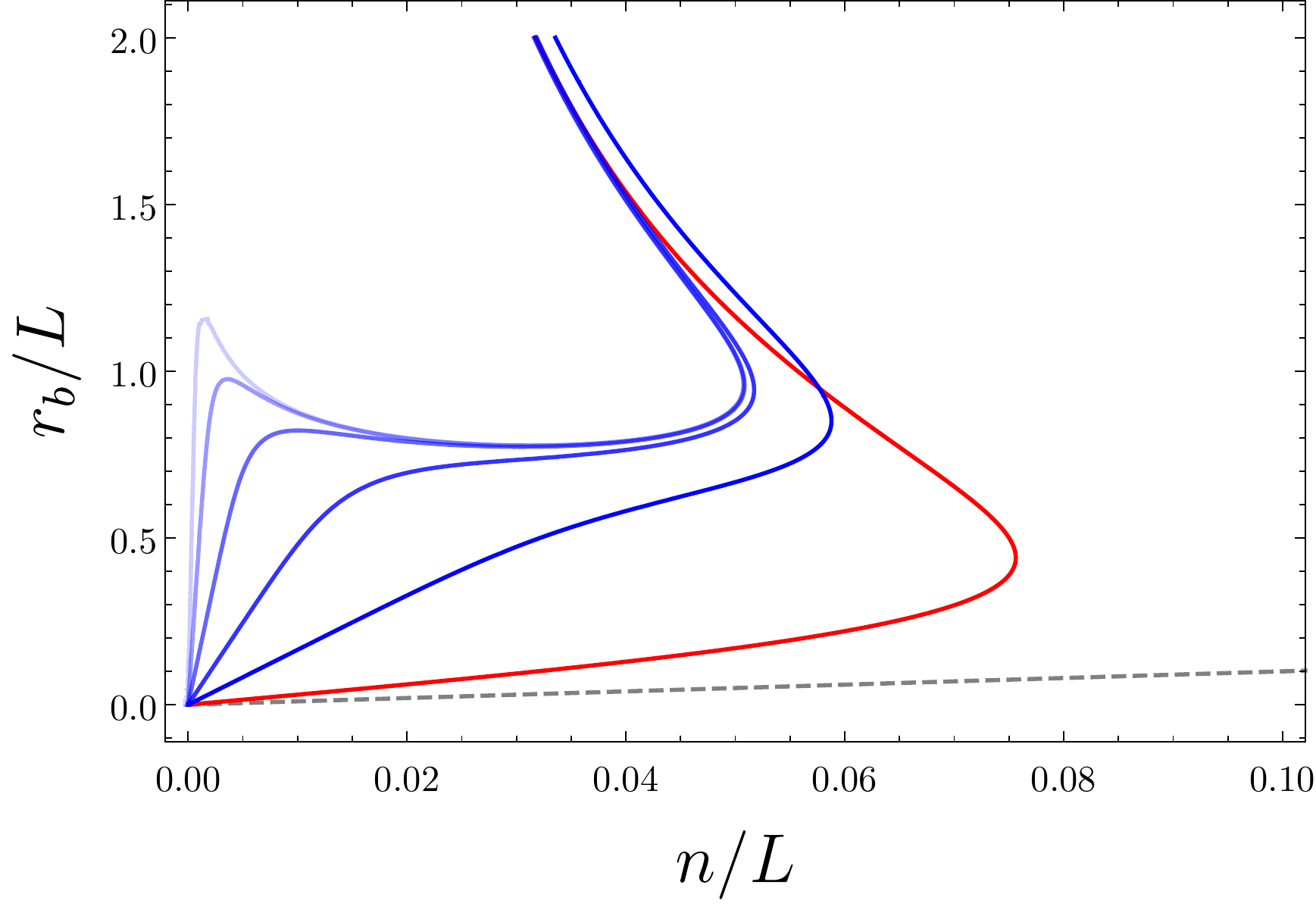}
	\caption{Top row: Plots of $\rh$ vs.~$n$ for fixed $\xi = -4 \times 10^{-5}$ with $\zeta \times 10^{5} = -20, -23.15, -23.8, -25, -38.03, -70$ (more to less opacity, or left to right for any horizontal slice through the plot). The right plot is a zoomed in version of the left, showing the interesting structure for bolt solutions. Bottom row: Plots of $\rh$ vs.~$n$ for positive $\zeta$. The left plot shows curves for $\zeta = 10^{-3}$ with $\xi = -10^{-3}, -10^{-4}, -10^{-5}, -10^{-6}, -10^{-7}, -10^{-10}$ (in order of decreasing opacity in the plot, or right to left along a horizontal slice through the plot). The right plot shows, for the same values of $\xi$, the result when $\zeta = 27/256 - \xi$, which corresponds to the critical limit. The behaviour when $\zeta > 0$ is all qualitatively similar.  In all plots, the red curve represents the Einstein gravity result, and the dashed, gray line represents the limiting circumstance of $\rh = n$.}
	\label{fig:quartic_CP_rh_vs_n}
\end{figure}


It is useful to understand the qualitative behaviour of the bolts in the various partitions of the parameter space shown in Fig.~\ref{fig:bolt_params_CP2}. We illustrate this in the top row of Fig.~\ref{fig:quartic_CP_rh_vs_n}, which represents a `vertical slice' through Fig.~\ref{fig:bolt_params_CP2} for $\xi = -4 \times 10^{-5}$. The plot on the right is a zoomed-in copy of the left, and the decreasing opacity of the blue curves (left to right) denotes $\zeta$ becoming more negative, while the red curve corresponds to the Einstein gravity result when both couplings vanish. We see that when $\xi$ and $\zeta$ are small (or, equivalently, when $\rh$ is large) the bolt radius reduces nicely to the Einstein gravity result. The interesting behaviour is observed for smaller bolt radius. The first curve corresponds to $\zeta = -20 \times 10^{-5}$ which is in the white region of Fig.~\ref{fig:bolt_params_CP2} and above the orange line. We see that in this case, the behaviour is similar to Einstein gravity, with two possible values for the bolt radius. As $\zeta$ is further decreased, the structure of the curve remains similar but a small `flattened' region begins to form, ultimately becoming vertical for $\zeta \approx  -23.16 \times 10^{-5}$ which corresponds to the point on the orange line of Fig.~\ref{fig:bolt_params_CP2}. Continuing to decrease $\zeta$ further, we see that a bump emerges, and as a result there are up to four values of $\rh$ for a given $n$. This behaviour continues until $\zeta < 6 \xi$, which corresponds to the blue shaded region of Fig.~\ref{fig:bolt_params_CP2}. At this point, the  structure of the curve changes drastically, and there are bolts for arbitrarily large $n$. Further, in the region where $\zeta < 6 \xi$ but remains above the region bounded by the black curve in Fig.~\ref{fig:bolt_params_CP2}, there are up to three bolts for a given value of $n$.  As $\zeta$ is further decreased we continue to see three bolts for a given $n$ until we reach the black curve of Fig.~\ref{fig:bolt_params_CP2}, which corresponds to $\zeta \approx      38.03 \times 10^{-5}$. At this point, the three bolts coalesce, and for values of $\zeta$ smaller than this there is only ever a single bolt for a given $n$.


It is also possible for $\zeta$ to take on positive values, provided that $\xi < 0$ and $\zeta \le 27/256 - \xi$. The bottom row of plots in Fig.~\ref{fig:quartic_CP_rh_vs_n} shows representative behaviour in this case. The qualitative shape of the curve is controlled by the ratio $\xi/\zeta$. When $\xi/\zeta \to 0^{-}$, a peak forms at small $n$. The overall behaviour is similar to Einstein gravity: there is a maximum value of $n$ beyond which bolts cannot exist. For $n$ smaller than this value, there are two values of $\rh$  for any given $n$.

The above discussion highlights the general trend in this parameter space. The lines of `critical points' mark the boundaries where there is a change in the maximum number of bolts for a given NUT charge. For a fixed $\xi$, the structure is (referring to Fig.~\ref{fig:bolt_params_CP2}): two bolts and Einstein-like structure in the white region above the orange line; up to four bolts in the white region below the orange line; up to three bolts in the blue shaded region above the black line, and one bolt in the blue shaded region below the black line. When $\zeta$ takes on positive values, the structure remains the same as in the white region above the orange line, but a peak forms at small $n$ as $\xi/\zeta \to 0^{-}$.

\begin{figure}
\includegraphics[scale=0.45]{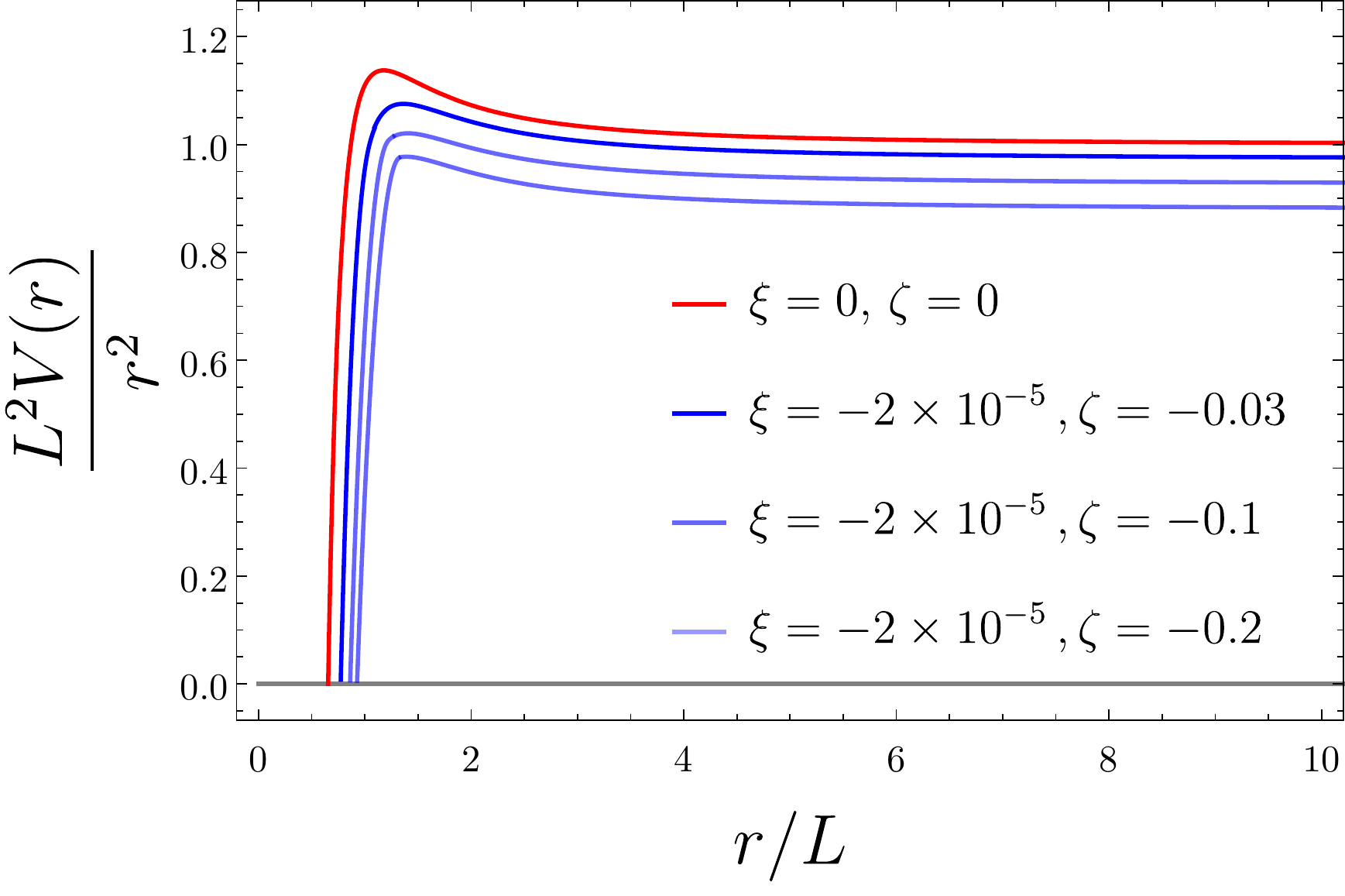}
\quad 
\includegraphics[scale=0.45]{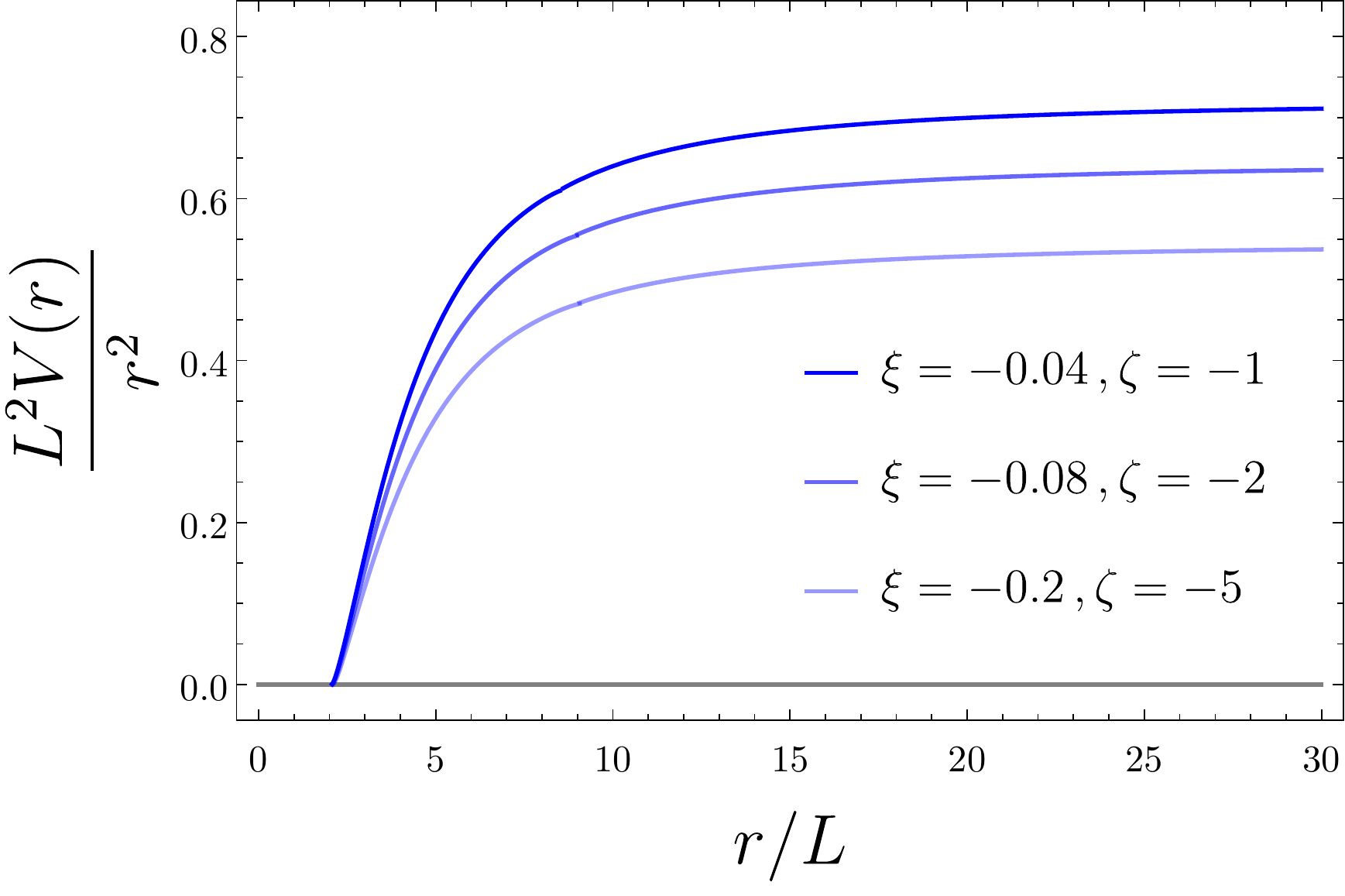}
\caption{The metric function $L^2 V_{\mathbb{CP}^2}(r)/r^2$ is plotted for bolt solutions of the quartic theories. The left plot is for different combinations of the quartic couplings with $n/L = 7/100$. For this value of the NUT parameter, the bolt solutions in the quartic theories can be compared to Einstein gravity solutions. In the right plot, the NUT parameter has been set to $n/L = 2$. For this value of the NUT parameter there are no bolt solutions in Einstein gravity and the existence of these solutions is purely because of the quartic curvature terms.}
\label{CP-bolt-numerics}
\end{figure}

So far, our study of the bolt solutions has focused on the properties of the near horizon solutions. It is important to verify that these near horizon solutions can be joined smoothly on to the asymptotic solution~\eqref{CP_asymp} that was presented at the beginning of this section. This can be shown by numerically solving the field equations, with some relevant examples shown in Fig.~\ref{CP-bolt-numerics}. The left plot shows example bolt solutions for $n/L = 7/100$. In this regime, both Einstein gravity and the quartic theories admit bolt solutions, and the two can be compared. The solutions are qualitatively similar but, of course, the solutions to the quartic solutions asymptote to $\fin r^2/L^2$ with $\fin \neq 1$. In the right plot, we show examples for $n/L = 2$ --- for this value of the NUT parameter, there are no bolt solutions in Einstein gravity.

 Finally, turning to the on-shell action, it can be computed using the same prescription as in the NUT case, but now evaluating for the bolt at $r = \rh$. In performing the calculation, we make use of the near horizon equation~\eqref{eqn:bolt_nh_cp2} to simplify the result. We find that,
\begin{align}
I_E &=  -\frac{ \pi^2}{54  L^2 G} \bigg[ 243 \rh^4 L^2 - 972 \rh^3(L^2 + 3 \rh^2) n -486 L^2 \rh^2 n^2  
\nonumber\\
&+ 972 \rh(3L^2 + 10 \rh^2)n^3 + 243 L^2 n^4 -14580 \rh n^5 + \frac{\zeta L^8}{n^3(\rh^2 - n^2)} (40 \rh n^2 + 24 n^3)
\nonumber\\
&- \frac{\xi L^8}{n^3(\rh^2-n^2)} (18 \rh^3 + 144 \rh^2 n  + 222 \rh n^2 )  \bigg] \, .
\end{align}
Making use of the chain rule and the second equation in~\eqref{eqn:bolt_nh_cp2}, we find that $E = \partial_\beta F = M$, justifying the terminology ``mass parameter'' used earlier.  The entropy is just given by $S = \beta E - I_E$ which reads
\begin{align}
S =& \frac{\pi^2}{54 \rh G} \bigg[243 \rh^5 - 486 \rh^3 n^2 - 2916 \rh^2 n^3 + 243 n^4 \rh - 2916 n^5 - \frac{4860(\rh^4 - 6 \rh^2 n^2 - 3 n^4) n^3}{L^2} 
	\nonumber\\ &+ \frac{8 \zeta L^6 (10 \rh^2 + 3 \rh n - 5 n^2)}{n (\rh^2 - n^2)} - \frac{6 \xi L^6(12 \rh^4 + 72 \rh^3 n + 65 \rh^2 n^2 - 48 \rh n^3 - 37 n^4)}{n^3(\rh^2 - n^2)} \bigg] \, .
\end{align}
We can study the extended thermodynamics of these bolts in the same manner as the NUTs. The extended first law has the same form as \eqref{eflawCP} but for the bolts the potentials are given by
\begin{equation}
V = \frac{6 \pi^2 \rh}{5} \left(3 \rh^4 - 10 \rh^2 n^2 + 15 n^4 \right) \, , \quad \Upsilon^{\cal S} = \frac{\pi \rh(3 \rh^2 + 24 \rh n + 37 n^2)}{108 n^4 G(\rh^2 - n^2)} \, , \quad \Upsilon^{\cal Z} = - \frac{\pi (5\rh - 3 n)}{81 n^2 G ( \rh^2 - n^2)} \, ,
\end{equation}
and we recall that here we are working with $\lambda_{\ssc \rm GB} = 0$. These quantities also satisfy the Smarr relation that follows from scaling, which has the same form here as in~\eqref{smarrCP}.  Again, the formula for the thermodynamic volume is unaltered from its form in Einstein gravity. Though, since $\rh$ implicitly depends on the higher-curvature couplings, the numerical value of the thermodynamic volume for fixed $n$, $\xi$ and $\zeta$ will in general differ from the Einstein gravity value. Contrast this with the situation for the NUTs where the thermodynamic volume is completely insensitive to the theory of gravity, so long as the theory belongs to the generalized quasi-topological class.

\begin{figure}[H]
	\centering 
	\includegraphics[scale=0.41]{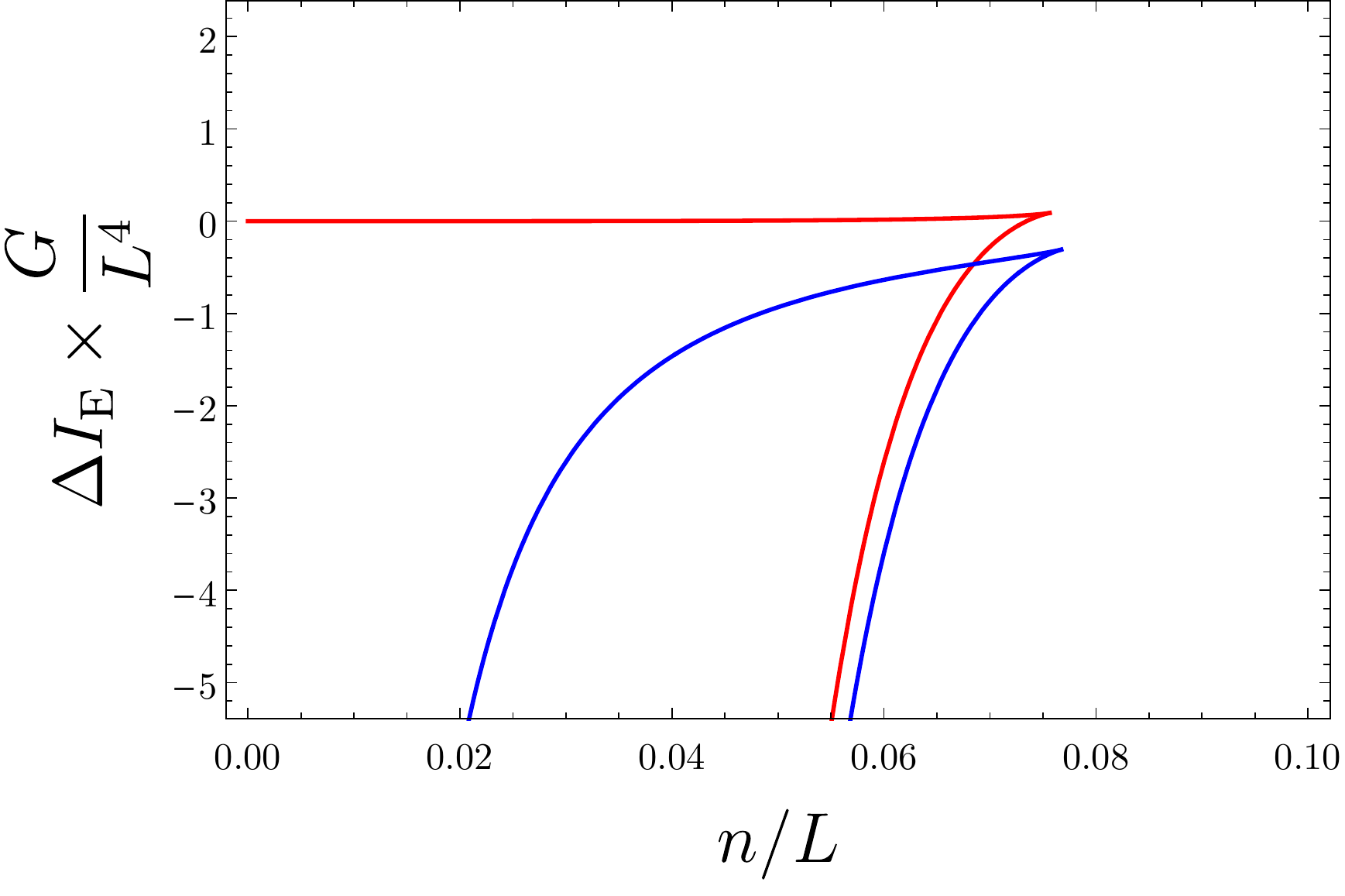}
	\includegraphics[scale=0.41]{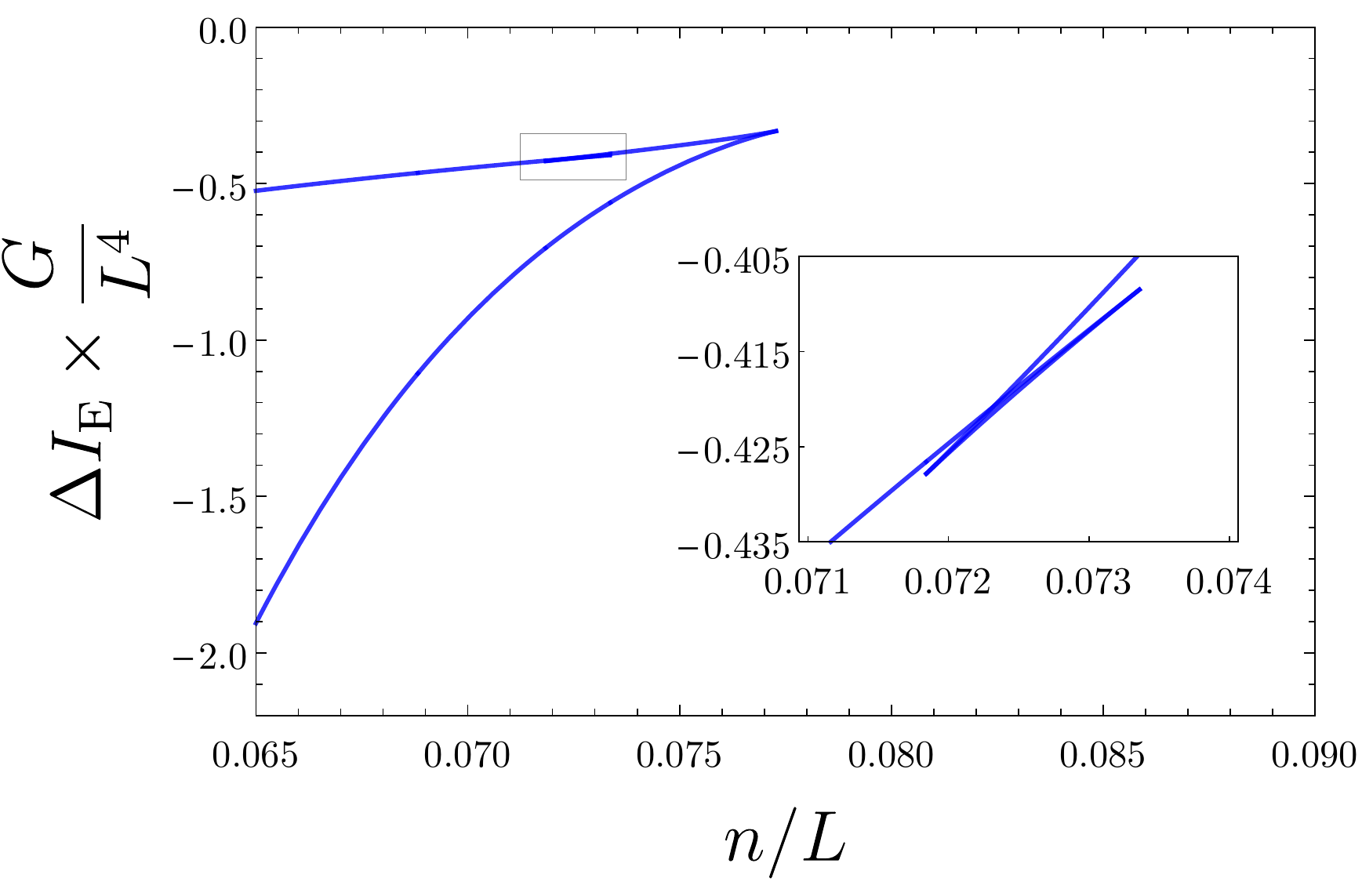}
	\includegraphics[scale=0.41]{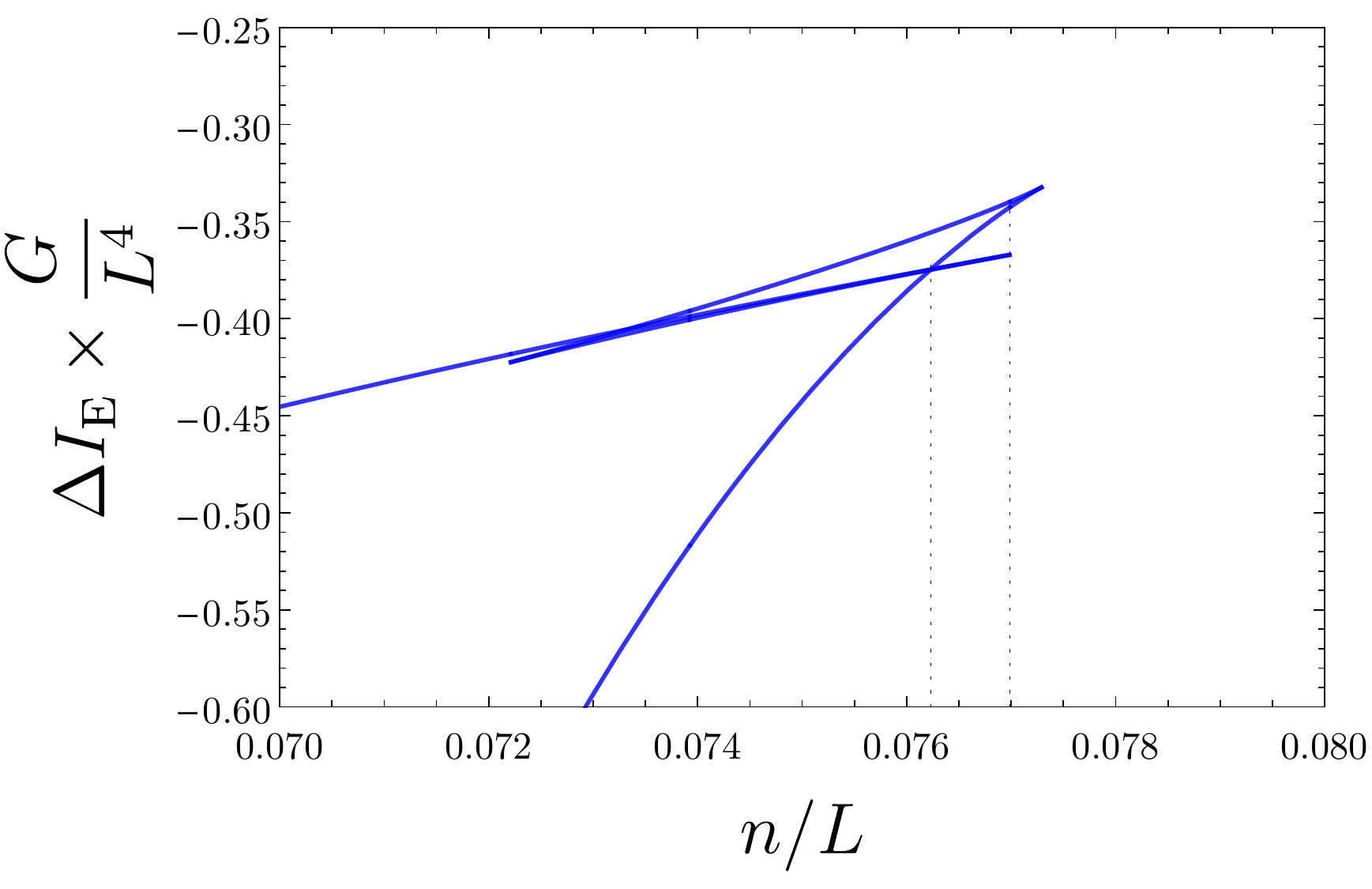}
	\includegraphics[scale=0.41]{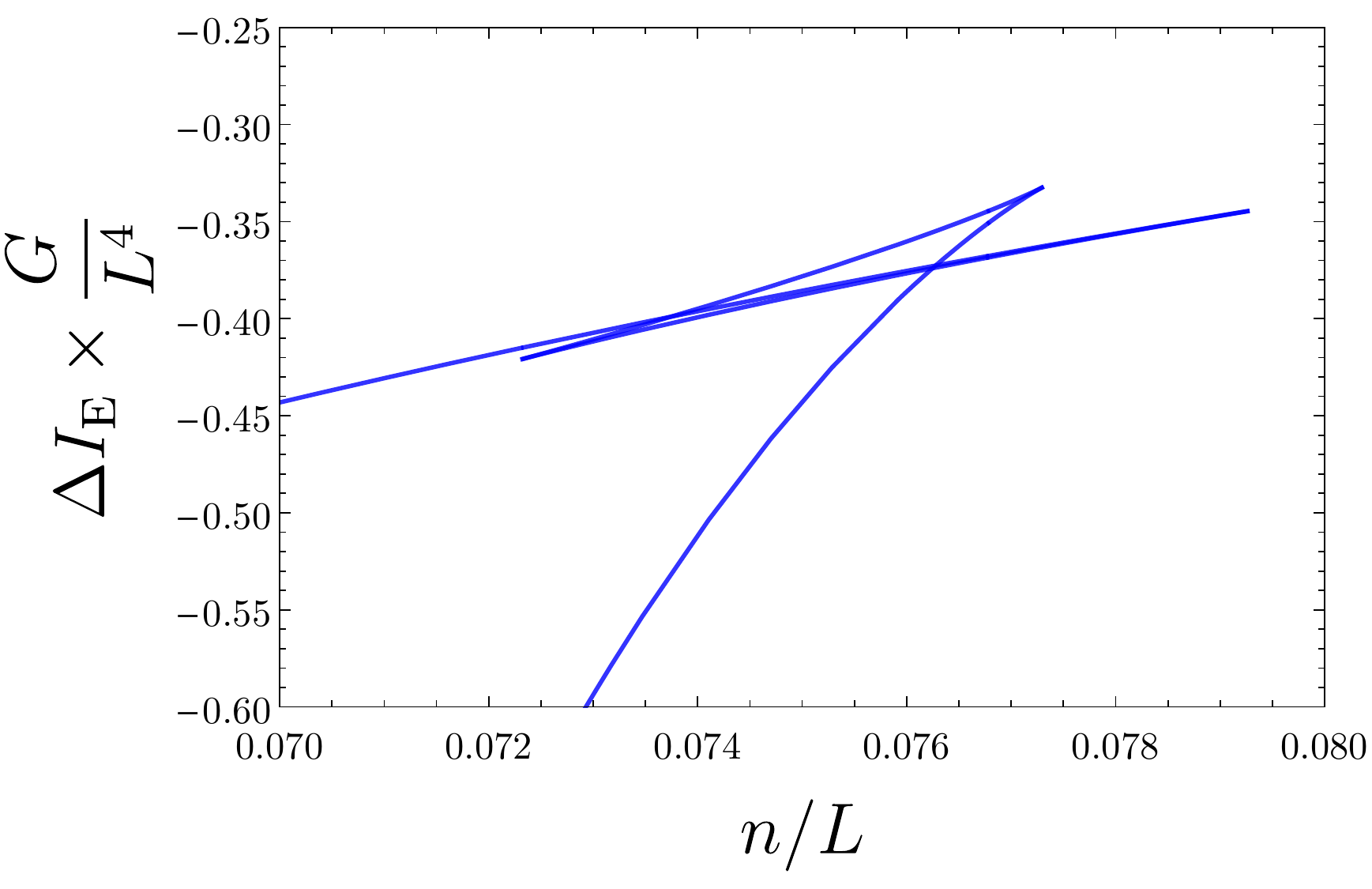}
	\includegraphics[scale=0.41]{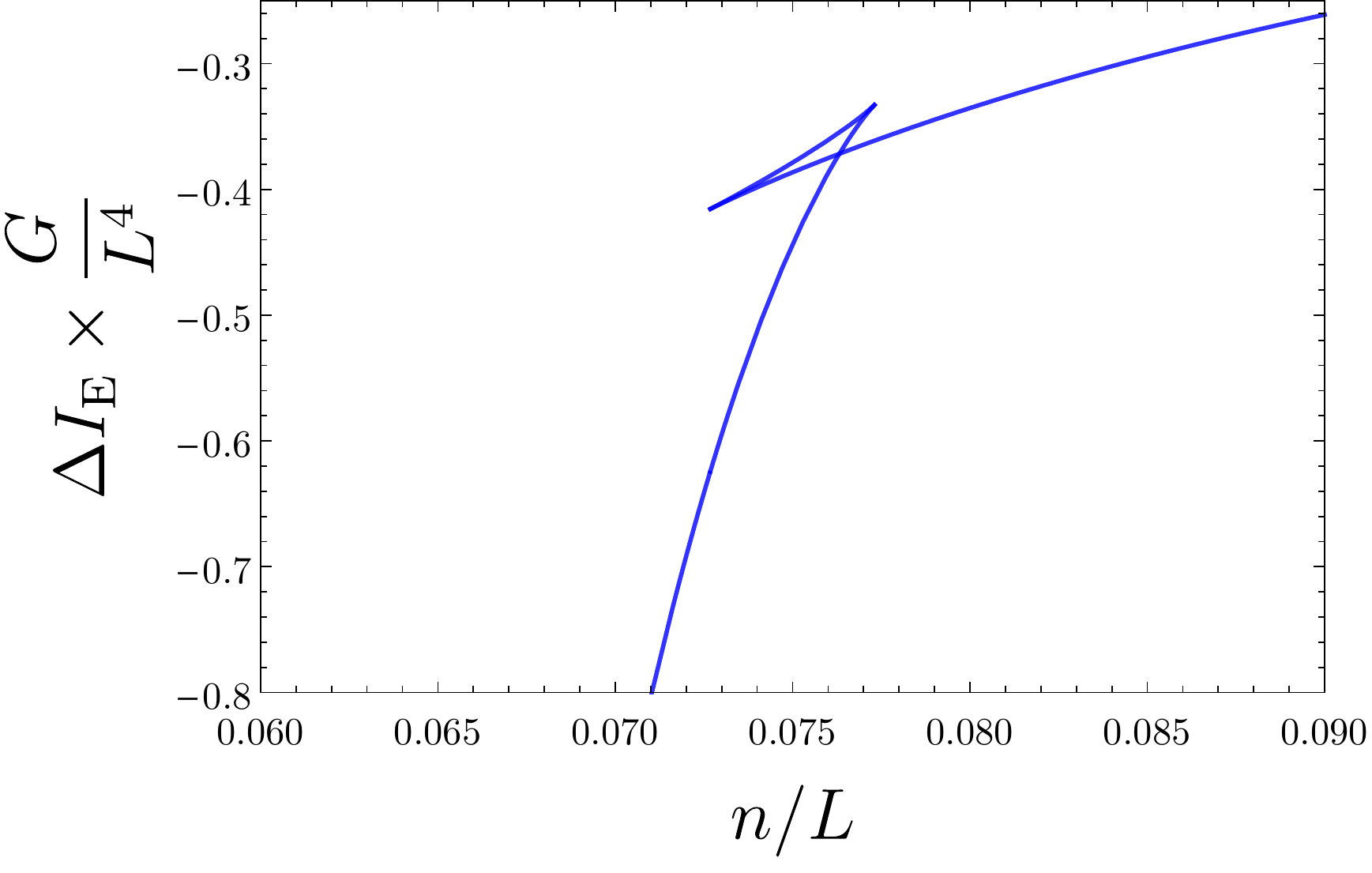}
	\includegraphics[scale=0.41]{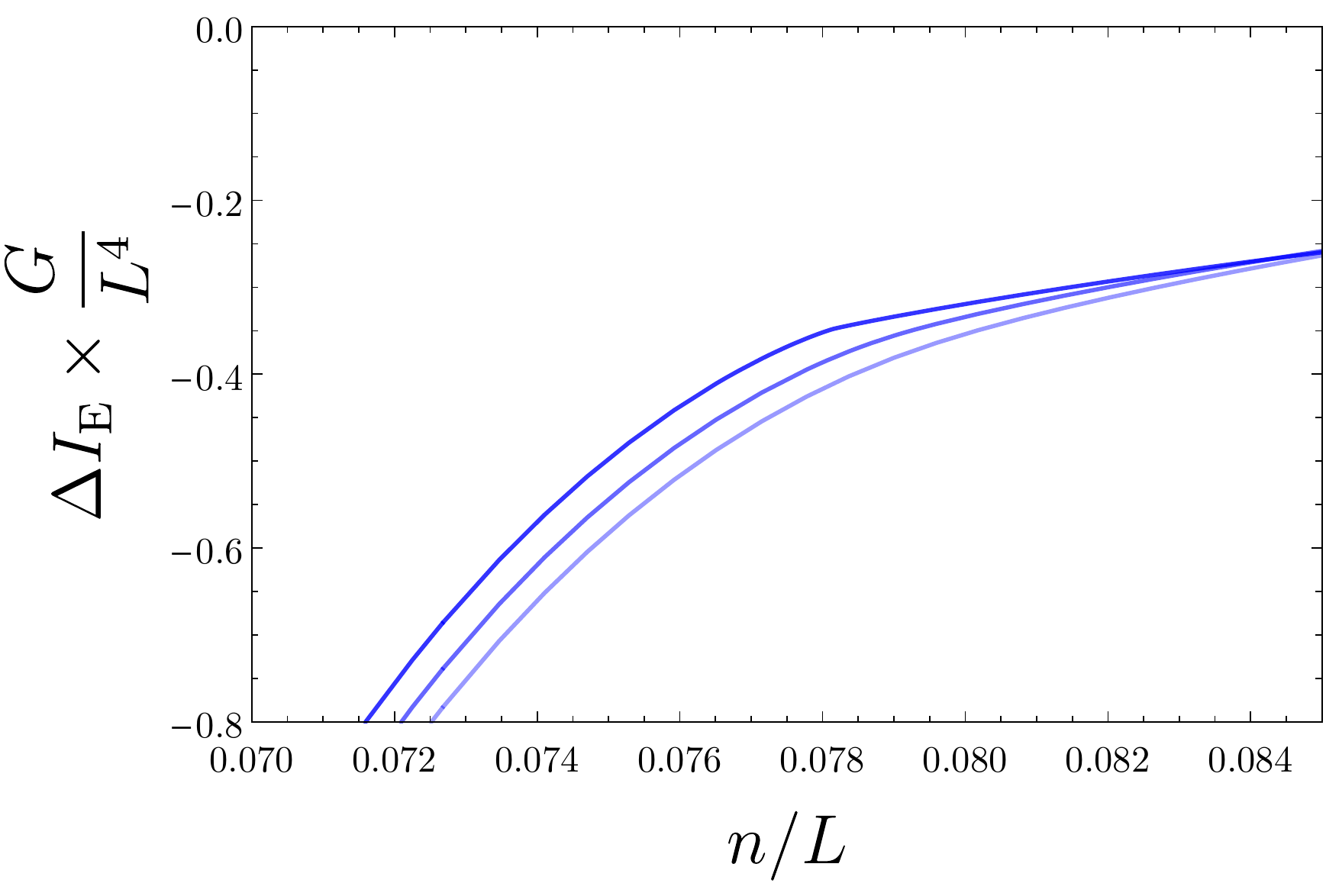}
	\caption{Euclidean on-shell action difference $\Delta I_{\rm E} = I_{\rm E}^{\rm bolt} - I_{\rm E}^{\rm NUT}$  for $\mathcal{B}=\mathbb{CP}^2$ solutions in the quartic theories. Red corresponds to Einstein gravity, while all blue curves have $\xi = -4 \times 10^{-5}$ for various values of $\zeta$. Top left: A comparison between Einstein gravity and the quartic theories with $\zeta = -10 \times 10^{-5}$, we see in both cases the action is a `cusp'.  Top right: Here $\zeta = -23.5 \times 10^{-5}$; the inset shows a zoomed-in plot of the boxed area, showing the swallowtail structure that has emerged. Center left: Here $\zeta = - 23.74 \times 10^{-5}$; the swallowtail now intersects the lower branch of the cusp. The vertical dotted lines correspond to a first order phase transition (leftmost line) and a zeroth-order phase transition (rightmost line). Center right: Here $\zeta = -23.82 \times 10^{-5}$. The swallowtail has elongated, and now extends past the cusp. Bottom left: Here $\zeta = -24.1 \times 10^{-5}$. Bolts now exist for all values of $n$, and there is a swallowtail structure present.  Bottom right: Here $\zeta = -38.026 \times 10^{-5} , -50 \times 10^{-5}$ and $-60 \times 10^{-5}$ (more to less opacity, respectively). Along the first curve, there is a critical point located at $n/L \approx 0.07815$, while the other two curves are smooth. The structure of the on-shell action is qualitatively similar to these last two curves for all $\zeta \lesssim -38.026 \times 10^{-5}$. } 
	\label{fig:CP_bolt_IE}
\end{figure}

The Euclidean on-shell action exhibits rich structure for the bolt solutions. In understanding the behaviour, it is helpful to once again refer to Fig.~\ref{fig:bolt_params_CP2}. As it turns out, this figure partitions the parameter space into regions where the behaviour is qualitatively similar.  Referring to Fig.~\ref{fig:bolt_params_CP2}, the most interesting changes in behaviour occur when the orange and black lines are crossed, which correspond to actual critical points in the thermal phase space marking the appearance/disappearance of swallowtail structures in the on-shell action.  Also when transitioning from the white-shaded to blue-shaded region, the action switches from terminating at a cusp at some finite $n$ to existing for all values of $n$. In the white region, in all cases but Einstein gravity there will be a zeroth order phase transition between bolt solutions and NUT solutions at the value of $n$ corresponding to the maximum value of $\rh$. In Einstein gravity there is also a phase transition at this point, but in that case it is first order. As an example that highlights the salient points pertaining to the bolts, let us once again consider the $\xi = -4 \times 10^{-5}$ slice through the parameter space for different values of $\zeta$ --- various relevant examples are shown in Fig.~\ref{fig:CP_bolt_IE}, where the Euclidean action of the NUT solutions has been subtracted off, $\Delta I_{\rm E} = I_{\rm E}^{\rm bolt} - I_{\rm E}^{\rm NUT}$. Though the discussion will make reference to numerical values in only this particular case, the qualitative features are general.

Particularizing now the discussion to $\xi = -4 \times 10^{-5}$, for positive $\zeta$ through to $\zeta  \approx  -23.1565 \times 10^{-5}$ (which corresponds to the orange line in Fig.~\ref{fig:bolt_params_CP2}), the behaviour is similar to Einstein gravity, with the on-shell action exhibiting two smooth branches that end at a cusp located at the maximum value of NUT charge. Precisely when $\zeta$ is chosen on the orange line shown in Fig.~\ref{fig:bolt_params_CP2} ($\zeta  \approx  -23.1565 \times 10^{-5}$ in this case),  the upper branch of $I_E$ develops a cusp, corresponding to a critical point in the system. As $\zeta$ is further decreased, a swallowtail emerges from the cusp on the upper branch, as shown in the second plot of Fig.~\ref{fig:CP_bolt_IE}. Further decreasing $\zeta$ elongates the swallowtail, and eventually it intersects the lower branch of $I_E$ --- for the particular case of $\xi = -4 \times 10^{-5}$, this intersection occurs for $\zeta \approx - 23.705 \times 10^{-5}$. This intersection then gives rise to a region where a re-entrant phase transition occurs as $n$ is increased, as shown in the  center-left plot of Fig.~\ref{fig:CP_bolt_IE}.   The two vertical black, dotted lines show the locations where these transitions occur. There is a first order phase transition from phase 1 to phase 2, followed by a zeroth order phase transition which returns the system back to the initial phase. It is in this sense that we have a re-entrant phase transition --- a monotonous variation of the NUT charge gives rise to two phase transitions with the final and initial phases coinciding. Let us note that re-entrant phase transitions were first observed in nicotine/water mixtures in~\cite{hudson1904mutual}. In the context of black hole physics, while somewhat exotic, they are well-establised --- see~\cite{Altamirano:2013ane} for an example in a rotating black hole spacetime, and~\cite{Frassino:2014pha, Hennigar:2015esa} for black hole examples in higher curvature theories of gravity. We believe this is the first instance observed for NUT charged solutions.  As $\zeta$ is further decreased, the swallowtail continues to elongate, and for $\zeta \approx - 23.753 \times 10^{-5}$, the tip of the swallowtail extends past the cusp --- this ends the region of parameter space for which re-entrant phase transitions occur. 

There is a drastic change in structure at $\zeta = 6 \xi$. Corresponding to the boundary of the blue-shaded region in Fig.~\ref{fig:bolt_params_CP2}, this condition yields the largest $\zeta$ for which there is a maximum NUT parameter for which bolts exist. From the perspective of the on-shell action, essentially what happens is, at this point, the swallowtail has now elongated ``to infinity''. Between $\zeta = 6 \xi$ and $\zeta \approx - 38.026 \times 10^{-5}  $ the action displays a swallowtail structure that is associated with a first-order phase transition. The swallowtail vanishes at a critical point when $\zeta \approx  - 38.026 \times 10^{-5} $ (the black line in Fig.~\ref{fig:bolt_params_CP2}). For $\zeta \lesssim  - 38.026 \times 10^{-5} $, the on-shell action displays only a single branch for all values of $n$.

Lastly, let us make some remarks regarding the critical points that are present at some points in the parameter space. As mentioned above, the lines of critical points appearing in Fig.~\ref{fig:bolt_params_CP2} are bonafide critical points in the thermodynamic parameter space. When in the region of the parameter space corresponding to the white region of Fig.~\ref{fig:bolt_params_CP2}, the action exhibits a cusp structure qualitatively similar to that shown in the top left plot of Fig.~\ref{fig:CP_bolt_IE}. We find that one of the critical points always occurs on the upper branch of this cusp (those corresponding to the orange curve in Fig.~\ref{fig:bolt_params_CP2}). These critical points will, therefore, not be realized since they do not comprise the dominant contribution to the partition function. The critical points that correspond to the points on the black curve shown in Fig.~\ref{fig:bolt_params_CP2} belong to the lower branch of the cusp in the white region or are on the single physical branch in the blue shaded region. These critical points are physically realized.

At the critical point, certain physical quantities blow up in power law fashion. To get a sense of the critical exponents governing these divergences, we can study the behaviour of the specific heat,
\be 
C = - T \frac{\partial^2 F}{\partial T^2}  \propto \left(1 - \frac{T}{T_c} \right)^{\tilde\alpha}
\ee
where $F = T I_E$ and $\tilde\alpha$ is the critical exponent governing this divergence\footnote{We use the notation $\tilde\alpha$ to avoid confusion with much of the black hole chemistry literature, e.g.~\cite{Kubiznak:2016qmn}, where $\alpha$ is exclusively used in reference to the specific heat at constant volume. }. Due to the complexity of the equations relating the bolt radius to the NUT parameter, it is difficult to perform an analytic study near the critical point. Instead, to make progress, we plot 
\be 
\log \bigg| \frac{1}{T} - \frac{1}{T_c} \bigg| \quad \text{vs.} \quad \log \bigg| \frac{\partial F}{\partial T} (T) -  \frac{\partial F}{\partial T} (T_c)  \bigg|
\ee 
numerically and extract the slope of this line via a linear fit. As an example, we find in the case $\xi = -4 \times 10^{-5}$ the following fit:
\be 
\log \bigg| \frac{1}{T} - \frac{1}{T_c} \bigg| = 2.941  \log \bigg| \frac{\partial F}{\partial T} (T) -  \frac{\partial F}{\partial T} (T_c)  \bigg| + {\rm constant}
\ee
which after some simple algebra yields 
\be 
\tilde\alpha = 0.659
\ee
which is consistent with $\tilde\alpha = 2/3$ to within the numerical precision. This value for the critical exponent is often observed for the divergence of the specific heat at constant pressure in black hole systems --- see, e.g.,~\cite{Chamblin:1999hg}. In this sense, it is not surprising to find that the same critical exponent governs the behaviour near the critical point for the bolts. A numerical survey of many critical points for different values of the couplings shows that they are all consistent with this result.

The red dot shown in Fig.~\ref{fig:bolt_params_CP2} represents a special point in the parameter space where two critical points merge. Because of this, one might hope to see novel critical exponents similar to how the coalescence of multiple critical points leads to non-mean field theory critical exponents for Lovelock black holes~\cite{Dolan:2014vba}. However, unfortunately, this is not the case here. The reason is that as the red dot is approached, there is one critical point on the upper branch of the cusp and one on the lower branch. When these critical points merge, they also meet at the cusp which acts as a phase boundary --- no solutions exist beyond the tip of the cusp. To within the accuracy of our calculation, the critical exponent associated with each critical point as the cusp is approached remains consistent with $\tilde\alpha = 2/3$.

\vspace{.01cm}

\subsection{$\mathcal{B}=\mathbb{S}^2\times \mathbb{S}^2$}

As mentioned at the beginning of this section, the case of $\mathcal{B} = \mathbb{S}^2\times \mathbb{S}^2$ is somewhat special as it only admits theories of the generalized quasi-topological type. Our action is then~\eqref{eqn:quart_6d_act} but now with $\zeta = 0$. 

The metric takes the form of Eq.~\eqref{FFnut}, with the following 1-form and base space metric, 
\begin{align}
A_{\mathbb{S}^2\times \mathbb{S}^2}^2=2 \cos\theta_1 d\phi_1 + 2\cos \theta_2 d\phi_2\, , \quad
d\sigma_{\mathbb{S}^2\times \mathbb{S}^2}^2=d\theta_1^2+\sin^2\theta_1 d\phi^2_1 +d\theta_2^2+\sin^2\theta_2 d\phi^2_2   \, .
\end{align}
The field equations are relatively complicated and have been listed in full detail above. The  $\mathbb{S}^2\times \mathbb{S}^2$ case is somewhat interesting because there is the appearance of a logarithm term in the integrated field equations. 

We first wish to consider the field equations asymptotically.   The vacua of the theory are determined by the equation $h(f_\infty) = 0$ with
\be
h(f_\infty) \equiv 1 - f_\infty + \lambda_{\rm \ssc GB}f_\infty^2 + \xi f_\infty^4 \, .
\ee
Again, the solution will consist of a homogenous part $g(r)$ and a particular part $V_p(r)$, given by
\begin{align} 
V_p(r) &= f_\infty \frac{r^2}{L^2} + \frac{1}{3} - \frac{3 f_\infty n^2}{L^2} -\frac{2}{9 L^2 r^2 \fin^2h'(\fin)} \big[-18 \fin^2(\lambda_{\rm \ssc GB} \fin^2 - 2) n^4 
\nonumber\\
&+ 6 L^2 \fin(3\fin^2\lambda_{\rm \ssc GB} - 3 \fin +2)n^2 - L^4(5\fin^2 \lambda_{\rm \ssc GB} - 6 \fin + 6) \big]
- \frac{C}{2 h'(\fin) r^3} + {\cal O}(r^{-4}) 
\end{align}
and obtained by performing a large-$r$ series solution of the field equation.
The homogeneous equation, at large $r$, is a second order differential equation for $g(r)$,
\be 
a(r) g''(r) + b(r) g'(r) + c(r) g(r) = 0
\ee
with the coefficients given by,
\begin{align}
a(r) &= \frac{16 \xi \fin r}{3} (18 \fin^2 n^4 - 6 L^2 \fin n^2 + L^4 ) \, ,
\nonumber\\
b(r) &= - 16 \xi \fin (18 \fin^2 n^4 - 6 L^2 \fin n^2 + L^4 ) \, ,
\nonumber\\
c(r) &= - 2 h'(\fin) r^3 \, .		
\end{align}
The differential equation can be solved exactly in terms of modified Bessel functions,
\be \label{grsolnS2S2}
g(r) = C_1 r^2 I_1 \left( r \sqrt{-\frac{c(r)}{4 a(r)}} \right)
+ C_2 r^2 K_1 \left( r \sqrt{-\frac{c(r)}{4 a(r)}} \right) \, .
\ee
The asymptotic form of this solution will consist of a super-exponentially growing and a super-exponentially decaying mode provided that $a(r) < 0$. If $a(r) > 0$, the asymptotic behaviour is pathological. 
Therefore, we must demand that $a(r) < 0$ and use the asymptotic AdS boundary conditions to set the growing mode ($C_1$) to zero. Demanding $a(r) < 0$ is equivalent to demanding that the coupling $\xi$ is negative, since the term in parentheses in the expression for $a(r)$ is always positive.

\subsubsection{Taub-NUT solutions}

Taub-NUT solutions with base space $\mathbb{S}^2\times \mathbb{S}^2$ are generically pathological, even in Einstein gravity, due to a curvature singularity at the NUT $r = n$. The situation is actually worse here, since the logarithm in the field equations --- see above --- develops an essential singularity at $r = n$.  Thus it seems that there are no well-behaved NUT solutions for $\mathcal{B} = \mathbb{S}^2\times \mathbb{S}^2$.

\subsubsection{Taub-bolt solutions}

Even in the case of Einstein gravity, Taub-NUT solutions are singular at the NUT when the base is either $\mathcal{B} = \mathbb{S}^2 \times \mathbb{S}^2$ or $\mathcal{B} = \mathbb{S}^2 \times \mathbb{T}^2$. This pathology leads to problems when solving the field equations for higher-curvature gravities.  These problems were first observed in the case of  Gauss-Bonnet gravity~\cite{Dehghani:2005zm} where it was found that NUT solutions with $\mathcal{B} = \mathbb{S}^2 \times \mathbb{S}^2$ do not exist for non-vanishing Gauss-Bonnet coupling. It was conjectured that this was due to the fact that the corresponding Einstein gravity solutions are singular~\cite{Dehghani:2005zm}.

In the higher-curvature gravities considered here, we are faced with a similar problem --- the field equations develop an essential singularity at $r = n$, and therefore the NUT solutions do no exist. However, luckily, the situation is better in the case of bolt solutions. In this case, we will proceed in the same manner as before noting that the periodicity of Euclidean time must be $\tau \sim \tau + 12 \pi n$ to ensure the absence of Dirac-Misner string singularities. This value is the same as the periodicity enforced by the field equations themselves in Einstein gravity. In a higher curvature theory, like the ones studied here, the field equations may not naturally ensure regularity and we must enforce it by hand. This approach was used in~\cite{Dehghani:2005zm} to successfully construct bolt solutions in Gauss-Bonnet gravity that limit smoothly to the Einstein gravity solutions as the Gauss-Bonnet coupling is turned off, and we will adopt the same approach here.

%
%
%

We begin by expanding the metric near the bolt
\be 
V(r) = \frac{1}{3 n} (r-\rh) + \sum_{i=2}^\infty (r-\rh)^i a_i 
\ee
The first two relationships are
\begin{align}\label{eqn:S2xS2-bolt-eq}
C &= \frac{2}{3 \rh L^2} \left(-3 \rh^6 + (15n^2- L^2)\rh^4 + (6 L^2 n^2 - 45 n^4)\rh^2 + 3 L^2 n^4 - 15 n^6 \right) 
\nonumber\\
&+ \frac{L^6\xi}{243 \rh n^4} \left(163 n^2 + 48 n \rh + 9 \rh^2 + 81 \rh n \log \left(\frac{\rh - n}{\rh+ n} \right)\right) \, ,
\nonumber\\
0 &= \frac{2 (\rh-n)^2(\rh + n)^2(-15 \rh^2 n + L^2 \rh - 3 L^2 n + 15 n^3) }{3 n L^2 \rh^2 } 
\nonumber\\
&- \frac{\xi L^8}{243 L^2 n^4 \rh^2(\rh^2-n^2)} (3  \rh^4 - 10 \rh^2 n^2 + 168 \rh n^3 - 163 n^4)	\, ,
\end{align}
which correspond to the ${\cal O}(1)$ and ${\cal O}\left((\rh -n)^1 \right)$ terms in the field equations. At the next order, one determines $a_3$ in terms of $a_2$, $n$, and $\rh$. This pattern continues, and one is left with the unfixed parameter $a_2$ that should be, as usual, determined by demanding that the field equations have the correct asymptotic behaviour.

The on-shell action can be computed in the same way as before and we find that the expression is given by
\begin{align} 
I_{\rm E} &= \frac{4 \pi^2}{ G L^2} \bigg[ -12 \rh^5 n + L^2 \rh^4 - 4 n (L^2 - 10 n^2) \rh^3 - 2 L^2 \rh^2 n^2 + 12 n^3(L^2 - 5 n^2) \rh + L^2 n^4 
\nonumber\\
&-\frac{2 \xi L^8 }{81 n^3 (\rh^2 - n^2)} \left(3 \rh^3 + 24 \rh^2 n + 82 \rh n^2 - 108 n^3 \right)- \frac{\xi L^8}{n^2} \log\left(\frac{\rh-n}{\rh + n} \right)\bigg] \, .
\end{align}
The standard thermodynamic relationships are now
\begin{align}
E &= - \frac{2 \pi C}{G} \, , 
\nonumber\\
S &= \frac{4 \pi^2}{G L^2} \bigg[ \rh^4 L^2 - 80 \rh^3 n^3 + 2 L^2 \rh^2 n^2 + (240 n^5 - 24 L^2 n^3)\rh - 3 L^2 n^4 
\nonumber\\
&- \frac{8 \xi L^8}{27 n^3 (\rh - n)^2(\rh+n)^2} \left(\rh^5 + 6 \rh^4 n + \frac{227}{12} n^2 \rh^3 - 19 n^3 \rh^2 - \frac{409}{12} n^4 \rh + 27 n^5 \right) 
\nonumber\\
&- \frac{3 \xi L^8}{n^2} \log \left(\frac{\rh - n}{\rh + n} \right)\bigg] 
\end{align}
where the second equation in~\eqref{eqn:S2xS2-bolt-eq} was used to simplify the results.

\subsection{$\mathcal{B}=\mathbb{S}^2\times \mathbb{T}^2$}
Next we consider the metric~\eqref{FFnut} with $\mathcal{B}=\mathbb{S}^2\times \mathbb{T}^2$. The potential and base space metric read
\begin{equation}
A_{\mathbb{S}^2\times \mathbb{T}^2}^2=2 \cos\theta d\phi + 2\eta d\zeta \, , \quad
d\sigma_{\mathbb{S}^2\times \mathbb{T}^2}^2=d\theta^2+\sin^2\theta d\phi^2 +d\eta^2 +d\zeta^2\, .
\end{equation}
We take the action to be~\eqref{eqn:quart_6d_act} but now with $\xi = 0$. The vacua of the theory are then determined as roots of the polynomial equation,
\be 
h(\fin) = 1 - \fin + \lambda_{\rm \ssc GB}\fin^2 + \zeta \fin^4 \, .
\ee
At large distances, the solution consists of a particular and homogenous part, $V(r) = V_p(r) + g(r)$. The particular solution is given, as usual, by performing a series expansion of the equation in $1/r$ and matching the coefficients to give a consistent solution order by order.  This yields
\begin{align} 
V_p(r) &= \fin \frac{r^2}{L^2} + \frac{1}{6} - \frac{3 \fin n^2}{L^2} - \frac{1}{36 L^2 r^2 \fin^2 h'(\fin)}\big[ -144 \fin^2(\fin^2\lambda_{\rm \ssc GB} - 2)n^4 
\nonumber\\
&+ 24 L^2 \fin(3 \fin^2 \lambda_{\rm \ssc GB} - 3 \fin + 2)n^2 + L^4 (5 \fin^2 \lambda_{\rm \ssc GB} - 6 \fin + 6) \big] - \frac{C}{2 r^3 h'(\fin)} + {\cal O}(r^{-4}) 
\end{align}
for the particular solution. The homogenous equation, at large $r$, is a second order differential equation for $g(r)$, 
\be 
a(r) g''(r) + b(r) g'(r) + c(r) g(r) = 0
\ee
with the coefficients given by,
\begin{align}
a(r) &= \frac{80 \fin n^2 \zeta}{81 r}(L^2-12 n^2\fin)^2 \, ,
\nonumber\\
b(r) &= - \frac{380 \fin n^2 \zeta }{81 r^2}(L^2 - 12 \fin n^2)^2
\nonumber\\
c(r) &= -2 h'(\fin) r^3 \, .			
\end{align}
The large-$r$ homogeneous solution has a known solution in terms of modified Bessel functions,
\be 
g(r) = C_1 r^{23/8} I_{23/24}\left(r \sqrt{-\frac{c(r)}{9 a(r)}}\right) + C_2 r^{23/8} K_{23/24}\left(r \sqrt{-\frac{c(r)}{9 a(r)}}\right)
\ee
The asymptotic form of this solution will consist of a super-exponentially growing and a super-exponentially decaying mode provided that $a(r) < 0$. If $a(r) > 0$, the asymptotic behaviour is pathological. 
Therefore, we must demand that $a(r) < 0$ and use the asymptotic AdS boundary conditions to set the growing mode ($C_1$) to zero. Demanding $a(r) < 0$ is equivalent to demanding that the coupling $\zeta$ is negative, since the term in parentheses in the expression for $a(r)$ is always positive.

\subsubsection{Taub-NUT solutions}

First, we note that whenever the base is $\mathbb{S}^2 \times \mathbb{T}^2$, the NUT solutions will necessarily possess curvature singularities at $r = n$. With this in mind, let us begin by discussing the situation for Einstein and Gauss-Bonnet gravity, as subtleties arise already in these cases. In the pure Einstein gravity case, the solution to the field equations reads
\be\label{eqn:ein_s2t2} 
V_{\mathbb{S}^2 \times \mathbb{T}^2}^{\rm E}(r) = \frac{6 r^6 + (L^2 - 30 n^2) r^4 + (90 n^4 - 6 L^2 n^2) r^2 + 3 C_{\mathbb{S}^2 \times \mathbb{T}^2} r L^2 - 3 L^2 n^4 + 30 n^6 }{6 L^2 (n+r)^2(r-n)^2} \, ,
\ee
where $C_{\mathbb{S}^2 \times \mathbb{T}^2}$ is the integration constant in the field equation \req{eq6d}. If we demand that this solution permits NUTs, then we must have $V(r = n) = 0$ and this fixes the integration constant uniquely as 
\be
C_{\mathbb{S}^2 \times \mathbb{T}^2} =  \frac{8n^3}{3 L^2}(L^2 - 12 n^2) \, ,
\ee
which yields \req{NUTE2} with $\kappa=0$.
Alternatively, we could have begun by expanding $V(r)$ as a series near $r = n$ as
\be 
V(r) = 4 \pi T (r-n) + \sum_{i=2}^\infty (r-n)^i a_i \, .
\ee
Then we would find that, near the NUT, the field equation takes the form,
\be \label{eqn:ein_s2t2_series} 
0 = \frac{8n^3}{3 L^2}(L^2 - 12 n^2) - C_{\mathbb{S}^2 \times \mathbb{T}^2} + \left(8 n (4 \pi T) - \frac{4}{3} \right)(r-n)^3 + {\cal O}\left((r-n)^4 \right) \, .
\ee
From either perspective, it is easy to check that, at the location of the NUT we have $V'(n)= 4 \pi T = 1/(6n)$. 

Next we turn on the Gauss-Bonnet term. The addition of this term gives rise to two exact (but messy) solutions for $V(r)$.  One of these solutions limits to the Einstein gravity result as $\lambda_{\rm \ssc GB} \to 0$ (this solution is given in~\cite{Dehghani:2005zm} --- see Eq.~(6) in that paper), and the other blows up in that limit. If we expand the field equations near $r = n$ using the same ansatz as before, we now find the first few terms to be\begin{align}
0 &= \frac{8n^3}{3 L^2}(L^2 - 12 n^2) - C_{\mathbb{S}^2 \times \mathbb{T}^2} + \left(\frac{4}{3} L^2 \lambda_{\rm \ssc GB} (4 \pi T)(1- 3 n (4 \pi T) ) \right) (r-n)^2 
\nonumber\\
& + \left( 8 n (4 \pi T) - \frac{4}{3} - \frac{2\lambda_{\rm \ssc GB} L^2}{3 n} \left[12 n^2 (4 \pi T) a_2 - 2 n a_2 + 4 \pi T \right]  \right)(r-n)^3
+ {\cal O}\left((r-n)^4 \right) \, .
\end{align}
This expansion of the field equations is the same as~\eqref{eqn:ein_s2t2_series} in the limit $\lambda_{\rm \ssc GB} \to 0$. However, we can see that the solution of these equations for the temperature does not limit to the Einstein gravity result. While the result for the integration constant is the same as before, the Gauss-Bonnet term now contributes at a lower order in the field equations, bringing a new term into the expansion that is not present in the case of pure Einstein gravity --- see the $\mathcal{O}((r-n)^2)$ term above. There are three possibilities for a solution at that order: (1) $\lambda_{\rm \ssc GB}  = 0$, in which case we recover the Einstein result quoted above, (2) $T = 0$, which when extended to the full non-perturbative solution corresponds to the Gauss-Bonnet branch that limits to the Einstein branch (the one discussed in~\cite{Dehghani:2005zm}), or (3) $4 \pi T = 1/(3n)$, which corresponds to the Gauss-Bonnet branch  that is singular in the limit $\lambda_{\rm \ssc GB} \to 0$.  We conclude that none of these possibilities for the temperature actually limits to the temperature of the Einstein gravity solution, even though the full non-perturbative solution corresponding to the $T=0$ branch does limit to the Einstein gravity solution. 

What we have arrived at here is an order of limits problem: performing the $\lambda_{\rm \ssc GB} \to 0$ before the $r \to n$ limit gives a different result than first performing the $r \to n$ limit followed by $\lambda_{\rm \ssc GB} \to 0$. The limit of the temperature expression is not continuous: for any non-zero $\lambda_{\rm \ssc GB}$ it should be $T = 0$, but when $\lambda_{\rm \ssc GB}$ is precisely zero, it ``jumps'' to $4 \pi T = 1/(6n)$. The origin of this incompatibility of limits would seem to be linked to the fact that the space is actually singular at $r = n$ for $\mathcal{B} = \mathbb{S}^2 \times \mathbb{T}^2$.

Having reviewed this structure, let us now consider the case with both $\zeta$ and $\lambda_{\rm \ssc GB}$ non-vanishing.  We expand the metric function as above and demand the field equations are satisfied order by order. The first terms in the expansion of the field equations are  
\begin{align}
0 &= - C_{\mathbb{S}^2 \times \mathbb{T}^2} + \frac{8 n^3}{3 L^2} (L^2-12 n^2) - \frac{(4 \pi T)^3 L^6 \zeta}{3} \left(24 n (4\pi T) - 11 \right)
\nonumber\\ 
&- \frac{(4 \pi T)^2 L^6 \zeta}{2 n} \left(64 n^2 (4 \pi T) a_2 - 36 n (4 \pi T)^2 - 22 n a_2 + 13 (4 \pi T)  \right)(r-n) + \mathcal{O}\left( (r-n)^2 \right)  
\end{align}
and we see here that the new higher curvature terms contribute at even lower order than the Gauss-Bonnet term. Again, we have two (non-trivial) possibilities for a solution here. The first possibility is to have $T = 0$. This case leads to a non-perturbative solution with the value of $a_2$ determined as a solution to the following equation:
\begin{align}
0 &= 1 -  \lambda_{\rm \ssc GB} L^2 a_2 - \frac{7}{3}  \zeta L^6  a_2^3 \, ,
\end{align}
with $a_3$, $a_4$ and so on given directly (and uniquely) as functions of the couplings and $a_2$. From this equation, it is clear that one of the three possible roots for $a_2$ will limit to the extremal Gauss-Bonnet solution when $\zeta \to 0$.
The second possibility is for
\be 
a_2 = \left(\frac{4 \pi T}{2 n}\right) \frac{36 n (4 \pi T) - 13}{32 n (4 \pi T) - 11}\, ,
\ee
and the rest of the constants are determined uniquely in terms of $T$, the couplings, and $n$. In this solution, $T$ is left as an arbitrary parameter and the solution limits to neither the Einstein result nor the Gauss-Bonnet case as $\zeta \to 0$. Based on our intuition from the Gauss-Bonnet situation analyzed above, the most reasonable conclusion would seem to be that the NUT solutions on $\mathcal{B} = \mathbb{S}^2 \times \mathbb{T}^2$ should be regarded as extremal solutions. That is, taking $T=0$ seems to be the most reasonable of the various options discussed above.

\subsubsection{Taub-bolt solutions}

For simplicity, we will at this point set $\lambda_{\rm \ssc GB} = 0$.  Then, we expand the metric function as
\be 
V(r) = 4 \pi T (r-\rh) + \sum_{i=2}^\infty (r-\rh)^i a_i\, ,
\ee
and demand the field equations are satisfied order by order. There will be no curvature singularities unless $\rh = n$. \ Since the NUT solutions for $\mathbb{B} = \mathbb{S}^2 \times \mathbb{T}^2$ are somewhat pathological, and should probably be regarded as zero temperature or extremal solutions, there is not a natural periodicity enforced on the bolts via regularity of the NUTs. If we wish the temperature to match the bolts from Einstein gravity, then we would require that $\beta = 24 \pi n$. Here, similar to what was done in~\cite{Dehghani:2005zm} for this base, we will keep the temperature explicitly present in the following analysis.  

The first few components of the field equations are given by
\begin{align}
 C_{\mathbb{S}^2 \times \mathbb{T}^2} &= -  \frac{2 \rh^5}{L^2} - \frac{(L^2 - 30 n^2)\rh^3}{3 L^2} + \frac{2 n^2 (L^2 - 15 n^2) \rh}{L^2} + \frac{n^4(L^2-10 n^2)}{\rh L^2} - \frac{20}{9 } \frac{L^6 n^2 \zeta (4 \pi T)^4 }{\rh}
\nonumber\\
0 &= \frac{(\rh - n)^2(\rh + n)^2( -10 \rh^2 + 2 L^2 \rh (4 \pi T) + 10 n^2 - L^2)}{\rh^2 L^2} 
\nonumber\\
&+ \frac{10 L^6 n^2 \zeta (4 \pi T)^3 (2 \rh^2 (4 \pi T) + 2 n^2 (4 \pi T) + \rh)}{9 \rh^2 (\rh - n)(\rh + n)}	\, .
\end{align}
The first equation determines the mass parameter, while the second gives a relation between the bolt radius and the NUT parameter. At the next order in the field equations, $a_3$ is determined by $a_2$, the bolt radius and the NUT parameter. This pattern continues in the usual way, and once $a_2$ is determined by demanding the appropriate asymptotics the full near horizon solution will be determined.

We can compute the free energy from the on-shell Euclidean action as before. Denoting the compactification length of the $\mathbb{T}^2$ as $l$, we find the following:
\begin{align}
F &= \frac{l \pi}{G} \bigg[- \frac{1}{3 L^2 (4\pi T)} \left( 12 \rh^5 - 3 L^2 \rh^4 (4 \pi T) + (2 L^2 - 40 n^2) \rh^3 + 6 L^2 \rh^2 n^2 (4 \pi T) \right.
\nonumber\\
&\left.- 6 n^2(L^2 - 10n^2) \rh - 3 L^2 n^4 (4 \pi T) \right) - \frac{10 \zeta L^6 n^2 (4 \pi T)^2 \left(1 + 4 \rh (4 \pi T)\right)}{9 (\rh - n)(\rh + n)}\bigg]\, ,
\end{align}
from which we obtain
\begin{align}
E &= - \frac{l C_{\mathbb{S}^2 \times \mathbb{T}^2}}{2 G} \, ,
\nonumber\\
S &= \frac{l \pi (\rh^2 - n^2)^2}{G} \bigg[1  + \frac{10 L^6 n^2 \zeta (4\pi T)^2\left[3 + 16 \rh (4\pi T) \right]}{(\rh^2 - n^2)^3} \bigg]	 
\end{align}
for the energy and entropy.

\subsection{$\mathcal{B}=\mathbb{T}^2\times \mathbb{T}^2$}
Let us finally consider the case of $\mathbb{T}^2\times \mathbb{T}^2$. In that case, the 1-form and the metric of the base space are given by 
\begin{align}
A_{\mathbb{T}^2\times \mathbb{T}^2}&= \frac{4}{L^2}\left(\eta_1 d\zeta_1+\eta_2 d\zeta_2\right) \, ,\quad d\sigma_{\mathbb{T}^2\times \mathbb{T}^2}^2=\frac{1}{L^2}\left(d\eta_1^2 +d\zeta_1^2+d\eta_2^2 +d\zeta_2^2\right)\, .
\end{align}
Each pair of coordinates $(\eta_1, \zeta_1)$ and $(\eta_2, \zeta_2)$ parametrize a $ \mathbb{T}^2$, and for simplicity we can assume that $\eta_1, \zeta_1$ have both periodicity $l_1$ and  $\eta_2, \zeta_2$ have periodicity $l_2$. The coordinate $\tau$ is also compact and its period $\beta_{\tau}$ is a free parameter, since there are no regularity conditions in this case. The function $V(r)$ is determined as usual by Eq. \req{eq6d}. 
Both densities $\mathcal{Z}$ and $\mathcal{S}$ can be introduced in this case, and for completeness we will also consider the Gauss-Bonnet term. As usual, let us start by determining the asymptotic behaviour of the solution, which we decompose as a particular solution plus a homogeneous one as $V(r)=V_{p}(r)+r^2g(r)/L^2$. The particular solution is found by performing a $1/r$ expansion, which reads
\begin{equation}\label{asympt2t2}
V_p(r)=f_{\infty}\frac{r^2}{L^2}-3f_{\infty}\frac{n^2}{L^2}-\frac{4n^4(2-\lambda_{\rm \ssc GB} f_{\infty}^2)}{h'(f_{\infty})L^2r^2}-\frac{C_{\mathbb{T}^2\times\mathbb{T}^2}}{2 h'(f_{\infty})r^3}+\mathcal{O}\left(\frac{1}{r^{4}}\right)\, .
\end{equation}
On the other hand, at linear order $g(r)$ satisfies the equation
\begin{equation}
a(r)g''(r)+b(r)g'(r)+c(r) g(r)=0\, ,
\end{equation}
where the functions $a(r)$, $b(r)$ and $c(r)$ take the following form asymptotically
\begin{eqnarray}
a(r)&=&-192\xi \frac{f_{\infty}^3 n^4 r^3}{L^2}+(12480 \xi-2560\zeta)\frac{f_{\infty}^3n^6 r}{9 L^2}+\mathcal{O}\left(\frac{1}{r}\right)\, ,\\
b(r)&=&-192\xi \frac{f_{\infty}^3 n^4 r^2}{L^2}+(9216 \xi+1920\zeta)\frac{f_{\infty}^3n^6 }{9 L^2}+\mathcal{O}\left(\frac{1}{r^2}\right)\, ,\\
c(r)&=&\frac{4h'(f_{\infty})r^5}{L^2}-\frac{8h'(f_{\infty})n^2 r^3}{L^2}+\mathcal{O}(r)\, .
\end{eqnarray}
The solution is qualitatively different depending on whether $\xi=0$ or $\xi\neq 0$. In the latter case, taking into account only the leading terms in $a$, $b$ $c$, we get the following solution asymptotically:
\begin{equation}
g(r)=c_1 J_0\left(\frac{\sqrt{-h'(f_{\infty})}r^2}{8n^2\sqrt{3\xi f_{\infty}^3}}\right)+c_2 Y_0\left(\frac{\sqrt{-h'(f_{\infty})}r^2}{8n^2\sqrt{3\xi f_{\infty}^3}}\right)\, .
\end{equation}
On the other hand, when $\xi=0$, the asymptotic solution reads instead
\begin{equation}
g(r)=c_1 r^{7/8}J_{7/24}\left(\frac{\sqrt{-h'(f_{\infty})}r^3}{8n^3\sqrt{10\zeta f_{\infty}^3}}\right)+c_2  r^{7/8}J_{-7/24}\left(\frac{\sqrt{-h'(f_{\infty})}r^3}{8n^3\sqrt{10\zeta f_{\infty}^3}}\right)\, .
\end{equation}
In these expressions, $J_k$ and $Y_k$  are Bessel functions of the first and second kinds respectively. Let us note that when the argument of the Bessel functions is real, their asymptotic behaviour is oscillatory, while for imaginary arguments they behave as real growing and decaying exponentials, which is the kind of behaviour we require in order to impose an appropriate boundary condition asymptotically. Noting that $h'(f_{\infty})<0$ for the physical vacuum, we find the condition $\xi+\delta_{\xi,0}\zeta\le 0$, where $\delta_{\xi,0}=1$ if $\xi=0$, and 0 otherwise. Then, by demanding that the solution has the right asymptotic behaviour, we fix one of the integration constants, as usual.

From \req{asympt2t2}, we see that the boundary metric when $r\rightarrow\infty$ takes the form
\begin{equation}\label{bdryplanar6D}
\frac{^{(5)}ds^2_{\infty}}{r^2}= \frac{f_{\infty}}{L^2}\left(d \tau+ n A \right)^2+\frac{d\eta_1^2 +d\zeta_1^2+d\eta_2^2 +d\zeta_2^2}{L^2}\, ,
\end{equation}
We also can determine the ADM energy, which is proportional to the coefficient of the $1/r^3$ term in \req{asympt2t2},
\begin{equation}\label{ADM6Dp}
E_{\rm ADM}=-\frac{l_1^2l_2^2  C_{\mathbb{T}^2\times\mathbb{T}^2}}{8\pi G L^4\sqrt{f_{\infty}}}\, .
\end{equation}
Let us now explore the regularity conditions which as usual will fix the remaining integration constant of the solution. 

\subsubsection{Taub-NUT solutions}
We start with NUT solutions which, as usual, are determined by the conditions $V_{\mathbb{T}^2\times\mathbb{T}^2}(n)=0$ and $V_{\mathbb{T}^2\times\mathbb{T}^2}'(n)=4\pi T$.
In the case of Einstein-Gauss-Bonnet gravity the solutions are forced to be extremal, \ie $T=0$ --- see \req{betaEin2} for the Einstein case. Here we find that this is also the only possibility if we want the solutions to reduce to those of EGB. Setting $ T=0$ and plugging an expansion of  $V_{\mathbb{T}^2\times\mathbb{T}^2}(r)$ around $r=n$ into the field equations, we obtain
\be
C_{\mathbb{T}^2\times\mathbb{T}^2}=-\frac{32 n^5}{ L^2}\, ,
\ee
and an infinite number of equations that fix the coefficients $a_{i}$ as functions of the couplings $\lambda_{\rm \ssc GB}$, $\xi$, $\zeta$. For example, the first two of them read
\begin{eqnarray}
90-36a_2 L^2+18a_2^2L^4 \lambda_{\rm  \ssc GB}+a_2^4L^8(\zeta-6\xi)&=&0\, ,\\
18+a_2^4L^8(66\xi-17\zeta)+6n L^2 a_3\left(3 -3a_2 L^2 \lambda_{\rm \ssc GB}+a_2^3L^6(18\xi+13\zeta)\right)&=&0\, .
\end{eqnarray}
Note that in this case there are not free parameters and the full series is univocally determined once we choose one of the roots of the first equation, which should be the one that reduces to the Einstein gravity one when $\lambda_{\rm  \ssc GB}$, $\xi$, $\zeta\rightarrow 0$. Therefore, at least near $r=n$ we have solved the equation by using the series expansion. However, when $\xi$, $\zeta\neq 0$ one also has to make sure that the asymptotic behaviour is the correct one. In this case it is unclear whether the solution constructed from the near-horizon expansion satisfies this property. Were this not the case, it would imply that no NUT solutions exist when we include the quartic corrections. 
In any case, assuming the solution exists globally, it is illustrative to compute the free energy as $F=T I_{E}$, the result being
\be
F=\frac{2l_1^2l_2^2 n^5}{3\pi G L^6\sqrt{f_{\infty}}}\, .
\ee
On the other hand, the ADM energy \req{ADM6Dp} reads 
\be
E_{\rm ADM}=\frac{4l_1^2l_2^2 n^5}{\pi G L^6 \sqrt{f_{\infty}}}=F+n\frac{\partial F}{\partial n}\, .
\ee
 As we can see, $n$ acts as a thermodynamical variable and it has to be taken into account when we compute the energy $E$ from the free energy $F$. This is not a coincidence and, as we show below, the same observation is valid for bolt solutions.

\subsubsection{Taub-bolt solutions}

Let us turn now to bolt solutions, which are more interesting. As usual, we assume $V_{\mathbb{T}^2\times\mathbb{T}^2}(\rh)=0$ and $V_{\mathbb{T}^2\times\mathbb{T}^2}'(\rh)=4\pi T$ for certain $\rh>n$. Then, we can Taylor expand the solution as
\begin{equation}
V_{\mathbb{T}^2\times\mathbb{T}^2}(r)=4\pi  T (r-\rh)+\sum_{i=2}^{\infty}(r-\rh)^i a_i
\end{equation}
and the equations of motion fix the value of the integration constant $C_{\mathbb{T}^2\times\mathbb{T}^2}$, the relation between $\rh$, $ T$ and $n$, and all coefficients of the expansion $a_{i>2}$ in terms of $a_2$. As usual, this constant is determined by the boundary condition at infinity and can be found, along with the full $V_{\mathbb{T}^2\times\mathbb{T}^2}(r)$, using numerical methods. We will focus on the thermodynamic properties, which can be obtained analytically.  The bolt radius $\rh$ is determined implicitly by $n$ and $ T$ through the equation
\be\label{rbolt6Dp}
5n^2-5\rh^2+4\rh\pi  L^2 T-\frac{L^8(4\pi T)^4}{18(\rh^2-n^2)^3}\left[3\xi(3\rh^4-10 n^2\rh^2-n^4)+20\zeta n^2(\rh^2+n^2)\right]=0\, ,
\ee
while the integration constant $C_{\mathbb{T}^2\times\mathbb{T}^2}$ is given by
\be\label{GM6Dp}
C_{\mathbb{T}^2\times\mathbb{T}^2}=-\frac{2}{L^2}\left(\frac{5 n^6}{\rh}+15n^4\rh-5n^2\rh^3+\rh^5\right)-\frac{L^6(4\pi T)^4}{9 \rh}\left[-3\xi(n^2+9\rh^2)+20 \zeta n^2\right]\, ,
\ee
from which we can obtain the ADM energy of the solution using \req{ADM6Dp}. Observe that while the Gauss-Bonnet term does not modify \req{rbolt6Dp} or \req{GM6Dp} with respect to the corresponding Einstein gravity expressions, the quartic theories produce important modifications.

The free energy can be computed from the Euclidean action analogously to the rest of cases, and it reads
\be
\begin{aligned}
	F=\frac{l_1^2l_2^2}{144\pi G L^6\sqrt{f_{\infty}}}\bigg[&36\rh^5-120n^2\rh^3+180n^4\rh-9 L^2(\rh^2-n^2)^2(4\pi T)\, ,\\
	&+\frac{L^8(4\pi T)^4\rh}{\rh^2-n^2}\left( 6\xi(n^2+3\rh^2)-40\zeta n^2\right)\bigg]\, .
\end{aligned}
\ee
The analysis now follows the same lines as in Section \ref{SecT2}: since we have an extended phase space, we must introduce the variable $\theta=1/n$, and the free energy should be interpreted as a function of the physical temperature $\hat T= T/\sqrt{f_{\infty}}$ and of $\theta$: $F=F(\hat T,\theta)$.  From this we obtain the entropy $S$ and the potential $\Psi$ as defined in \req{SPsi} (but with respect to $\hat T$):
\begin{eqnarray}
S&=&\frac{l_1^2l_2^2}{4 G L^4}\left[\frac{(\rh^2\theta^2-1)^2}{\theta^4}+L^6\frac{(4\pi \sqrt{f_{\infty}} \hat T)^3\rh}{9(\rh^2\theta^2-1)}\left(24\xi(1+3\rh^2\theta^2)-160\zeta \right)\right]\, ,\\
\Psi&=&\frac{l_1^2l_2^2}{\pi G L^6\sqrt{f_{\infty}}}\bigg[\frac{5(3+6\rh^2\theta^2-\rh^4\theta^4)}{12\rh\theta^7} \nonumber \\ 
&+&L^8\frac{(4\pi \sqrt{f_{\infty}} \hat T)^4}{72\rh\theta^3(\rh^2\theta^2-1)}\left(3\xi(1+11\rh^2\theta^2)-20\zeta(1+2\rh^2\theta^2)\right)\bigg]\, ,
\end{eqnarray}
where we used \req{rbolt6Dp} in order to simplify these expressions.
Using the expression for the ADM energy \req{ADM6Dp} and \req{GM6Dp}, we check that $E_{\rm ADM}=F+\hat TS+\theta \Psi$, and hence we obtain the first law
\begin{equation}
dE_{\rm ADM}=\hat T dS+\theta d\Psi\, .
\end{equation}

 To ensure consistency of the Smarr relation, we consider $\Lambda$ and the couplings as thermodynamic parameters and find that the following potentials satisfy the extended first law
\begin{equation}
V = \frac{l_1^2l_2^2 \rh (3 \rh^4\theta^4 - 10 \rh^2 \theta^2 + 15)}{15 L^4 \theta^4} \, , \quad \Upsilon^{\cal Z} =  -\frac{5 l_1^2l_2^2 \rh (4 \pi T)^4}{18 L^4 \pi G (\rh^2\theta^2-1)} \, , \quad \Upsilon^{\cal S} = \frac{l_1^2 l_2^2 \rh (3 \rh^2 \theta^2 + 1)}{24 \pi G L^4 (\rh^2 \theta^2 - 1)} \ .  
\end{equation}
With these thermodynamic potentials, the Smarr formula that follows from scaling holds --- this is of the same form as Eq.~\eqref{smarrCP}, but now we must include an additional $4\theta\Psi$ term.

Even though it is known the first law should hold in general theories, it is remarkable that this can be explicitly checked in these very non-trivial examples.






\section{Concluding Remarks}\label{conclusions}

In this paper we have constructed new Taub-NUT and Bolt solutions for several  higher-curvature gravities for various base spaces in $D=4$ and $D=6$. In particular, the solutions constructed in Section \ref{ECGs} for Einsteinian cubic gravity are the first examples of four-dimensional  higher-curvature generalizations of the Einstein gravity Taub solutions. In all cases, the solutions generalize the Einstein gravity (and Gauss-Bonnet) ones, and reduce to them as the  higher-curvature couplings are set to zero. Also in all cases, and in analogy to the new classes of black holes constructed in \cite{Hennigar:2016gkm,PabloPablo2,Hennigar:2017ego,PabloPablo3,Ahmed:2017jod,Feng:2017tev,PabloPablo4}, the solutions are always characterized by a single base-space-dependent metric function $V_{\mathcal{B}}(r)$. Even though we cannot compute this function analytically in most cases, the thermodynamic properties of the solutions can be accessed in a fully analytic form, as we have shown. As we have seen, turning on the  higher-curvature couplings notably modifies the structure and thermodynamic properties of the solutions with respect to the Einstein gravity case. In particular, they typically modify the multiplicity of solutions as the NUT charge is varied --- \eg new bolt solutions exist for values of $n$ that are forbidden in Einstein gravity ---  and drastically modify the phase spaces --- \eg for NUT solutions with $\mathcal{B}=\mathbb{CP}^{1,2}$, the free energy generically diverges as $n\rightarrow 0$, instead of going to zero like it happens for Einstein gravity. Remarkably, the phase space of $D=6$ solutions with $\mathcal{B}=\mathbb{CP}^{2}$ present re-entrant phase transitions,  these being the first examples of this kind observed for NUT-charged solutions. It would be interesting to better understand this phase structure from the perspective of extended phase space thermodynamics.  

 The cases studied here (see also appendix \ref{quarticAp}) are just the simplest possible within the Generalized quasi-topological family. We expect additional solutions of the form \req{FFnut} to exist in $D=4,6$ for even higher curvature theories of this class, and similarly for $D\geq 8$. It would be interesting to construct them. In particular, in $D=4$ one could consider the invariants presented in \cite{PabloPablo4}. It is possible that a closed form expression --- valid for arbitrarily high curvature terms --- for the equation determining the metric function $V_{\mathcal{B}}(r)$, analogous to the one found in \cite{PabloPablo4} for the black hole solutions, can be found. This would allow for a characterization of the solutions for infinitely many theories. 

 One of the motivations for this work, and an obvious application of our results, can be found in the holographic context. There, Taub solutions with $\mathbb{CP}^{\frac{D-2}{2}}$ base spaces generically dominate the semiclassical partition function for holographic theories on a particularly interesting class of squashed spheres \cite{Chamblin:1998pz,Hartnoll:2005yc,Anninos:2012ft,Bobev:2016sap,Bobev:2017asb}. The fact that their thermodynamic properties can be computed analytically makes our solutions particularly appealing from a holographic perspective. In particular, they can be used to study the properties of squashed-sphere partition functions for a class of theories much broader than the one available so far.  
	As it turns out, our results here can be used to identify new universal properties, presumably valid for general CFTs  (holographic or not) \cite{Bueno:2018yzo}.

 Although we have only considered asymptotically AdS solutions here, the corresponding asymptotically flat counterparts can be easily obtained by taking $L\rightarrow\infty$ while keeping the dimensionful higher-curvature couplings (such as $\mu L^4\equiv\hat\mu$) finite. Asymptotically flat Taub-NUT solutions are the main constituents of Kaluza-Klein monopoles, which arise when constructing lower-dimensional solutions from compactifications of higher-dimensional theories \cite{Sorkin:1983ns,Gross:1983hb,Mann:2005gk}. It would be interesting to explore the new solutions from this perspective.


\section*{Acknowledgments}
We wish to thank Nikolay Bobev, Carlos Hoyos, Gabor Kunstatter, Hugo Marrochio and C. S. Shahbazi for useful discussions and comments.
The work of PB was supported by a postdoctoral fellowship from the National Science Foundation of Belgium
(FWO). The work of PAC is funded by Fundación la Caixa through a “la Caixa - Severo Ochoa''
International pre-doctoral grant and partially by the MINECO/FEDER, UE grant FPA2015-66793-P,
and from the “Centro de Excelencia Severo Ochoa'' Program grant SEV-2016-0597. PAC also
thanks the Perimeter Institute “Visiting Graduate Fellows'' program.  The work of R. Hennigar and R.B. Mann was 
supported in part by the Natural Sciences and Engineering Research Council of Canada.

\appendix
\section{Four dimensions: Quartic generalized quasi-topological term}\label{quarticAp}
As shown in \cite{PabloPablo4}, besides ECG, there are infinitely many theories involving terms of arbitrarily high order in curvature which allow for black hole solutions with $g_{tt}g_{rr}=-1$ in four dimensions --- this is also expected to be the case in higher dimensions. Hence, it is reasonable to expect that some of these theories will also possess NUT-charged solutions characterized by a single function $V_{\mathcal{B}}(r)$, \ie of the form \req{FFnut}. In this appendix we show this to be the case when we supplement the ECG action \req{ECG} with a particular quartic term belonging to the GQTG class \cite{Ahmed:2017jod}. In particular, we study how the ECG Taub-NUT solutions with $\mathcal{B}=\mathbb{S}^2$ constructed in section \ref{S2cubic4d} are modified by the introduction of this term.
Let us then consider the Euclidean action
\begin{equation}\label{QuarticT}
	I_{\rm E}=-\frac{1}{16\pi G}\int d^4x \sqrt{|g|}\left[\frac{6}{L^2}+R-\frac{\mu L^4}{8} \mathcal{P}-\frac{\xi L^6}{16}\mathcal{Q} \right]\, ,
\end{equation}
where 
\begin{equation}
	\begin{aligned}
		\mathcal{Q}=&-44R^{abcd }R_{ab }^{\ \ ef }R_{c\ e}^{\ g \ h}R_{d g f h}-5 R^{abcd }R_{ab }^{\ \ ef }R_{ce}^{\ \ gh}R_{d f g h}+5 R^{abcd }R_{abc }^{\ \ \ \ e}R_{f g h d}R^{f g h}_{\ \ \ \ e}\\
		&+24 R^{ab }R^{cd ef }R_{c\ e a}^{\ g}R_{d g f b}\, ,
	\end{aligned}
\end{equation}
is a particular GQTG density.

\begin{figure}[t]
	\centering \vspace{0.2cm}
	\includegraphics[scale=0.6]{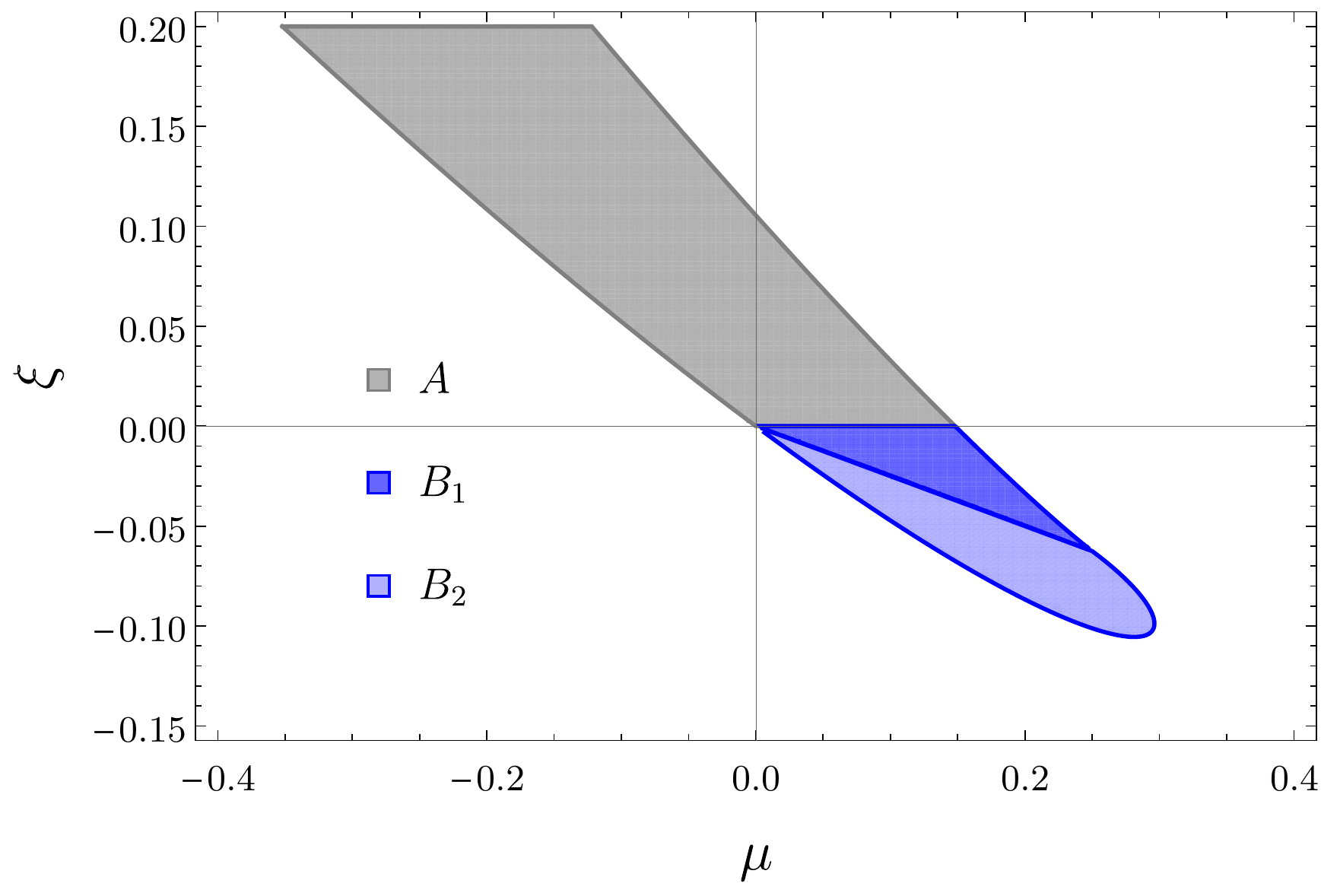}
	\caption{Region of the parameter space for which there is at least one physical AdS vacuum.}
	\label{VacuaQ4}
\end{figure}

Let us start by determining the AdS vacua of \req{QuarticT}. As usual, we write the relation between the action scale $L$ and the AdS radius $\tilde{L}$ as $\tilde L^2=L^2/f_{\infty}$. Then, the possible values of $f_{\infty}$ are determined by the positive roots of the polynomial
\begin{equation}\label{Qvacua}
	h(f_{\infty})\equiv1-f_{\infty}+\mu f_{\infty}^3+\xi f_{\infty}^4=0\, .
\end{equation}
For a given vacuum, the effective gravitational constant can be computed as $G_{\rm eff}=-G/h'(f_{\infty})$. Hence, in order to get a positive energy graviton, we must demand $h'(f_{\infty})<0$, the critical case corresponding to $h'(f_{\infty})=0$. Just like for ECG, there is an additional constraint coming from imposing the existence of positive-energy solutions. This reads $\mu+2f_{\infty}\xi\ge 0$ and, interestingly, it is equivalent to $h''(f_{\infty})\ge 0$ (assuming $f_{\infty}>0$). Therefore, we need to identify solutions to \req{Qvacua} satisfying $f_{\infty}>0$, $h'(f_{\infty})<0$ and $h''(f_{\infty})\ge 0$.  All these conditions bound the space of parameters $(\mu,\,\xi)$, and we can write the allowed set as
\begin{align}\label{cass}
\begin{rcases}
&\mu = \alpha^2(3+\beta)-4\alpha^3\\ 
&\xi = 3\alpha^4-(2+\beta)\alpha^3
\end{rcases}\quad  \text{where} \quad \alpha\ge 0,\, \beta\ge 0,\, 2\alpha-\beta\ge 1\, .
\end{align}
If the parameters belong to this set, there exists at least one AdS vacuum satisfying all the aforementioned constraints with $f_{\infty}=1/\alpha$, $G_{\rm eff}=G/\beta$. Remarkably, we do not find any other allowed vacuum, so in this region of the parameter space the vacuum exists and it is unique. In Fig. \ref{VacuaQ4} we show the region defined by \req{cass}.
It is convenient to divide it into  three different zones. Zone $A$ is the one with $\xi\ge 0$, and in this case the allowed AdS vacua is the second largest real root of $h$. The largest root has $h'(f_{\infty})>0$ so it is not allowed. Zone $B_1$ corresponds to $\xi<0$. There, a third root appears which becomes the largest one. This one has $h'(f_{\infty})<0$ but $h''(f_{\infty})<0$, so it is not suitable. At this point the physical vacuum is the third largest root of $h$. If $\xi$ is negative enough, the two roots larger than the physical one disappear. They coalesce for $(\mu,\, \xi)=(3\alpha^2-4\alpha^3,3\alpha^4-2\alpha^3)$, $0\le\alpha\le 1/2$, in which case there appears a special critical point. This line is the one which separates zones $B_1$ and $B_2$ in Fig. \ref{VacuaQ4}.  Below the line, in zone $B_2$, the physical vacuum is the largest root of $h$, which has interesting consequences for the Taub-NUT solutions, as we will see.

There is a one-parameter family of critical theories, \ie for which $h'(f_{\infty})=0$. We can use $f_{\infty}$ to parametrize the value of the couplings in that case, namely
\begin{equation}
	\mu_{\rm cr}=\frac{3 f_{\infty}-4}{f_{\infty}^3}\, ,\quad \xi_{\rm cr}=\frac{3-2f_{\infty}}{f_{\infty}^4}\, .
\end{equation}
Of course, if we impose $\xi_{\rm cr}$ to be zero, we recover critical ECG, for which $f_{\infty}=3/2$ and $\mu_{\rm cr}=4/27$.

When evaluated on the ansatz \req{FFnut} for $\mathcal{B}=\mathbb{S}^2,\mathbb{T}^2,\mathbb{H}^2$, we find again that the field equations of \req{QuarticT} reduce to a single equation for the function $V_{\mathcal{B}}$. As before, we find that this equation allows for an integrable factor $(1-n^2/r^2)$, and we can write it as in \req{mastere}, namely
%
\begin{equation}
	\label{eqV4}
	\begin{aligned}
		&V \left(\frac{2 n^2}{r}-2 r\right)+\frac{2 \left(k L^2 \left(n^2+r^2\right)-3 n^4-6 n^2 r^2+r^4\right)}{L^2 r}\\
		&+\mu L^4 \Bigg[\frac{6 V^3 n^2 \left(n^2+9 r^2\right)}{r \left(n^2-r^2\right)^3}
		+\left(V'\right)^2 \left(\frac{3 V n^2}{n^2 r-r^3}-\frac{3 k}{2 r}\right)-\frac{\left(V'\right)^3}{2}+V' \left(\frac{3 V^2 \left(17 n^2+r^2\right)}{\left(n^2-r^2\right)^2}+\frac{3 V k}{n^2-r^2}\right)\\
		&+V'' \left(-\frac{3 V^2 \left(4 n^2+r^2\right)}{r^3-n^2 r}+\frac{3 V V'}{2}+\frac{3 V k}{r}\right)\Bigg]+\xi L^6 \Bigg[\frac{V^4 \left(22 n^6+270 n^4 r^2+36 n^2 r^4\right)}{r \left(r^2-n^2\right)^5}\\
		&-\frac{4 V^3 k n^2 \left(n^2+9 r^2\right)}{r \left(n^2-r^2\right)^4}+\left(V'\right)^3 \left(\frac{k}{n^2-r^2}-\frac{V \left(15 n^2+r^2\right)}{2 (n-r)^2 (n+r)^2}\right)+\frac{\left(V'\right)^4 \left(9 n^2+3 r^2\right)}{8 n^2 r-8 r^3}\\
		&+\left(V'\right)^2 \left(\frac{3 V^2 \left(13 n^4+30 n^2 r^2+r^4\right)}{r \left(r^2-n^2\right)^3}-\frac{3 V k \left(n^2+r^2\right)}{r \left(n^2-r^2\right)^2}\right)+V'' \Bigg(\frac{24 V^3 n^2 \left(n^2+r^2\right)}{r \left(r^2-n^2\right)^3}\\
		&-\frac{6 V^2 k n^2}{r \left(n^2-r^2\right)^2}+\frac{3 V \left(V'\right)^2 \left(3 n^2+r^2\right)}{2 \left(r^3-n^2 r\right)}+V' \left(-\frac{3 V^2 \left(11 n^2+r^2\right)}{\left(n^2-r^2\right)^2}-\frac{3 V k}{n^2-r^2}\right)\Bigg)+\\
		&V' \left(-\frac{6 V^3 \left(43 n^4+21 n^2 r^2\right)}{\left(n^2-r^2\right)^4}-\frac{36 V^2 k n^2}{\left(n^2-r^2\right)^3}\right)\Bigg]=4C\, ,
	\end{aligned}
\end{equation}
where $k=+1,0,-1$ for $\mathcal{B}=\mathbb{S}^2$, $\mathbb{T}^2$  and $\mathbb{H}^2$, respectively. 

Let us start by determining the asymptotic behaviour in this case. As usual, we can separate the solution as the sum of a particular solution plus a homogeneous one. The particular solution can be obtained by performing a $1/r$ expansion, which yields
\begin{equation}\label{VpQ}
	V_p(r)=f_{\infty}\frac{r^2}{L^2}+k-5f_{\infty}\frac{n^2}{L^2}+\frac{2C}{h'(f_{\infty})r}+\mathcal{O}(r^{-2})\, ,
\end{equation}
where $h'(f_{\infty})=-1+3\mu f_{\infty}^2+4\xi f_{\infty}^3<0$, according to the unitarity constraint.  From this asymptotic expansion, and using the fact that $G_{\rm eff}=-G/h'(f_{\infty})$ \cite{NewTaub},  we see that for a spherical base space, $C=GM$, where $M$ is the ADM mass \cite{Arnowitt:1960es,Arnowitt:1960zzc}, or more appropriately, the Abbott-Deser energy \cite{Abbott:1981ff, Deser:2002jk, Tekin4, Adami:2017phg}.   For the rest of topologies, $C$ is also proportional to the total energy, but the proportionality constant is different. If we now consider $V(r)=V_p(r)+\frac{r^2}{L^2}g(r)$ and expand linearly in $g$, we obtain the following differential equation keeping only the leading terms when $r\rightarrow\infty$\footnote{For example, we are neglecting a term $g'/r^3$ against $g''/r$.}
\begin{equation}
	-\frac{3L^2 C h''(f_{\infty})}{2h'(f_{\infty}) r}g''(r)+2 h'(f_{\infty}) g(r)=0\, .
\end{equation}
Just like for ECG, the solution is again given in terms of Airy functions,
\begin{equation}\label{asymptg}
	g(r)=A \textrm{AiryAi}\left[\left(\frac{4h'(f_{\infty})^2}{3L^2 C h''(f_{\infty})}
	\right)^{1/3}r\right]+B \textrm{AiryBi}\left[\left(\frac{4h'(f_{\infty})^2}{3L^2 C h''(f_{\infty})}
	\right)^{1/3}r\right]\, ,
\end{equation}
and the analysis is analogous. If $C h''(f_{\infty})>0$ there is a growing mode and a decaying one, so by eliminating the former we obtain an asymptotically AdS solution. If $C h''(f_{\infty})<0$ all solutions except the trivial one are pathological at infinity. Then, in order to ensure the existence of solutions of positive mass, $C>0$, we demand that $ h''(f_{\infty})>0$, which is the constraint anticipated before. 

\subsection{$\mathcal{B}=\mathbb{S}^2$}

From this point on, we focus on the case $\mathcal{B}=\mathbb{S}^2$. Then, the Taub-NUT metric takes the form \req{taub}, where $V_{\mathbb{S}^2}(r)$ satisfies \req{eqV4} with $k=1$. As usual, the period of $\tau$ is fixed to $\beta_{\tau}=8\pi n$, which removes the Dirac-Misner string.
\subsubsection*{Taub-NUT solutions}
Assuming $V_{\mathbb{S}^2}(n)=0$ and the regularity condition $V_{\mathbb{S}^2}'(n)=1/(2n)$, we can write an expansion around $r=n$ as
\begin{equation}
	V(r)=\frac{r-n}{2n}+\sum_{i=2}^{\infty} (r-n)^i a_i\, .
	\label{near2}
\end{equation}
If we introduce this expansion in \req{eqV4}, we obtain a series of relations that must be satisfied order by order in $(r-n)$. From the first one we read the mass of the solution, which is given by
\begin{equation}\label{mmas}
	GM=n-\frac{4 n^3}{ L^2}-\frac{\mu  L^4}{16 n^3}-\frac{\xi L^6 }{64 n^5}\, .
\end{equation}
Naturally, this generalizes the ECG result \req{masss} and reduces to it for $\xi=0$.
Also analogously to the ECG case, the following term in the expansion gives a relation between $a_3$ and $a_2$ from where we obtain $a_3(a_2)$, the next fixes $a_4(a_2)$, and so on. Therefore, once again, the complete series is determined by a single free parameter which must be chosen so that the condition $B=0$ in \req{asymptg} is met.
	
Let us now compute the Euclidean on-shell action of the solutions. For that, we use the generalized action \req{assd}, where the charge $a^*$ is given in this case by $a^*=(1+3\mu f_{\infty}^2+2\xi f_{\infty}^3)\tilde{L}^2/(4G)$. Then, we can write the full action as
\begin{equation}\label{EuclideanQ}
\begin{aligned}
I_E=-\int \frac{d^4x \sqrt{g}}{16\pi G} \left[\frac{6}{L^2}+R-\frac{\mu L^4}{8} \mathcal{P}-\frac{\xi L^6}{16}\mathcal{Q} \right]
-\frac{a^*}{2 \pi \tilde{L}^2}\int_{\partial \mathcal{M}}d^3x\sqrt{h}\left[K-\frac{2\sqrt{f_{\infty}}}{L}-\frac{L}{2\sqrt{f_{\infty}}}\mathcal{R}\right]\, ,
\end{aligned}
\end{equation}
The evaluation of all terms in \req{EuclideanQ} is analogous to the one performed in detail for ECG in appendix \ref{freeen}. We observe that the divergent terms coming from the various contribution cancel, and we are left with the following fininte answer
\begin{equation}
	I_E=\frac{4\pi}{G}\left[n^2-\frac{2n^4}{L^2}+\frac{\mu L^4}{16 n^2}+\frac{\xi L^6}{128 n^4}\right]\, ,
\end{equation}
which generalizes the ECG result \req{freeee1}.
Taking into account that $\beta=8\pi n$, we can obtain the energy and the entropy $E=\partial I_E/\partial \beta$, $S=\beta E-I_E$. The first exactly coincides with the ADM mass in \req{mmas}, $E=M$, whereas for the entropy we find
\begin{equation}
	S=\frac{4\pi}{G}\left[n^2-\frac{6n^4}{L^2}-\frac{3\mu L^4}{16 n^2}-\frac{5\xi L^6 }{128 n^4}\right]\, ,
\end{equation}
which generalizes the ECG answer \req{entR}.

\subsubsection*{Taub-bolt solutions}
Let us now assume that $V_{\mathbb{S}^2}$ vanishes for some $r=\rh>n$. In order to avoid a conical singularity we demand again that $V_{\mathbb{S}^2}'(\rh)=1/(2n)$, so that $V_{\mathbb{S}^2}(r)$ should be Taylor-expanded as
\begin{equation}
	V_{\mathbb{S}^2}(r)=\frac{r-\rh}{2n}+\sum_{i=2}^{\infty} (r-\rh)^i a_i\, .
	\label{near3}
\end{equation}
Plugging this expansion into \req{eqV4},  we find that the mass of the bolt is given by
\begin{equation}
	GM=\frac{\left(n^2+\rh^2\right)}{2\rh}-\frac{3 n^4+6 n^2 \rh^2-\rh^4}{2L^2 \rh}-\frac{\mu  L^4 (6 n+\rh)}{64n^3 \rh}-\frac{ \xi L^6 \left(9 n^2+16 n \rh+3 \rh^2\right)}{512n^4 \rh \left(\rh^2-n^2\right)}\, ,
\end{equation}
where $\rh$ is implicitly related to $n$ through 
\begin{equation}\label{bolt4Q}
	\begin{aligned}
		&\frac{6(\rh^2-n^2)^2}{L^2\rh^2}-\frac{(\rh^2-n^2)(\rh-2 n)}{ n \rh^2}-\frac{3 \mu  L^4 \left(n^2+n \rh+\rh^2\right)}{8 n^2 \rh^2(\rh^2-n^2) }\\
		&-\frac{L^6 \xi  \left(9 n^4+48 n^3 \rh+30 n^2 \rh^2+16 n \rh^3+\rh^4\right)}{128 n^4 \rh^2 \left(\rh^2-n^2\right)^2}=0\, .
	\end{aligned}
\end{equation}
Just as for ECG, the rest of equations fix the coefficients $a_{i>2}$ in terms of $a_2$, which must be chosen so that the solution is asymptotically AdS, condition which selects a unique value of $a_2$.

\begin{figure}[t]
	\centering 
	\includegraphics[scale=0.47]{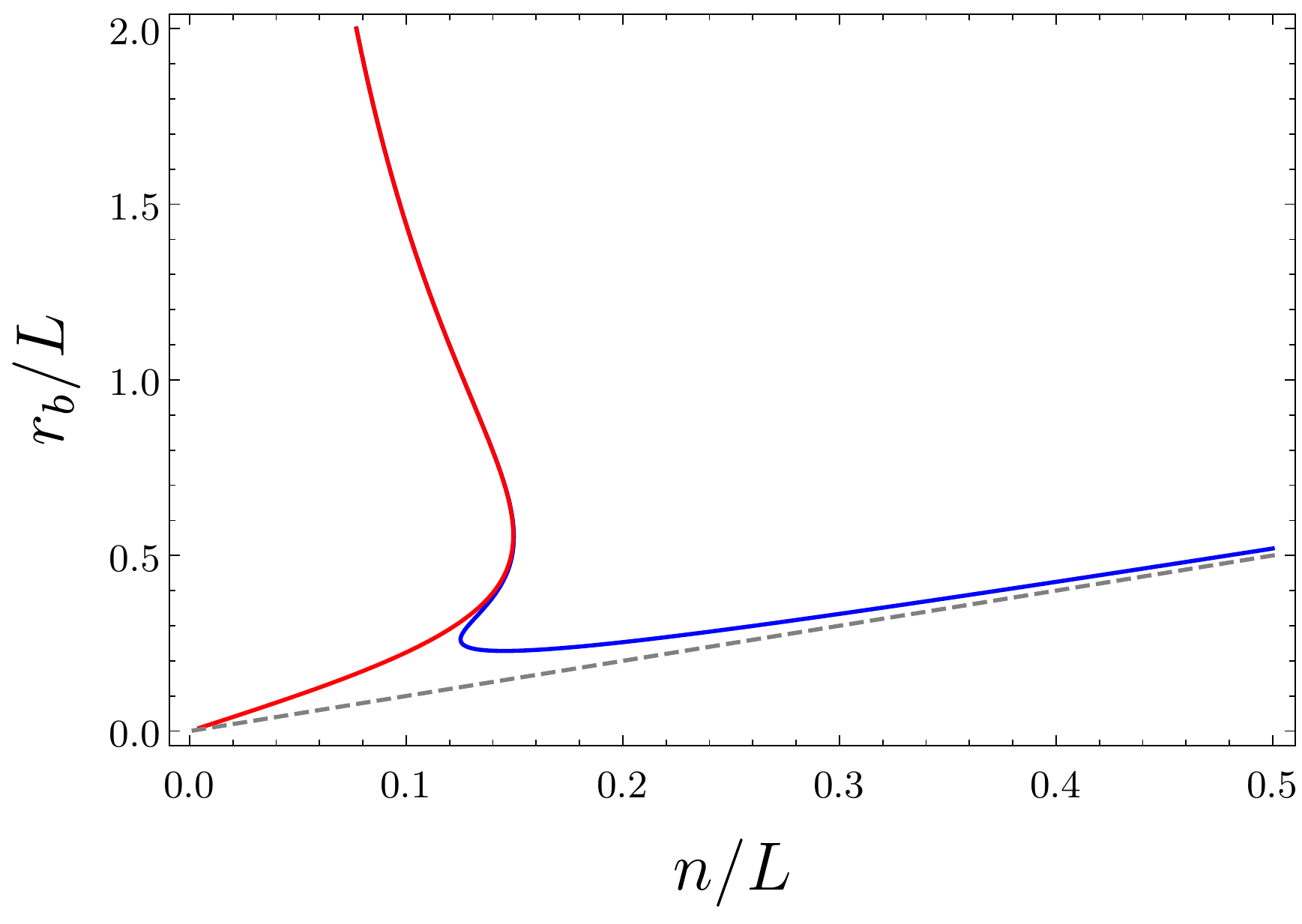}
	\includegraphics[scale=0.47]{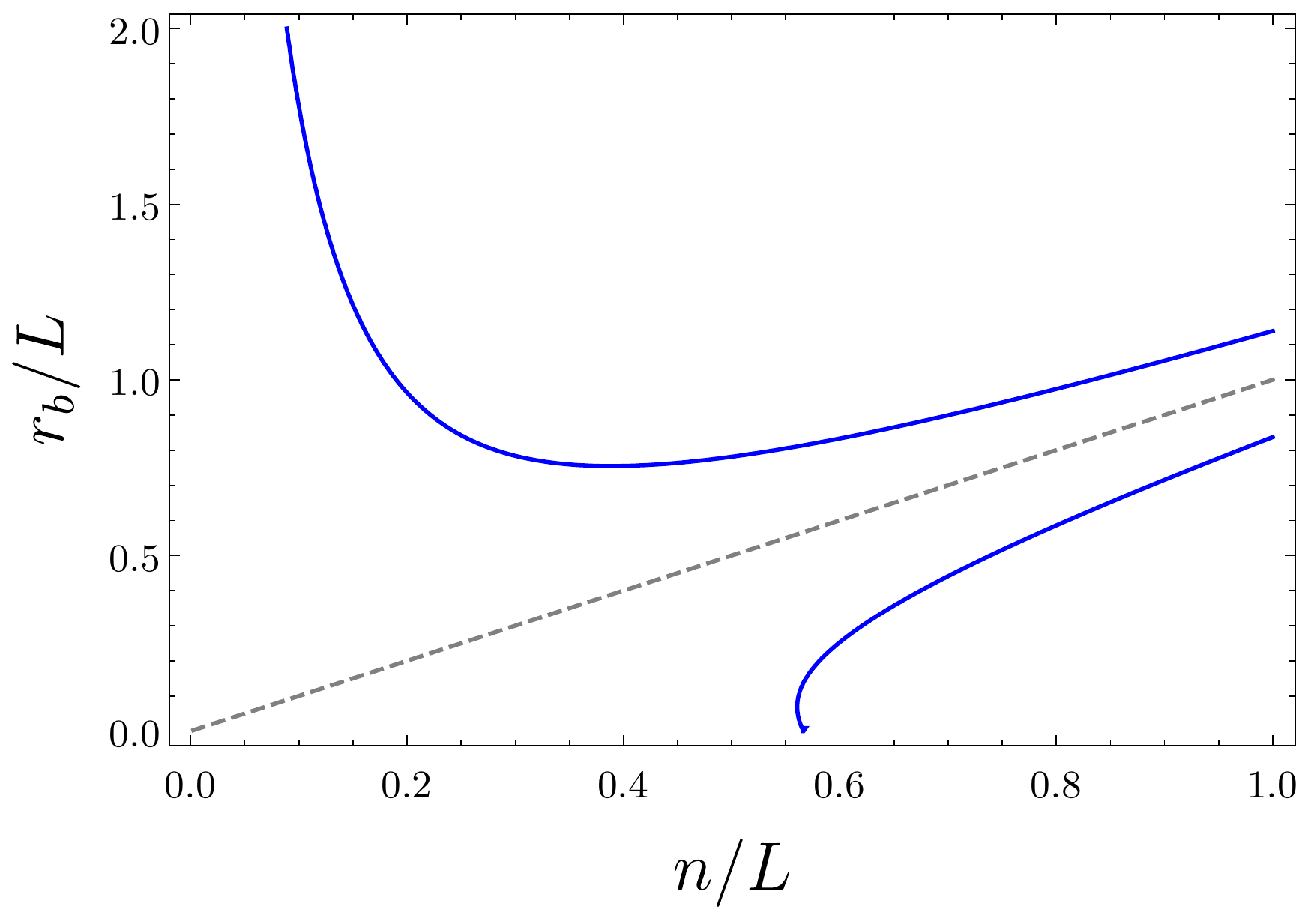}
	\includegraphics[scale=0.47]{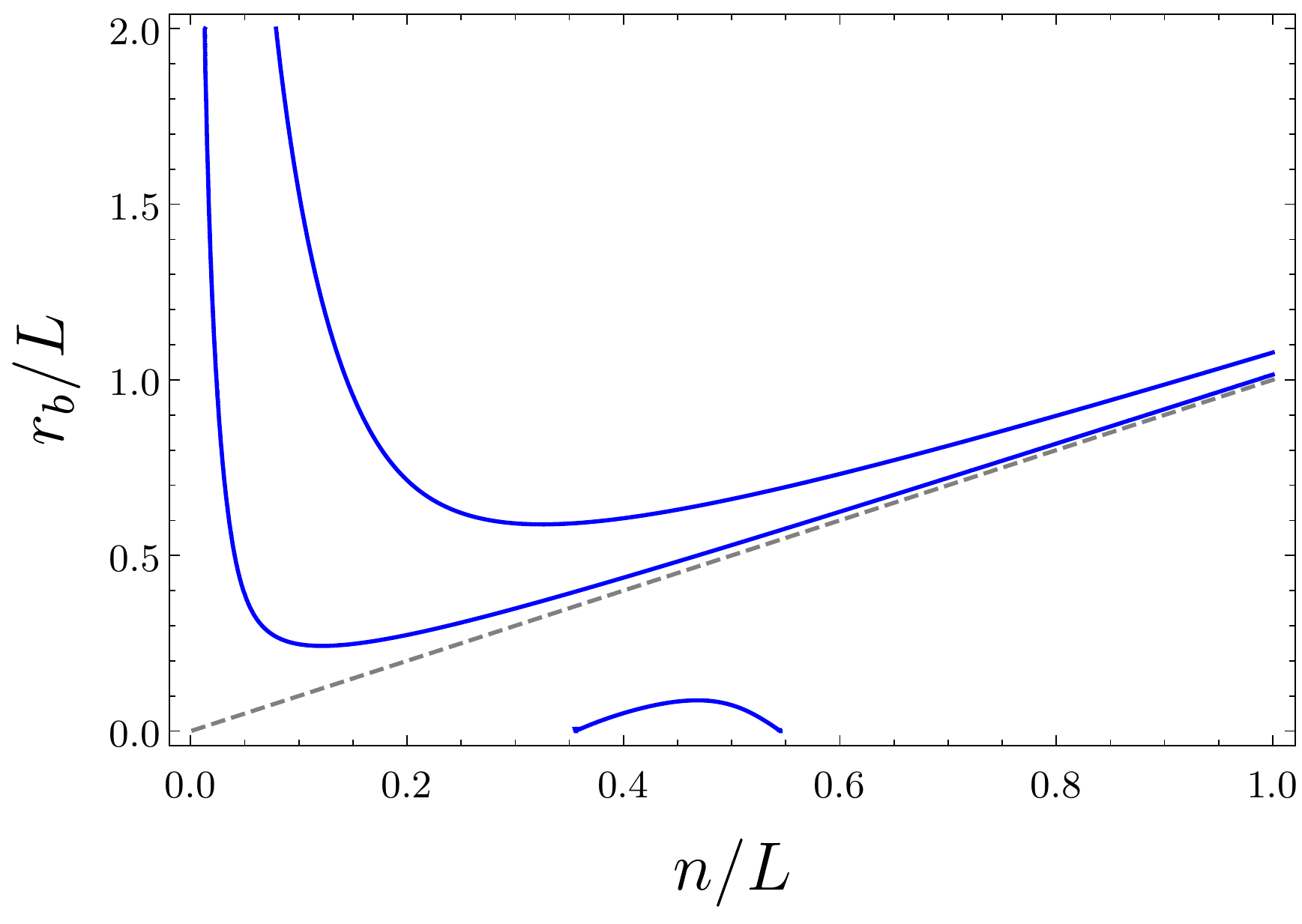}
	\includegraphics[scale=0.47]{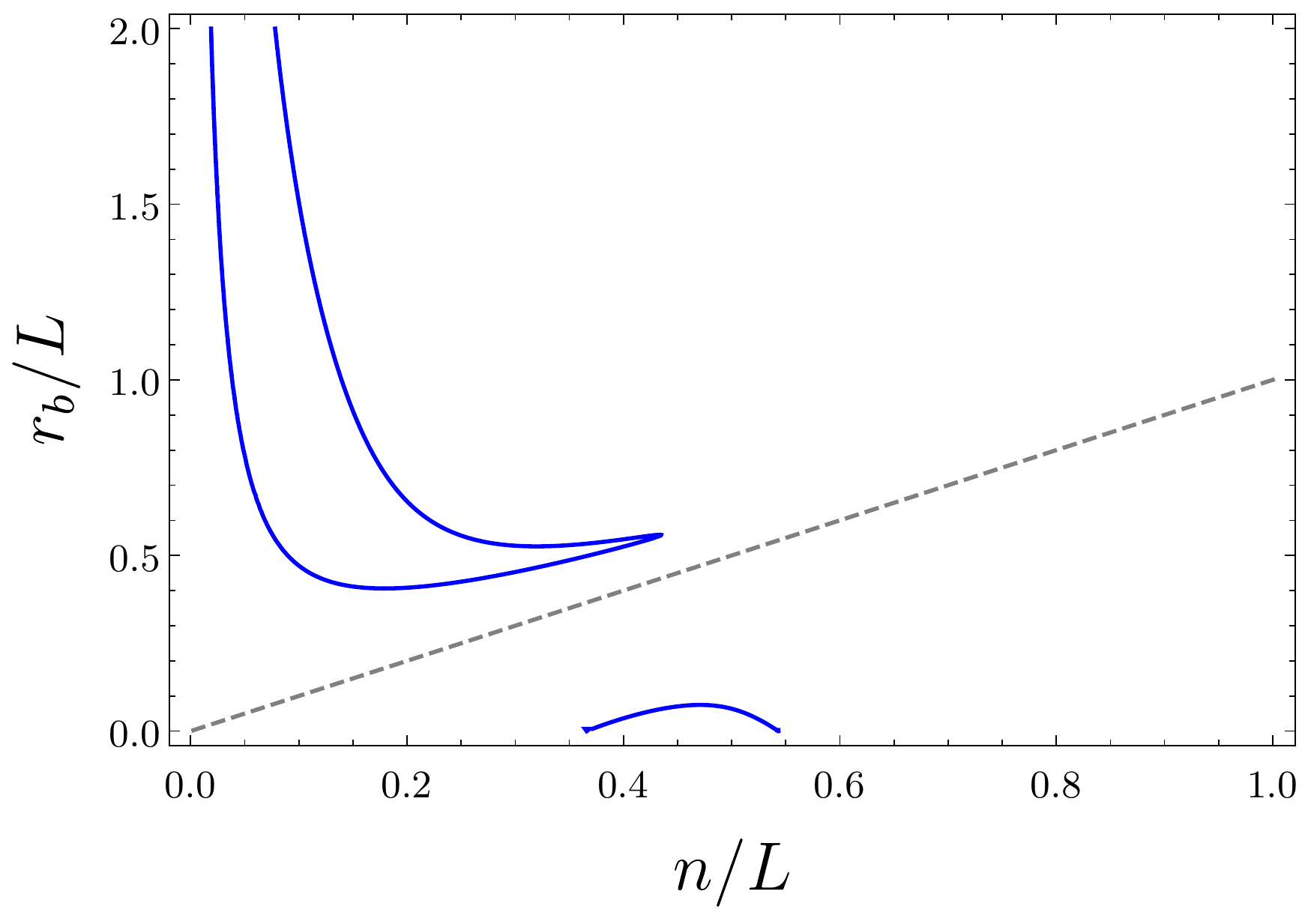}
	\caption{Roots of the equation \req{bolt4Q} for several values of the parameters. Only the roots above the reference dashed line $\rh=n$ admit the construction of bolt solutions. Upper left: behaviour in region $A$ when the parameters are very small ($\mu=\xi=10^{-5}$ in this case); there is a range of $n$ with three different bolt solutions. Upper right: $\mu=0.05$, $\xi=0.05$; this represents the typical case for region $A$. Lower left: region $B_1$ ($\mu=0.05$, $\xi=-0.002$); there are two roots for every value of $n$. Lower right: region $B_2$ ($\mu=0.05$, $\xi=-0.0052$); there are two roots if $n$ is smaller than certain value; both diverge as $n\rightarrow 0$. }
	\label{boltq4}
\end{figure}

The roots of \req{bolt4Q} behave in different ways depending on the values of the parameters. We can characterize several qualitative features depending on the region of the parameter space shown in Fig. \ref{VacuaQ4}. First, recall that in the case of Einstein gravity, this is, $\mu=\xi=0$, there are two allowed roots  when $n/L<\left((2-\sqrt{3})/12 \right)^{1/2}$ --- see \req{Einss} --- and no solutions otherwise. One of the roots goes to zero for $n\rightarrow 0$ and the other one diverges. When $\mu\neq 0$ or $\xi\neq 0$ there is no root going to $0$ for $n\rightarrow 0$. In fact, in this limit we can expand $\rh$ as
\begin{equation}
	\rh=\frac{c_0L^2}{n}+c_1 n+\mathcal{O}(n^3)\, , \quad \text{with} \quad c_0^3(6c_0-1)=\frac{\xi}{128}\, ,\quad c_1=\frac{256 c_0^3-48c_0^2+\mu}{8 c_0^2(8c_0-1)}\, ,
\end{equation}
where we must demand $c_0>0$. The first equation gives us some information about the roots, depending on the region. If $\xi>0$, there is a unique value of $c_0$, so there is a single solution for $n\rightarrow 0$. Indeed, we observe that there is a unique branch in the diagram $(\rh,n)$ if $\xi>0$ and that there is a solution for every value of $n$, including large values. When $-1/16<\xi<0$, there are two different roots $c_0$, so there are two different solutions for $n\rightarrow 0$. We see that if $\xi\in B_1$, then these solutions extend to every $n$, while for $\xi\in B_2$, the solutions only exist for $n$ smaller than certain value. Finally, if $\xi<-1/16$, we find that there are no bolt solutions. In Fig. \ref{boltq4} we summarize the different possibilities.

In all possible cases in which solutions exist, the mass when $n\rightarrow 0$ is given by
\begin{equation}
	GM=\frac{L^4(-256 c_0^3+48 c_0^2-\mu)}{64 n^3}+\mathcal{O}(n)\, ,
\end{equation}
which can be shown to be positive as long as the parameters lie in the allowed region.

Then, the Euclidean on-shell action can be computed along the same lines as in the NUT case using \req{EuclideanQ}. The final result reads
\begin{equation}
		I_E=\frac{\pi}{G}\Bigg[\frac{ 4 n \rh \left(\rh^2-3 n^2\right)}{L^2}+n^2+4 n \rh-\rh^2+\frac{\mu  L^4 \left(5 n^2+12 n \rh+\rh^2\right)}{16n^2 (\rh^2-n^2)}
		+\frac{\xi L^6  \left(8 n^3+9 n^2 \rh+8 n \rh^2+\rh^3\right)}{64n^3 \left(\rh^2-n^2\right)^2}\Bigg]\, ,
\end{equation}
where $\rh$ is a function of $n$ given implicitly by \req{bolt4Q}. In the $n\rightarrow 0$ limit, we can write explicitly
\begin{equation}
	I_E=\frac{\pi L^2}{G}\left[\frac{(256 c_0^3-48 c_0^2+\mu)L^2}{16 n^2}+\frac{3\mu}{4 c_0}+12 c_0(8 c_0-1)+\mathcal{O}(n^2)\right]\, .
\end{equation}

\section{Explicit on-shell action calculation}\label{freeen}
In this appendix we present a explicit calculation of the on-shell action corresponding to the Einsteinian cubic gravity Taub-NUT solution with $\mathcal{B}=\mathbb{S}^2$ presented in Section \ref{S2cubic4d}.

For the configuration \req{taub}, the Lagrangian is a total derivative, so the bulk part of the action can be integrated exactly,
\begin{equation}
I_{\rm bulk}=-\frac{4\pi \beta}{16 \pi G}\int_{n}^{L^2/\delta}dr(r^2-n^2)\mathcal{L}=-\frac{\beta}{4G}F(r)\Big|_{n}^{L^2/\delta}\, ,
\end{equation}
introducing a UV cutoff $\delta$, 
where $\beta=8\pi n$ is the periodicity of the Euclidean time and 
\begin{equation}
\begin{aligned}
F(r)&=\left(n^2-r^2\right) V'(r)-2 r V(r)+\frac{2 r \left(L^2-3 n^2+r^2\right)}{L^2}\\
&+\mu L^4 \Bigg[\left(-\frac{6 n^2 V(r)}{\left(n^2-r^2\right)^2}-\frac{21 n^2 \left(n^2+r^2\right) V(r)^2}{\left(n^2-r^2\right)^3}\right) V'(r)-\frac{\left(5 n^2+r^2\right) V'(r)^3}{4 n^2-4 r^2}\\
&-\frac{6 n^2 r \left(5 n^2+r^2\right) V(r)^3}{\left(n^2-r^2\right)^4}-\frac{6 n^2 r V(r)^2}{\left(n^2-r^2\right)^3}+\left(\frac{3 r}{2 \left(r^2-n^2\right)}-\frac{3 \left(9 n^2 r+r^3\right) V(r)}{2 \left(n^2-r^2\right)^2}\right) V'(r)^2\Bigg]\, .
\end{aligned}
\end{equation}
 Using \req{rasymp} to compute $F(L^2/\delta)$, we find
\begin{equation}
\begin{aligned}
I_{\rm bulk}=\frac{2\pi n}{G}\left[F(r\rightarrow n)-\left(\frac{2L^4}{\delta^3}-\frac{6 n^2}{\delta}\right)\left(1-2 f_{\infty}-2 f_{\infty}^3 \mu\right)-2G_{\rm eff}M\left(1+3 f_{\infty}^2 \mu\right)+\mathcal{O}(\delta/L^2)\right]\, .
\end{aligned}
\end{equation}
Now, for the boundary contributions, we use the trace of the extrinsic curvature at $r=L^2/\delta$, and the Ricci scalar of the induced metric, respectively given  by
\begin{equation}\label{K4D}
K=\frac{2(L^2/\delta)}{(L^2/\delta)^2-n^2}V(L^2/\delta)^{1/2}+\frac{1}{2}\frac{ V'(L^2/\delta)}{V(L^2/\delta)^{1/2}}\, , \quad 	\mathcal{R}=\frac{2(L^2/\delta)^2-2(1+V(L^2/\delta))n^2}{(n^2-(L^2/\delta)^2)^2}\, .
\end{equation}
Then, using the asymptotic expansion \req{rasymp} we find the boundary contribution
\begin{equation}
I_{\rm boundary}=\frac{2\pi n}{G}\left(1+3\mu f_{\infty}^2\right)\left[-f_{\infty}\left(\frac{2L^4}{\delta^3}-\frac{6 n^2}{\delta}\right)+ 2 G_{\rm eff}M\right]+\mathcal{O}(\delta/L^2)\, .
\end{equation}
Adding up bulk and boundary contributions, we find 
\begin{equation}
I_E=\frac{2\pi n}{G}\left[F(r\rightarrow n)-\left(\frac{2L^4}{\delta^3}-\frac{6 n^2}{\delta}\right)\left(1- f_{\infty}+ f_{\infty}^3 \mu\right)+\mathcal{O}(\delta/L^2)\right]=\frac{2\pi n}{G}F(r\rightarrow n)+\mathcal{O}(\delta/L^2)\, ,
\end{equation}
where in the last equality we used the defining equation of $f_{\infty}$, \req{finn}. Remarkably, all contributions coming from the boundary cancel out, including constant terms. Finally, taking the limit $\delta\rightarrow 0$ and using the expansion \req{near}, we are left with the simple result
\begin{equation}\label{freeee}
I_E=\frac{4\pi}{G}\left[n^2-\frac{2n^4}{L^2}+\frac{\mu L^4}{16 n^2}\right]\, .
\end{equation}

\section{Remarks on numerical methods}\label{methods}

We have presented in our investigation a number of numerical solutions for the NUT and bolts. Here we provide some details on how these solutions were obtained.  The differential equations solved here are in general stiff, which results in difficulties in the numerical scheme. All numerical solutions presented in this work were obtained using Mathematica, utilizing the \emph{ImplicitRungeKutta} method of \emph{NDSolve}.  This method satisfies A-stability, making it a suitable method for stiff differential equations. High \emph{WorkingPrecision} was used in the numerical solver, ranging between 20 and 50 on a case by case basis.

\begin{figure}[t] 
\centering
\includegraphics[scale=0.47]{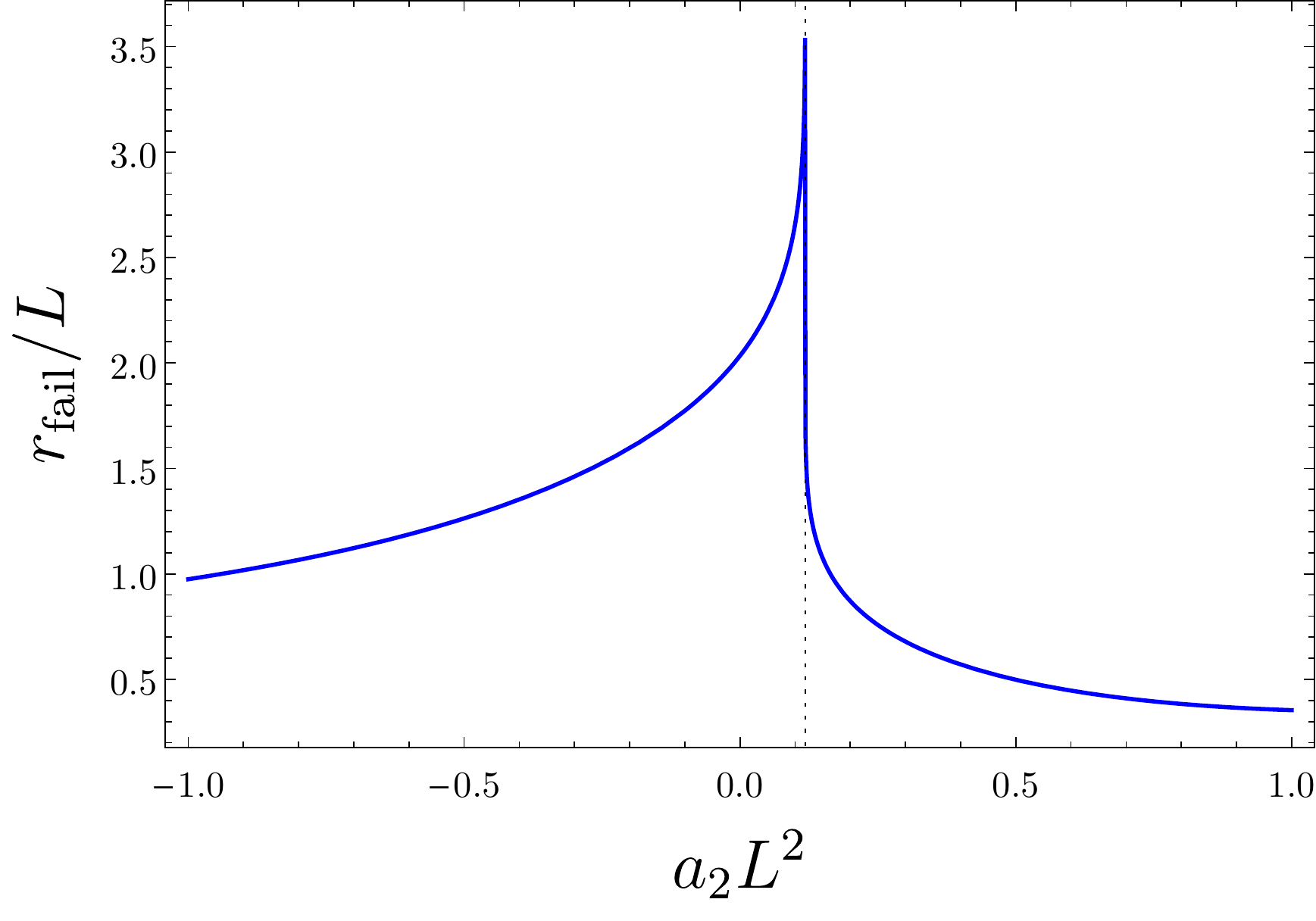}
\quad
\includegraphics[scale=0.47]{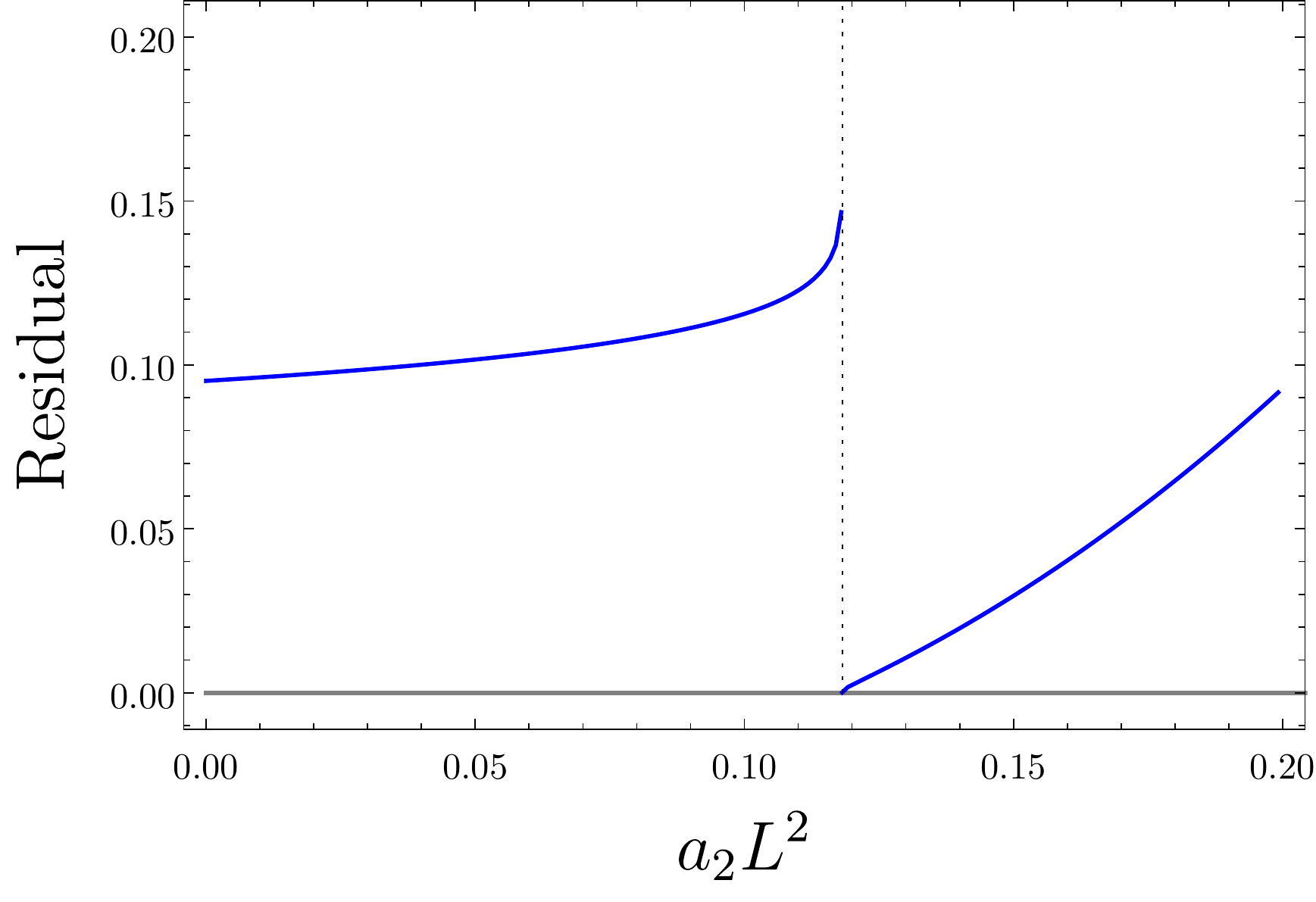} 
\caption{Left: A plot of $r_{\rm fail}$ the $r$ value at which the numerical scheme fails vs.~the shooting parameter $a_2$. Right: A plot of the residual as a function of the shooting parameter $a_2$. In both plots, the dotted line corresponds to $a_2 L^2 = 0.1181855186708097$. Both plots are for the case of NUT solutions in the quartic generalized quasi-topological theory for $\mathcal{B} = \mathbb{CP}^2$ with $\xi = -10$, $n/L = 1/3$ and $\epsilon = 10^{-2} L$.  A working precision of 40 was used in producing these particular plots. }
 \label{numerics-a2}
\end{figure}

Let us make some remarks on the details of the numerical scheme, focusing on the $\mathcal{B} = \mathbb{CP}^k$ bases. The metric function $V_{\mathcal{B}}(r)$ was expanded near a NUT or a bolt as 
\be\label{numericV}
V_{\mathbb{CP}^k}(\epsilon) = \frac{\epsilon}{(k+1)n} + a_2 \epsilon^2 \, ,
\ee
where $\epsilon = (r-n)$ for the NUTs or $\epsilon = (r-\rh)$ for the bolts is taken to be some small, positive quantity --- typically $10^{-2}L-10^{-3}L$ in this work. The parameter $a_2$ is not fixed by the near horizon solution, and must be determined via the shooting method. Specifically, for a given choice of $a_2$, Eq.~\eqref{numericV} is used to generate initial data for the differential equation, namely $V_{\mathbb{CP}^k}(\epsilon)$ and ${V'}_{\mathbb{CP}^k}(\epsilon)$.  Finding a numerical solution then reduces to finding a sensible value of $a_2$. 

\begin{figure}[t]
\centering 
\includegraphics[scale=0.65]{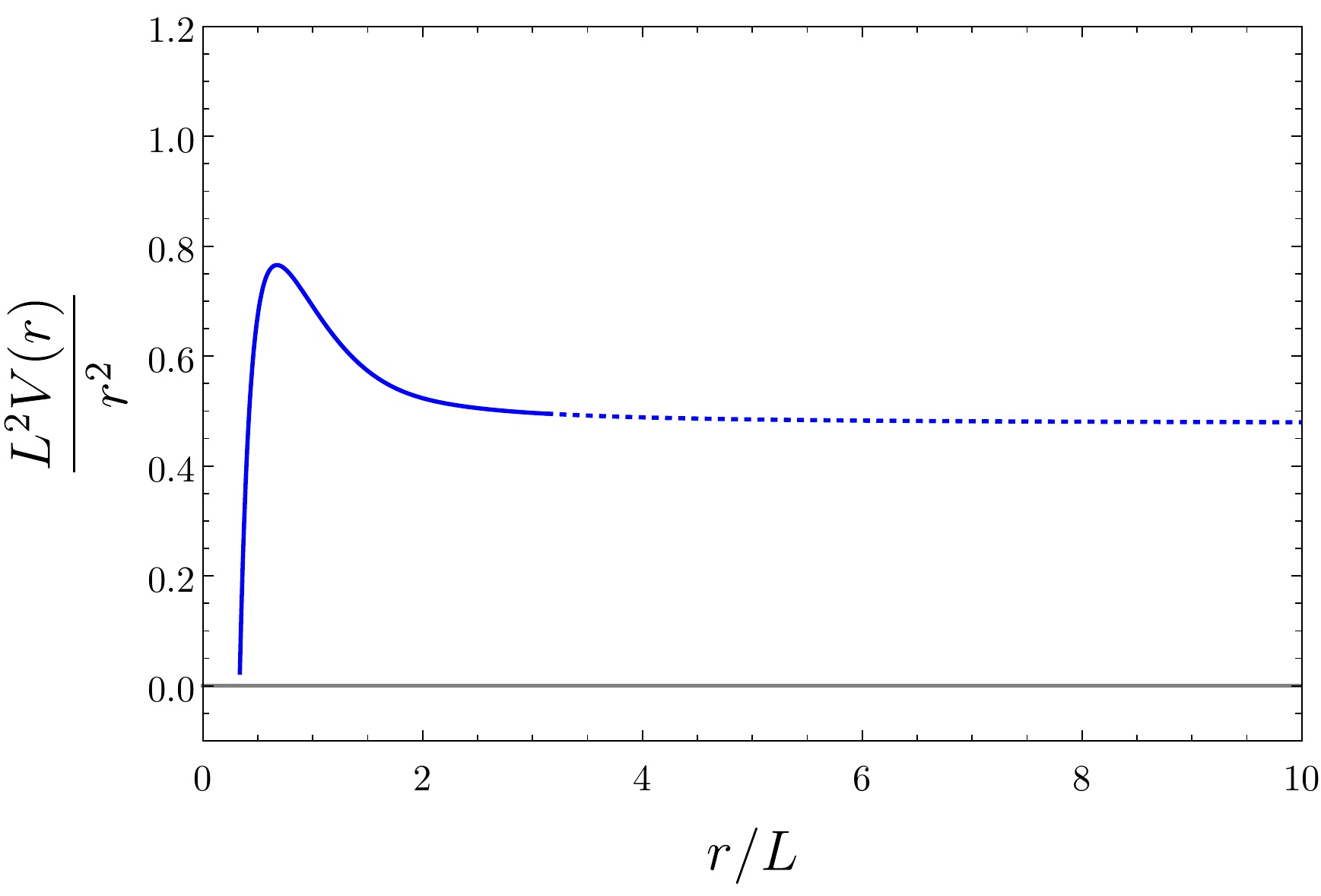} 
\caption{An example of a numeric solution. Here the solid blue curve corresponds to the result of the numeric integration, while the dotted curve corresponds to the asymptotic expansion. The plot is for the case of NUT solutions in the quartic  quasi-topological theory for $\mathcal{B} = \mathbb{CP}^2$ with $\zeta = -10$, $n/L = 1/3$, $a_2 L^2 = 0.1181855186708097$ and $\epsilon = 10^{-2} L$.  A working precision of 40 was used in producing this plot. }
\label{numericEx}
\end{figure}

A generic choice of $a_2$ will lead to the excitation of the growing modes that appear in the asymptotic expansion of the metric function. The correct choice of $a_2$ will result in a numerical solution that approaches the $1/r$ part of the asymptotic expansion at sufficiently large $r$.  Regardless of the choice of $a_2$, the numerical scheme will eventually breakdown because of the accumulation of errors due to finite working precision. It is useful to study the point at which the numerical solution fails as a function of the shooting parameter $a_2$ --- an example of this is shown in the left plot of Fig.~\ref{numerics-a2}. This figure makes clear that there is a special value of $a_2$ that allows the solution to be integrated the furthest. It also appears that this is the unique value of $a_2$ that joins the numerical solution smoothly onto the asymptotic expansion --- see Fig.~\ref{numericEx}.

While with the proper choice of $a_2$ the solution can be visually seen to join onto the asymptotic expansion smoothly, it is nice to have quantitative confirmation of this. In the right plot of Fig.~\ref{numerics-a2} we show a residual that measures how closely the numerical solution matches the asymptotic expansion in a region where they overlap. The residual shown was calculated according to
\be 
{\rm Residual} = \int_{0.9 r_{\rm fail}}^{r_{\rm fail}} \frac{L \big|  V_{\rm numeric}(r) - V_{1/r}(r) \big|}{r^2}  dr \, ,
\ee
where again $r_{\rm fail}$ is the point at which the numerical solution breaks down. In performing this calculation, terms up to ${\cal O}(r^{-3})$ where included in the asymptotic expansion. The plot shows that the error blows up $a_2 L^2 = 0.1181855186708097$ is approached from the left, while it goes to zero when approached from the right. This confirms that the numerical solution is indeed becoming arbitrarily close to the asymptotic solution, and the asymptotic solution can be used to continue the solution to infinity.

On the contrary, there are some regions in the parameter space (for example, when the mass is negative in $D=4$) for which we argued that the solutions do not exist due to a bad asymptotic behaviour. In those cases we are not able to match the numerical solution with the asymptotic expansion, and this confirms that those solutions do not exist. 

Several strategies may be used in order to improve the precision of the numerical methods. For example, instead of working with the function $V(r)$ one may work with $f(r)=L^2 V(r)/r^2$, which should approach the constant $f_{\infty}$ at infinity. Also, more terms can be included in the expansion (\ref{numericV}), so that one does not need to choose a very small $\epsilon$ (we recall that the full expansion depends only on $a_2$).  Let us close by mentioning that the numerical problem is considerably more stiff in $D=6$ than in $D=4$. In the latter case we do not require to increase substantially the \emph{WorkingPrecision} and the parameter $\epsilon$ can be chosen as small as $10^{-3}L$.  The numerical integration in $D=6$ is less stable and requires a larger value of $\epsilon$ and higher values of \emph{WorkingPrecision}.

\renewcommand{\leftmark}{\MakeUppercase{Bibliography}}
\phantomsection
\bibliographystyle{JHEP}
\bibliography{Gravities}
\label{biblio}

\end{document}